\documentclass[11pt,oneside,letterpaper]{article}
\usepackage{jheppub} % for details on the use of the package, please
\usepackage{amsfonts}
\usepackage{amsmath}
\allowdisplaybreaks[4]        
\usepackage{amssymb}
\usepackage{euscript}     
\usepackage{xcolor}         
\usepackage{tensor}        
\usepackage{amsthm}
\usepackage{graphicx}
\usepackage{subfigure}
\usepackage{bbm}
\usepackage[header,title,page,titletoc]{appendix}  
\usepackage[numbers]{natbib}  
\usepackage{float}

\newcommand{\be}{\begin{equation}}
\newcommand{\bea}{\begin{eqnarray}}
\newcommand{\eea}{\end{eqnarray}}
\newcommand{\ba}{\begin{array}}
\newcommand{\ea}{\end{array}}
\newcommand{\ee}{\end{equation}}
\newcommand{\bes}{\begin{equation*}}
\newcommand{\beas}{\begin{eqnarray*}}
\newcommand{\eeas}{\end{eqnarray*}}
\newcommand{\bas}{\begin{array*}}
\newcommand{\eas}{\end{array*}}
\newcommand{\ees}{\end{equation*}}

\title{\boldmath Generalized Volume-Complexity For Two-Sided Hyperscaling Violating Black Branes}

\author[]{Farzad Omidi}

\affiliation[]{School of Physics, Institute for Research in Fundamental Sciences (IPM),\\
	P.O. Box 19395-5531, Tehran, Iran}

\emailAdd{farzad@ipm.ir}

\abstract{
In this paper, we investigate generalized volume-complexity $\mathcal{C}_{\rm gen}$ for a two-sided uncharged HV black brane in $d+2$ dimensions. This quantity which was recently introduced in \cite{Belin:2021bga}, is an extension of volume in the Complexity=Volume (CV) proposal, by adding higher curvature corrections with a coupling constant $\lambda$ to the volume functional. We numerically calculate the growth rate of $\mathcal{C}_{\rm gen}$ for different values of the hyperscaling violation exponent $\theta$ and dynamical exponent $z$. It is observed that $\mathcal{C}_{\rm gen}$ always grows linearly at late times provided that we choose $\lambda$ properly. Moreover, it approaches its late time value from below.
For the case  $\lambda=0$, we find an analytic expression for the late time growth rate for arbitrary values of $\theta$ and $z$. 
However, for $\lambda \neq 0$, the late time growth rate can only be calculated analytically for some specific values of $\theta$ and $z$. We also examine the dependence of the growth rate on $d$, $\theta$, $z$ and $\lambda$. Furthermore, we calculate the complexity of formation obtained from volume-complexity and show that it is not UV divergent. We also examine its dependence on the thermal entropy and temperature of the black brane. At the end, we also numerically calculate the growth rate of $\mathcal{C}_{\rm gen}$ for the case where the higher curvature corrections are a linear combination of the Ricci scalar, square of the Ricci tensor and square of the Riemann tensor. We show that for appropriate values of the coupling constants, the late time growth rate is again linear. 
}
\arxivnumber{2207.05287}
\begin{document} 

\begin{flushright}
  IPM/P-2022/30 \\	
\end{flushright}

\maketitle
\flushbottom

%\newpage
%\tableofcontents

%%%%%%%%%%%%%%%%%%%%%%%%%%%%
\section{Introduction}
\label{Sec: Introduction}
%%%%%%%%%%%%%%%%%%%%%%%%%%%%

In the last two decades, we witnessed that quantum information quantities such as entanglement entropy and quantum complexity have been very fruitful in understanding the AdS/CFT correspondence (See \cite{Nishioka:2009un,Rangamani:2016dms,Susskind:2018pmk} for a review). Quantum complexity is defined as the minimum number of simple unitary operations required to prepare a given state form an initial reference state \cite{Aaronson:2016vto}. One of the intriguing features of
%motivations to study 
quantum complexity is that it continues to grow long after the time at which entanglement entropy saturates \cite{Susskind:2014rva,Stanford:2014jda,Susskind:2014moa,Brown:2015bva,Brown:2015lvg,Susskind:2018pmk}. Therefore, it is widely believed that it can provide information about the interior of black holes. In the context of AdS/CFT \cite{Maldacena:1997re}, there are two main proposals for the calculation of quantum complexity: Complexity=Volume (CV) \cite{Susskind:2014rva,Stanford:2014jda} and Complexity=Action (CA)  \cite{Brown:2015bva,Brown:2015lvg}. It should be pointed out that volume-complexity
%%%%%%%%%%%%%%%%%%%%%%%%%%
\footnote{We call the holographic complexity in the CV proposal {\it volume-complexity}.}
%%%%%%%%%%%%%%%%%%%%%%%%%%
 was also introduced for some subregions in the bulk spacetime such as in a region \cite{Alishahiha:2015rta,Ben-Ami:2016qex} enclosed by a (H)RT hypersurface \cite{Ryu:2006bv,Hubeny:2007xt} or inside a Wheeler-DeWitt (WDW) patch \cite{Couch:2016exn}.
%%%%%%%%%%%%%%%%%%%%%%%%%%%
\footnote{This proposal is also called CV2.0.}
%%%%%%%%%%%%%%%%%%%%%%%%%%%
Various aspects of the CV and CA proposals have been explored extensively in the literature \cite{Susskind:2014jwa,Susskind:2018tei,Cai:2016xho,Yang:2019gce,Carmi:2017jqz,Chapman:2018dem,Yang:2017czx,Auzzi:2018zdu,Auzzi:2018pbc,Chapman:2018lsv,Goto:2018iay,Alishahiha:2018tep,Bernamonti:2019zyy,Caceres:2019pgf,Bernamonti:2020bcf,Jorstad:2022mls,Auzzi:2021nrj,Auzzi:2022bfd,Hashemi:2019aop,Qaemmaqami:2017lzs,Swingle:2017zcd,Belin:2021bga,An:2019opz,Cai:2017sjv,Brown:2016wib,Susskind:2020gnl,Susskind:2019ddc,Brown:2018kvn,Flory:2019kah,Flory:2018akz,ChowdhuryRoy:2022dgo,Mounim:2021bba,Baiguera:2021cba,Avila:2021zhb,Ebrahim:2018uky,HosseiniMansoori:2018gdu}
%%%%%%%%%%%%%%%%%%%%%%%%%%%%
%\footnote{See also \cite{Chapman:2018dem,Auzzi:2018zdu,Auzzi:2018pbc,Chapman:2018lsv,Goto:2018iay,Alishahiha:2018tep,Bernamonti:2019zyy,Caceres:2019pgf,Bernamonti:2020bcf,Jorstad:2022mls,Auzzi:2021nrj,Auzzi:2022bfd,Hashemi:2019aop,Swingle:2017zcd,Belin:2021bga} for recent developments which is not an exhaustive list.}
%%%%%%%%%%%%%%%%%%%%%%%%%%%%
such as: 
%growth rate
%,Alishahiha:2021thv,Akhavan:2018wla
%\cite{Brown:2015lvg,Carmi:2017jqz,Alishahiha:2018tep,Swingle:2017zcd}, 
subregion complexity \cite{Alishahiha:2017cuk,Alishahiha:2018lfv,Abt:2017pmf,Auzzi:2019fnp,Agon:2018zso,Auzzi:2019mah,Auzzi:2019vyh,Auzzi:2021ozb,Braccia:2019xxi,Bakhshaei:2017qud,Borvayeh:2020yip,Bhattacharya:2021jrn,Bhattacharya:2021nqj,Abt:2018ywl,Bhattacharya:2021dnd,Chen:2018mcc}, complexity of formation \cite{Chapman:2016hwi,Carmi:2017jqz,Akhavan:2019zax,Bernamonti:2021jyu}, in Jackiw-Teitelboim gravity \cite{Brown:2018bms,Alishahiha:2018swh,Alishahiha:2022kzc}, for higher derivative gravity \cite{Alishahiha:2017hwg,An:2018dbz,Mandal:2022ztj}, for Kerr-AdS black holes \cite{Bernamonti:2021jyu,Mounim:2021ykr}, at finite radial cutoff \cite{Akhavan:2018wla,Alishahiha:2019lng,Alishahiha:2021thv,Alishahiha:2018swh,Alishahiha:2019cib,Hashemi:2019xeq}, structure of the UV divergent terms \cite{Carmi:2016wjl,Reynolds:2016rvl,Braccia:2019xxi,Omidi:2020oit,Akhavan:2019zax}, regularization methods \cite{Akhavan:2019zax,Omidi:2020oit} and covariant counterterms on spacelike/timelike \cite{Lehner:2016vdi,Carmi:2017jqz,Reynolds:2016rvl,Akhavan:2018wla,Akhavan:2019zax,Omidi:2020oit,Alishahiha:2019lng,Alishahiha:2018swh,Alishahiha:2019cib,Alishahiha:2021thv,Hashemi:2019xeq} or null \cite{Lehner:2016vdi,Carmi:2017jqz,Reynolds:2016rvl,Akhavan:2019zax,Omidi:2020oit} boundaries or joint points \cite{Yang:2017amx} of the WDW patch. For a detailed review refer to \cite{Susskind:2018pmk}. Moreover, quantum complexity was also investigated form the QFT point of view \cite{Brown:2017jil,Jefferson:2017sdb,Chapman:2017rqy,Hackl:2018ptj,Camargo:2018eof,Yang:2018cgx,Guo:2018kzl,Chapman:2018hou,Camargo:2019isp,Bernamonti:2019zyy,Caceres:2019pgf,Bernamonti:2020bcf,DiGiulio:2020hlz,DiGiulio:2021oal,Brown:2019whu,Flory:2020eot,Flory:2020dja,Doroudiani:2019llj,Bhattacharyya:2021fii,Camargo:2020yfv,Erdmenger:2021wzc,Auzzi:2020idm,Brown:2021rmz,Chagnet:2021uvi,Basteiro:2021ene,Yang:2020tna,Yang:2019iav,Yang:2018tpo,Haque:2021hyw,Haque:2021kdm,Bhattacharyya:2020iic,Bhattacharyya:2020art,Bhattacharyya:2020kgu,Bhattacharyya:2020rpy,Bhattacharyya:2019txx,Ali:2019zcj,Ali:2018fcz,Khan:2018rzm,Bhattacharyya:2019kvj,Moghimnejad:2021rqe,Brown:2022phc,Brown:2021euk}.
\\According to the CV proposal \cite{Stanford:2014jda,Susskind:2014rva}, the complexity of a state on a constant time slice $\Sigma_\tau$ in a holographic CFT is given by the volume $V (\Sigma(\tau))$ of an extremal codimension-one hypersurface $\Sigma(\tau)$ in the bulk spacetime 
%%%%%%%%%%%%%%%%%%%%%%%%%%%%
\footnote{Here, we assume that the bulk spacetime is $d+2$ dimensional.}
%%%%%%%%%%%%%%%%%%%%%%%%%%%%
\bea
\mathcal{C}_{\rm V} = \max_{\partial \Sigma (\tau) = \Sigma_\tau}
%{\mathop {\rm max}_{\scriptscriptstyle{\partial \Sigma (\tau) = \Sigma_\tau}} }
\Bigg[ \frac{V (\Sigma(\tau))}{ \ell G_N}\Bigg] = \max_{\partial \Sigma (\tau) = \Sigma_\tau}  \Bigg[ \frac{1}{G_N L} \int_{ \Sigma ({\tau})} d^{d+1} \sigma \sqrt{h} 
\Bigg].
\label{CV} 
\eea 
Here $\Sigma(\tau)$ is anchored on the left and right asymptotic boundaries of the bulk spacetime at times $t_L$ and $t_R$ such that $\partial \Sigma (\tau) = \Sigma_\tau$ (See Figure \ref{fig: Penrose-Extremal-Surface}). Moreover, $h$ is the induced metric on $\Sigma(\tau)$ and $\ell$ is a dimensionful constant which can be either the AdS radius $L$ or the black hole horizon radius $r_h$. In the following, we set $\ell=L$. 
%Moreover, we call the holographic complexity in the CV proposal {\it volume-complexity}.
%second version of the volume-complexity proposal, dubbed CV2.0 \cite{Couch:2016exn}. 
\\Recently, CV proposal was generalized in ref. \cite{Belin:2021bga} and the following functional was introduced as the holographic dual of quantum complexity
%\\One can generalize the CV proposal and introduce the following functional as the holographic dual of the quantum complexity \cite{Belin:2021bga}
\bea
O _{F_1, \Sigma_{F_2}} \left( \Sigma_{\tau} \right) = \frac{1}{G_N L} \int_{ \Sigma_{F_2}} d^{d+1} \sigma \sqrt{h} \; F_1 \left( g_{\mu \nu} ; X^\mu \right).
\label{O-F1}
\eea 
Here the integral is taken on a codimension-one hypersurface $\Sigma_{F_2}$ in the bulk spacetime which is anchored on the boundary of the bulk spacetime such that $\partial \Sigma_{F_2} = \Sigma_{\tau}$ where $\Sigma_{\tau}$ is a constant time slice in the boundary where the dual CFT lives.
Moreover, it is parametrized by the coordinates $X^{\mu} (\sigma^i)$ and $\sigma^i$ are the coordinates in the bulk spacetime. $F_{1,2} \left( g_{\mu \nu} , X^\mu \right)$ are arbitrary scalar functions of the bulk metric $g_{\mu \nu}$ and $X^{\mu} (\sigma^i)$. Furthermore, $h$ is the determinant of the induced metric on $\Sigma_{F_2}$, and the hypersurface $\Sigma_{F_2}$ is determined by extremizing the following functional \cite{Belin:2021bga}
\bea
\delta_{X} \left( \int_{\Sigma_{F_2}} d^{d+1} \sigma \sqrt{h} \; F_2 \left( g_{\mu \nu } ; X^\mu \right) \right) = 0.
\label{Sigma}
\eea 
%where $F_2 \left( g_{\mu \nu} , X^\mu \right)$ is another scalar function made out of the curvature invariants. 
Therefore, $F_2$ determines the hypersurface $\Sigma_{F_2}$ on which the integral in eq. \eqref{O-F1} is taken and $F_1$ determines $O _{F_1, \Sigma_{F_2}}$. It should be pointed out that $F_1$ and $F_2$ are two independent functions. However, for $F_1 = F_2$, eq. \eqref{O-F1} can be simplified as follows
\bea
\mathcal{C}_{\rm gen} (\tau) =  \max _{ \partial \Sigma (\tau) = \Sigma_\tau} \Bigg[ 
\frac{1}{G_N L} \int_{\Sigma (\tau ) } d^{d+1} \; \sigma \; \sqrt{h} \; F_1 \left(  g_{\mu \nu} ; X^\mu ( \sigma) \right) 
\Bigg],
\label{C-gen}
\eea 
where $\tau= 2 t_L = 2 t_R $
%%%%%%%%%%%%%%%%%%%%%%%%%%
\footnote{In the following, we consider symmetric hypersurfaces $\Sigma_\tau$ for which $t_L = t_R$ (See Figure \ref{fig: Penrose-Extremal-Surface}). Since $\tau= t_L +t_R$, one has $\tau= 2 t_L = 2 t_R $}
%%%%%%%%%%%%%%%%%%%%%%%%%%
and $t_{L,R}$ are the time coordinates in the left and right CFTs. 
%Notice that for $\lambda =0$, it reduces to the volume-complexity in the CV proposal, i.e. eq. \eqref{CV}. 
Moreover, for $F_1 = F_2 = 1$, the functional in eq. \eqref{O-F1} reduces to the volume functional in the CV proposal, i.e. eq. \eqref{CV}. Since $\mathcal{C}_{\rm gen}$ is a generalization of the volume-complexity, it is dubbed {\it generalized} volume-complexity.
%In this case, one can rewrite eq. \eqref{O-F1} as follows
%\bea
%\mathcal{C}_{\rm gen} (\tau) = {\rm max} _{ \partial \Sigma (\tau) = \Sigma_\tau} \Bigg[ 
%\frac{1}{G_N L} \int_{\Sigma (\tau ) } d^{d+1} \; \sigma \; \sqrt{h} \; F_1 \left(  g_{\mu \nu} ; X^\mu ( \sigma) \right) 
%\Bigg],
%\label{C-gen}
%\eea 
%where $\tau= 2 t_L = 2 t_R $
 It should also be emphasize that the extremal hypersurface $\Sigma (\tau )$ approaches a constant-r hypersurface at $r=r_f$ at late times, i.e. $\tau \rightarrow \infty$, which is called the {\it final} slice \cite{Belin:2021bga,Stanford:2014jda} (See Figure \ref{fig: Penrose-Extremal-Surface}).  Moreover, it was shown in ref. \cite{Belin:2021bga} that finding the extremal hypersurface $\Sigma( \tau)$ in eq. \eqref{C-gen}, is analogous to study the motion of a classical non-relativistic particle in an effective potential $\hat{U} (r)$ with the action $S \propto \mathcal{C}_{\rm gen}$. In this manner, extremizing equations for $\mathcal{C}_{\rm gen}$ are equivalent to the equations of motion of the particle (See eqs. \eqref{r-dot} and \eqref{v-dot}).
%%%%%%%%%%%%%%%%%%%%%%%%%%%%
\footnote{The same prescription also works for volume-complexity (See e.g. ref. \cite{Carmi:2017jqz}). }
%%%%%%%%%%%%%%%%%%%%%%%%%%%%
Furthermore, the growth rate of $\mathcal{C}_{\rm gen}$ behaves as follows \cite{Belin:2021bga}
\bea
\frac{d \mathcal{C}_{ \rm gen} }{d \tau} = \frac{1}{2} P_t \mid_{ \partial \Sigma (t)},
\label{dC-dtau-Pt}
\eea 
where $ P_t$ is the momentum conjugate to the time coordinate $t$. Therefore, to have a linear growth rate at late times, $P_t$ has to be a constant at late times. It was also shown in ref. \cite{Belin:2021bga} that, it happens whenever the effective potential has a local maximum {\it inside} the horizon. Therefore, investigating the local maxima of $\hat{U}(r)$ is crucial.
%the effective potential $\hat{U} (r)$ has a local maximum at $r=r_f$ \cite{Belin:2021bga}.
\\The aim of the paper is to explore the generalized volume-complexity for a two-sided Hyperscaling Violating (HV) black brane \cite{Alishahiha:2012qu} in $d+2$ dimensions. Different aspects of this background have been studied in the literature such as: Page curve \cite{Omidi:2021opl}, Volume-Complexity \cite{Swingle:2017zcd}
%%%%%%%%%%%%%%%%%%%%%%%%%%%%
\footnote{It should be emphasized that our HV black brane solution is connected to that investigated in ref. \cite{Swingle:2017zcd}. More precisely, by changing the radial coordinate as $r_{\rm there} = \frac{L^2}{r_F^{\theta_e}}  r_{\rm here}^{ \theta_e -1 }$, the two solutions transform to each other. We would like to thank the referee for her/his helpful comments in this regard.}
%%%%%%%%%%%%%%%%%%%%%%%%%%%%
 and Action-Complexity \cite{Alishahiha:2018tep,Alishahiha:2019lng,Swingle:2017zcd}. Similar to ref. \cite{Belin:2021bga}, we consider the following case
%$F_1=F_2 = 1 + \lambda L^4 $.
\bea
F_1 \left(  g_{\mu \nu} ; X^\mu ( \sigma) \right) = F_2 \left(  g_{\mu \nu} ; X^\mu ( \sigma) \right) = 1 + \lambda L^4 C^2,
\label{F1=F2}
\eea 
where $C^2 = C_{\alpha \beta \mu \nu} C^{\alpha \beta \mu \nu}$ is the square of the Weyl tensor in the bulk and $L$ is a length scale which reduces to the AdS radius when $\theta= 0$ and $z=1$. Moreover, $\lambda$ is a dimensionless coupling constant and it can have any value. However, we will see that asking for $\mathcal{C}_{\rm gen}$ to have a linear growth at late times, imposes a constraint on the value of $\lambda$. We calculate generalized volume-complexity for the two cases $\lambda =0$ and $\lambda \neq 0$, separately. 
%For, $\lambda=0$, it reduces to the celebrated volume-complexity. 
For both cases, we numerically obtain the growth rate of $\mathcal{C}_{\rm gen}$. For $\lambda=0$, we find analytically the growth rate. However, for $\lambda \neq 0$, we can find it analytically for some specific cases:  1) $\theta=0$ and arbitrary values of $d$ and $z$, 2) $d=2$, $\theta=1$ and $z=1$. We also prove that $\frac{d \mathcal{C}_{\rm gen}}{d \tau}$ reaches its late time value from below similar to AdS black holes. Moreover, we write the growth rate in terms of the effective potential of a classical non-relativistic particle and investigate its maxima in detail. On the other hand, we calculate the complexity of formation $\Delta \mathcal{C}_V$ obtained from volume-complexity and show that it is a UV finite quantity. Furthermore, for $\theta =0$, it is proportional to the thermal entropy $S$ of the black brane. For $\theta \neq 0$, there are two cases $d_e \neq \theta_e$ and $d_e = \theta_e$ which have different kinds of UV divergent terms in $\mathcal{C}_V$, and hence have to be considered separately. For the former, $\Delta \mathcal{C}_V$ is proportional to $S T^{- \frac{\theta_e}{z}}$ where $T$ is the temperature of the black brane. However, for the latter, it depends on $\log S$.
\\The outline of the paper is as follows: in Section \ref{Sec: Hyperscaling Violating Black Brane}, we briefly review HV black branes. In Section \ref{Sec: Generalized Volume Complexity-C-gen}, we calculate generalized volume-complexity for this background. In Section \ref{Sec: Complexity of Formation}, we calculate the complexity of formation $\Delta \mathcal{C}_V$ obtained from volume-complexity. In Section \ref{Sec: Discussion}, we summarize our results and discuss about future problems. Moreover, in Appendix \ref{Appendix A: Relating the Radial cutoffs}, we investigate the connection between the radial cutoffs $r_{\rm max}$ of the black brane and vacuum. These results are necessary for the calculation of the complexity of formation.
%To calculate the complexity of formation, we need to find the connection between the radial cutoffs $r_{\rm max}$ of the black brane and vacuum. These calculations are included in Appendix \ref{Appendix A: Relating the Radial cutoffs}. 
In Appendix \ref{Appendix B: Other Higher Derivative Functionals}, we study the generalized complexity for the case where $F_{1} = F_2$ such that they are an arbitrary combination of Ricci scalar, square of the Ricci tensor and square of the Riemann tensor. 
%\\The outline of the paper is as follows: in Section \ref{Sec: Hyperscaling Violating Black Brane}, we review HV black branes. In Section \ref{Sec: Generalized Volume Complexity-C-gen}, we calculate generalized volume-complexity for this background. We consider the two cases $\lambda =0$ and $\lambda \neq 0$, separately. For, $\lambda=0$, it reduces to the celebrated volume-complexity. For both cases, we numerically obtain the growth rate of $\mathcal{C}_{\rm gen}$ and analytically find its late time value. In Section \ref{Sec: Complexity of Formation}, we calculate the complexity of formation $\Delta \mathcal{C}_V$ obtained from volume-complexity and show that it is a UV finite quantity. Moreover, for $\theta =0$, it is proportional to the thermal entropy $S$. For $\theta \neq 0$, there are two cases $d_e \neq \theta_e$ and $d_e = \theta_e$. For the former, it is proportional to $S T^{- \frac{\theta_e}{z}}$ where $T$ is the temperature of the black brane. However, for the latter, it depends on $\log S$. In Section \ref{Sec: Discussion}, we summarize our results and discuss about future problems.

%%%%%%%%%%%%%%%%%%%%%%%%%%%%
\section{Hyperscaling Violating Black Brane}
\label{Sec: Hyperscaling Violating Black Brane}
%%%%%%%%%%%%%%%%%%%%%%%%%%%%

%In the following, we want to calculate the generalized volume-complexity given in eq. \eqref{C-gen} for a two-sided HV black brane. 
In this section, we review Hyperscaling Violating (HV) black branes which are solutions to the Einstein-Maxwell-Dilaton gravity theory. The corresponding action is given by \cite{Alishahiha:2012qu}
\bea
I_{\rm HV}= - \frac{1}{16 \pi G_N} \int d^{d+2} x \sqrt{-g} \left[ R - \frac{1}{2} \left( \partial \phi \right)^2 + V(\phi) - \frac{1}{4} e^{\lambda \phi} F^2 \right].
\label{action-HV}
\eea 
%The gauge field $A$ breaks Lorentz invariance and introduces the dynamical exponent $z$. 
%Moreover, the non-trivial potential $V(\phi)$ breaks the scaling symmetry and introduces the Hyperscaling Violation exponent $\theta$. 
%By solving the equations of motion, one can find 
The scalar field $\phi$, the potential $V(\phi)$ and the field strength of the gauge field are as follows \cite{Alishahiha:2012qu}
\bea
%e^\phi &=& e^{\phi_0} \; r^{\sqrt{2 d_e (z-1-\theta_e)}},
\phi &=& \phi_0 + \beta \ln r, \;\;\;\;\;\;\;\;\;\;\;\;\; 
\cr && \cr
V(\phi) &=& (d_e + z -1)(d_e +z ) e^{\gamma ( \phi - \phi_0) },
\cr && \cr
F_{ rt}  &=& \sqrt{2(z-1) (d_e + z)} e^{\frac{(d_e + \theta_e )  \phi_0}{ \beta }} \; r^{d_e +z-1} ,
%F_{1 \; rt} &=& \sqrt{2(z-1) (d_e + z)} e^{\frac{(d_e + \theta_e)}{\sqrt{2 d_e (z-1-\theta_e)}} \phi_0} \; r^{d_e +z-1} ,
%\cr && \cr
%F_{2 \; rt} &=& q_b \sqrt{ \frac{ (d_e-1) (d_e +z -3)}{2} } e^{ - \frac{\beta  \phi_0}{2 (d_e-1)}}  r^{- (d_e + z -2) },
%%F_{2 \; rt} &=& q_b \sqrt{2 d_e (d_e +z -2)} e^{ - \frac{\beta  \phi_0}{2 d_e}}  r^{- (d_e + z -1) },
%%F_{2 \; rt} &=& q_b \sqrt{2 d_e (d_e +z -2)} e^{- \sqrt{\frac{z-1- \theta_e}{2 d_e}} \phi_0}  r^{- (d_e + z -1) }.
\label{dilaton-gauge-fields}
\eea 
where 
\bea
\beta = \sqrt{2 d_e (-\theta_e + z -1)}, \;\;\;\;\;\;\;\;\;\;\;\;\; 
\lambda = - \frac{2 ( d_e + \theta_e ) }{\beta}, 
%\;\;\;\;\;\;\;\;\;\;\;\;\; \lambda_2 = \frac{\beta}{d_e-1}, 
\;\;\;\;\;\;\;\;\;\;\;\;\;
%%\cr && \cr
\gamma= \frac{2 \theta_e}{\beta}.
%\;\;\;\;\;\;\;\;\;\;\;\;\; %V_0 = (d_e + z -1)(d_e +z) e^{- \gamma \phi_0},
\eea 
Moreover, $d_e= d - \theta$ is the effective dimension \cite{Dong:2012se} and $\theta_e = \frac{ \theta}{d}$. The metric is given by \cite{Alishahiha:2012qu,Pedraza:2018eey} 
%%%%%%%%%%%%%%%%%%%%%%%%%%
\footnote{Notice that there are two dimensionful parameters $r_F$ and $r_f$ which have very similar notations. Here, $r_F$ is the dynamical scale in the HV geometry and the metric in eq. \eqref{metric-BB} is only well-defined for $r > { r_F}$ \cite{Shaghoulian:2011aa,Dong:2012se}. On the other hand, $r_f$ is the radial coordinate of the final r-constant hypersurface that the extremal surface $\Sigma (\tau)$ approaches it at late times (See Figure \ref{fig: Penrose-Extremal-Surface}).}
%%%%%%%%%%%%%%%%%%%%%%%%%%
\bea
ds^2 = \left( \frac{r}{r_F} \right)^{- 2 \theta_e} \left( -  \left( \frac{r}{L} \right)^{2z} f(r) dt^2  + \frac{L^2}{r^2 f(r)} dr^2 + \frac{r^2}{L^2} \sum_{i=1}^{d} dx_i^2 \right),
\label{metric-BB}
\eea
%Furthermore, the blackening factor is as follows
where
\bea
f(r) = 1- \left( \frac{r_h}{r}\right)^{d_e +z}.
\label{f(r)}
\eea 
It should be emphasized that the solution has two parameters $z$ and $\theta$ which are called the dynamical and hyperscaling violation exponents, respectively. 
%Although the action is Lorentz invariant, the solution does not respect the Lorentz symmetry. 
Furthermore, under the following coordinate transformation \cite{Dong:2012se}
\bea
t \rightarrow \lambda^z t \;\;\;\;\;\;\;\;\;\;\;\;\;\;\; x^i \rightarrow \lambda x^i \;\;\;\;\;\;\;\;\;\;\;\;\;\;\; r \rightarrow \lambda^{-1} r,
\label{coord-transformation}
\eea 	
the metric \eqref{metric-BB} behaves as follows
\bea
ds \rightarrow \lambda^{\theta_e} ds,
\eea 
and hence it is not invariant. However, for $\theta =0$, the metric is scale invariant. Moreover, there is an anisotropy among the $t$ and $x^i$ directions in eq. \eqref{coord-transformation}. Therefore, the Lorentz symmetry is also broken for $z \neq 1$. In other words, for $\theta=0$ and $z=1$, the scaling and Lorentz symmetries are restored in the dual QFT. In this case, 
%In this case, the scalar field becomes a constant and the gauge field equals to zero. Moreover, $V(\phi)$ plays the rule of the cosmological constant $\Lambda$, and hence 
the solution reduces to a d+2 dimensional planar AdS-Schwarzschild black hole.
%%%%%%%%%%%%%%%%%%%%%%%%%%%
\footnote{Notice that for $z=1$, the gauge field is zero. Moreover, for $\theta=0$, the scalar field becomes a constant. In this case, $V(\phi)$ plays the role of the cosmological constant.}
%%%%%%%%%%%%%%%%%%%%%%%%%%%
%, and it is a consequence of having a non-zero gauge field in eq. \eqref{dilaton-gauge-fields} for $z \neq 1$.
Another important point is that the null energy condition imposes the following constraints on $z$ and $\theta$ \cite{Dong:2012se}
\bea
d_e ( d (z-1) - \theta) \geq 0, \;\;\;\;\;\;\;\;\;\;\;\;\;\;\;\;\; (z-1)(d_e +z) \geq 0.
\label{NEC}
\eea 
Moreover, it was argued in ref. \cite{Alishahiha:2012qu} that the solution is not valid for $d = \theta$. On the other hand, it was shown that the solution is not stable for $d <  \theta$ \cite{Dong:2012se}. Therefore, we restrict ourselves to the following cases
\bea
z \geq 1, \;\;\;\;\;\;\;\;\;\;\;\;\;\;\;\;\;\;\;\;\;\;\; d > \theta.
\eea  
On the other hand, the temperature and entropy of the black brane are given by \cite{Alishahiha:2012qu,Pedraza:2018eey}
\bea
T = \frac{( d_e +z )  r_h^z}{4 \pi L^{z+1}}, \;\;\;\;\;\;\;\;\;\;\;\;\;\;\; S = \frac{V_d r_h^{d_e} r_F^{ \theta} }{4 G_N L^d}.
\label{T-S}
\eea 
Moreover, the mass is as follows \cite{Pedraza:2018eey}
%%%%%%%%%%%%%%%%%%%%%%%%%%%%
\footnote{ \label{footnote-7}
In the literature, the metric of the transverse directions in eq. \eqref{metric-BB} is written as $d \Omega_d^2 = \frac{1}{L^2} \sum_{i=1}^{d} dx_i^2$. Notice that $\Omega_d = \frac{V_d}{L^d}$, where $\Omega_d$ and $V_d$ are the volumes of the metrics $d \Omega_d^2$ and $\sum_{i=1}^d d x_i^2$, respectively. Having said this, the entropy and mass of a HV black brane can be written as $S = \frac{\Omega_d r_h^{d_e} r_F^{ \theta} }{4 G_N}$ and $M= \frac{d_e V_d }{16 \pi G_N } \frac{r_h^{d_e +z} r_F^{\theta} }{ L^{z+1}}$ \cite{Pedraza:2018eey}. It should also pointed out that $\Omega_d$ is dimensionless and $V_d$ has the dimension of $\rm lenght^d$.}
%%%%%%%%%%%%%%%%%%%%%%%%%%%%
\bea
M= \frac{d_e}{d_e +z} T S=\frac{d_e V_d }{16 \pi G_N } \frac{r_h^{d_e +z} r_F^{\theta} }{ L^{d+z+1}}.
\label{M}
\eea 
Furthermore, the tortoise coordinate is simply given by 
\bea
r^\ast (r) = \int \frac{L^{z+1} }{r^{z+1} f(r) } dr,
\label{r-star}
\eea 
which is a hypergeometric function of $r$ \cite{Omidi:2021opl}. However, we do not need the explicit form of $r^\ast(r)$ here. It should be pointed out that this theory has a vacuum solution which is obtained by setting $f(r)=1$ in eq. \eqref{metric-BB}  \cite{Charmousis:2010zz,Dong:2012se}
\bea
ds^2 = \left( \frac{r}{r_F} \right)^{- 2 \theta_e} \left( -  \left( \frac{r}{L} \right)^{2z} dt^2  + \frac{L^2}{r^2 } dr^2 + \frac{r^2}{L^2} \sum_{i=1}^{d} dx_i^2 \right),
\label{metric-vacuum}
\eea
whose entropy and temperature are zero. In Section \ref{Sec: Complexity of Formation}, we need the volume-complexity of the vacuum to calculate the complexity of formation.

%%%%%%%%%%%%%%%%%%%%%%%%%%%%
\section{Generalized Volume-Complexity $\mathcal{C}_{\rm gen}$}
\label{Sec: Generalized Volume Complexity-C-gen}
%%%%%%%%%%%%%%%%%%%%%%%%%%%%

To calculate the generalized volume-complexity in eq. \eqref{C-gen}, we proceed in the same manner as refs. \cite{Carmi:2017jqz,Belin:2021bga}. In the infalling Eddington-Finkelstein coordinate $v= t + r^*$, the metric in eq. \eqref{metric-BB} can be written as follows
\bea
d s^2 = \left( \frac{r}{r_F} \right)^{- 2 \theta_e} \left( - \left( \frac{r}{L} \right)^{2z} f(r) d v^2 + 2 \left( \frac{r}{L} \right)^{z-1} d v \; d r + \frac{r^2}{L^2} \sum_{i=1}^{d} dx_i^2 \right).
\label{metric-BB-Eddington-Finkelstein}
\eea 
%%%%%%%%%%%%%%%%%%%%%%%%%%%%%%%%%
\begin{figure}
	\begin{center}
	\includegraphics[scale=1.1]{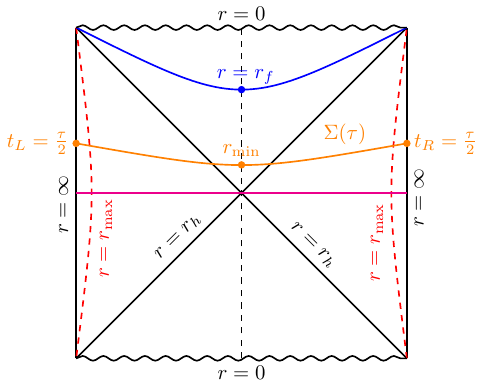}
	\end{center}
	\caption{The extremal codimension-one bulk hypersurface $\Sigma(\tau)$ denoted in orange, starts from the left boundary at time $t_L$, then goes inside the horizon and reaches the minimal radius $r_{\rm min}$. Next, it goes outside the horizon and reaches the right asymptotic boundary at time $t_R$. Here, we consider symmetric hypersurfaces where $t_L=t_R= \frac{\tau}{2}$. At late times, i.e. $\tau \rightarrow \infty$, it approaches a constant-r hypersurface dubbed {\it final} slice at $r=r_f$ which is denoted in blue. The red dashed curves are the radial cutoffs at $r = r_{\rm max}$. Moreover, the magenta straight line denotes the extremal hypersurface anchored at $t_L = t_R =0$ which is applied in the calculation of the complexity of formation in Section \ref{Sec: Complexity of Formation}.}
	\label{fig: Penrose-Extremal-Surface}
\end{figure}
%%%%%%%%%%%%%%%%%%%%%%%%%
Due to the translational symmetry along the transverse directions $x^{i=1 \cdots , d}$, we parametrize the spacelike hypersurface $\Sigma ( \tau) $ as follows
\bea
\Sigma (\tau) : \left( v( \sigma ), r( \sigma) , \vec{x} \right) \; .
\eea 
As mentioned before, here we consider symmetric configurations for which one has $t_L = t_R$ (See Figure \ref{fig: Penrose-Extremal-Surface}). Then one can write the generalized volume-complexity as follows
\bea
\mathcal{C}_{ \rm gen} = \frac{V_d}{ G_N L} \int d \sigma \frac{r^{d_e - \theta_e}}{ r_F^{ - \theta_e (d+1) } L^d } \sqrt{ - \left( \frac{r}{L} \right)^{2 z} f(r) \dot{v}^2 + 2 \left( \frac{ r}{L} \right)^{z-1} \dot{v} \dot{r} } \; a(r),
\label{C-gen-HV}
\eea 
where 
\bea
a(r) = 1 + \lambda L^4 C^2,
\label{ar}
\eea 
and $\lambda$ is a {\it dimensionless} coupling constant. Moreover, "." denotes the derivative with respect to the coordinate $\sigma$ and $V_d$ is the volume of the spatial directions $x^{i=1, \cdots d}$. Since $\mathcal{C}_{ \rm gen}$ is invariant under the diffeomorphism $ \sigma \rightarrow g( \sigma)$, we choose the following gauge (See also \cite{Belin:2021bga})
\bea
\sqrt{2 \dot{v} \dot{r} - \left( \frac{r}{L} \right)^{z+1} f(r) \dot{v}^2 } = \left( \frac{r}{L} \right)^{ \frac{z-1}{2}} \frac{r^{d_e - \theta_e}}{r_F^{- \theta_e (d+1)} L^d} a(r).
\label{gauge}
\eea 
Now, one can compare the problem with that of the motion of a classical non-relativistic particle with action $\mathcal{C}_{\rm gen}$ \cite{Belin:2021bga}. In this manner, extremizing $\mathcal{C}_{\rm gen}$ gives the corresponding equations of motion for the particle. Having said this, since the metric \eqref{metric-BB} is stationary, the momentum $P_v$ conjugate to the infalling time $v$ is conserved
%%%%%%%%%%%%%%%%%%%%%%%%%%
\footnote{It should be emphasized that $P_v$ is a function of the boundary time $\tau$ and for a given value of $\tau$ it is a constant which is fixed by eq. \eqref{tau}.}
%%%%%%%%%%%%%%%%%%%%%%%%%%
\bea
P_v (\tau) = \dot{r} - \left( \frac{r}{L} \right)^{z+1} f(r) \dot{v}  = const.
\label{P-v}
\eea 
Next, by solving eqs. \eqref{gauge} and \eqref{P-v}, one has
\bea
\dot{r} &=& \pm \sqrt{P_v^2 + \frac{r^{ 2 (d+z - \theta_e (d+1) )}}{L^{2 (d+z)}} r_F^{2 \theta_e (d+1) } a(r)^2 f(r) },
\label{r-dot}
\\
%\cr && \cr
\dot{v}&=& \left( \frac{L}{r} \right)^{z+1} \frac{1}{f(r)} \left( - P_v  \pm \sqrt{ P_v^2 + \frac{r^{2 (d +z - \theta_e (d+1))}}{L^{2 (d+z)}} r_F^{2 \theta_e (d+1)} a^2(r) f(r)} \right),
\label{v-dot}
\eea 
which extremize $\mathcal{C}_{\rm gen}$. In this case, the position of the particle is given by the coordinate $r$. On the other hand, the effective potential $\hat{U}(r)$ is defined by
\bea
\dot{r}^2 + \hat{U} (r) = P_v^2.
\label{Hamiltonian}
\eea 
Then from eq. \eqref{r-dot} and \eqref{Hamiltonian}, one can easily find the effective potential $ \hat{U}(r)$ as follows
\bea
\hat{U}(r) = - \frac{r^{2 (d+z - \theta_e (d+1))}}{ L^{2 (d+z)}} r_F^{2 \theta_e (d+1) } a^2(r) f(r).
\label{U-r}
\eea 
From the above expression and eqs. \eqref{ar} and \eqref{C-2}, it is obvious that $\hat{U}(r)$ becomes zero at $r=r_h$. Moreover, for $r<r_h$, one has $f(r) < 0$, and hence $\hat{U}(r)$ is positive in the range $0 < r < r_h$. 
%%%%%%%%%%%%%%%%%%%%%%%%%%%%
\footnote{It should be pointed out that for the case $\lambda=0$, one has $a(r) =1$ and $\hat{U}(0)=0$. However, for the case $\lambda \neq 0$, $a(r)$ is divergent at $r=0$. Therefore, $\hat{U}(r)$ is also divergent at $r=0$.}
%%%%%%%%%%%%%%%%%%%%%%%%%%%%
On the other hand, it is straightforward to show that for every value of the boundary time $\tau$, the conserved momentum $P_v$ is fixed by the following constraint
\bea
\tau= 2 t_R = - 2 \int_{r_{\rm min}}^{ \infty} dr \frac{P_v}{ \left( \frac{r}{L} \right)^{z+1} f(r) \sqrt{P_v^2 - \hat{U} (r)} },
\label{tau}
\eea 
where $r_{ \rm min}$ is the minimal radius on the timelike surface $t=0$ (See Figure \ref{fig: Penrose-Extremal-Surface}) at which one has \cite{Belin:2021bga,Carmi:2017jqz}
\bea
\dot{r}\bigr|_{r=r_{\rm min} }=0.
% \;\;\;\;\;\;\;\;\;\;\;\;\;\;\;\;\;\; \hat{U} (r_{ \rm min}) = P_v^2.
\label{r-min}
\eea 
In other words, $r_{ \rm min}$ is the turning point of the particle moving in the potential $\hat{U} (r)$. On the other hand, from Figure  \ref{fig: Penrose-Extremal-Surface}, it is clear that $r_{\rm min}$ is inside the horizon and at late times, one has $r_{ \rm min} \rightarrow r_f$ \cite{Carmi:2017jqz,Belin:2021bga}. From eqs. \eqref{r-dot}, \eqref{U-r} and \eqref{r-min}, one arrives at
\bea
P_v^2  &=& - \frac{r_{\rm min}^{2 (d+z - \theta_e (d+1))}}{ L^{2 (d+z)}} r_F^{2 \theta_e (d+1) } a^2(r_{\rm min}) f(r_{\rm min}) 
\cr && \cr
&=& \hat{U} ( r_{\rm min}).
\label{Pv-U-rmin}
\eea 
On the other hand, at the maximum $r= r_f$ of $\hat{U}(r)$, one has \cite{Belin:2021bga,Carmi:2017jqz}
\bea
\hat{U}(r_f) = P_{ \infty}^2, \;\;\;\;\;\;\;\;\;\;\;\;\;\;\; \hat{U}'(r_f)=0, \;\;\;\;\;\;\;\;\;\;\;\;\;\;\; \hat{U}''(r_f) \leq 0,
\label{r-f-U}
\eea 
where $P_{ \infty} = \lim_{ \tau \rightarrow \infty} P_v ( \tau)$. Notice that the first equation is a result of the fact that $r_f$ also satisfies eq. \eqref{r-min} and hence eq. \eqref{Pv-U-rmin} (See also ref. \cite{Carmi:2017jqz}).
%\\In the following we choose 
%\bea
%F_1 \left(  g_{\mu \nu} ; X^\mu ( \sigma^i ) \right) = a(r) = 1 + \lambda L^4 C^2,
%\label{F1}
%\eea 
%where $C^2 = C_{\mu \nu \rho} C^{\mu \nu \rho}$ is the square of the Weyl tensor. 
For later convenience, we change the radial coordinate as follows
\bea
w = \left( \frac{r}{r_h} \right)^{d_e +z}.
\label{w-r-coordinate}
\eea 
For the metric \eqref{metric-BB-Eddington-Finkelstein}, the square of the Weyl tensor is simply given by
\bea
C^2 = \frac{(d-1)}{(d+1) L^4} \left( \frac{r_h}{r_F} \right)^{4 \theta_e } w^{ \frac{4 \theta_e }{ d_e +z} -2} \big( d_e (d_e - z +2) - 2 z (z-1) w \big)^2.
\label{C-2}
\eea 
Then the effective potential \eqref{U-r} can be written as follows
\bea
\hat{U}(w) &=&  - \frac{r_h^{2 (d_e + z - \theta_e )}}{ L^{2(d+z)}} r_F^{2 \theta_e (d+1)} w^{2 (1 - \frac{\theta_e  } {d_e +z} ) }\left( 1 - \frac{1}{w} \right)
\cr && \cr
&& 
\times \left( 1 + \lambda \frac{ (d-1)}{ (d+1)} \left( \frac{r_h}{r_F} \right)^{4 \theta_e} w^{ \frac{4 \theta_e}{ d_e +z } -2 } \left( d_e (d_e -z +2 ) - 2 z (z-1) w \right)^2 \right)^2.
\label{U-w}
\eea 
Since there is a factor of $f(r)$ in the effective potential, it becomes zero at the horizon where $w=1$. Moreover, it is also zero on the singularity at $w=0$. 

%%%%%%%%%%%%%%%%%%%%%%%%%%
\subsection{Growth Rate}
\label{Sec: Growth Rate}
%%%%%%%%%%%%%%%%%%%%%%%%%%

In this section, we calculate the growth rate of $\mathcal{C}_{\rm gen}$. To this end, we proceed in the same way as ref. \cite{Carmi:2017jqz}. By changing the integral variable in eq. \eqref{C-gen-HV} from $\sigma$ to $r$ and applying eqs. \eqref{gauge} and \eqref{r-dot}
%%%%%%%%%%%%%%%%%%%%%%%%%%%%%%
\footnote{We choose the positive sign in eq. \eqref{r-dot}.}
%%%%%%%%%%%%%%%%%%%%%%%%%%%%%%
, one has
\bea
\mathcal{C}_{\rm gen}&=& 2 \frac{V_d}{G_N L}\int_{r_{\rm min}}^{ r_{\rm max} } \frac{dr}{ \dot{r}} \frac{r^{d_e - \theta_e}}{ r_F^{ - \theta_e (d+1) } L^d } \sqrt{ - \left( \frac{r}{L} \right)^{2 z} f(r) \dot{v}^2 + 2 \left( \frac{ r}{L} \right)^{z-1} \dot{v} \dot{r} } \; a(r)
\cr && \cr
&=&
\frac{2 V_d}{G_N L} \int_{r_{\rm min}}^{ r_{\rm max}} \frac{dr}{ \dot{r}} \frac{r^{2 (d_e - \theta_e)}}{ r_F^{ -2  \theta_e (d+1) } L^{2 d } } \left( \frac{r}{ L} \right)^{(z-1 )} a(r)^2 
\cr && \cr
&=&
\frac{2 V_d}{G_N L} \int_{r_{\rm min}}^{ r_{\rm max}} \frac{dr}{ \sqrt{P_v^2 + \frac{r^{ 2 (d+z - \theta_e (d+1) )}}{L^{2 (d+z)}} r_F^{2 \theta_e (d+1) } a(r)^2 f(r) }} \frac{r^{2 (d_e - \theta_e)}}{ r_F^{ -2  \theta_e (d+1) } L^{2 d } } \left( \frac{r}{ L} \right)^{(z-1 )} a(r)^2.
\nonumber
\\
\label{C-gen-HV-1}
\eea 
On the other hand, by integrating $d v = dt + dr^\ast$ and applying eqs. \eqref{r-dot} and \eqref{v-dot}, one has
%%%%%%%%%%%%%%%%%%%%%%%%%%%%%%%%%
%\footnote{In the last line, we changed the upper limit of the integral from $\infty$ to $r_{\rm max}$. Moreover, $r^\ast_\infty = r^\ast (\infty)$.}
%%%%%%%%%%%%%%%%%%%%%%%%%%%%%%%%%
\bea
t_R + r^\ast_{ \infty} - r^\ast (r_{ \rm min}) &=& \int_{v_{\rm min}}^{v_{\infty} } dv = \int_{r_{\rm min}}^{\infty} dr \frac{\dot{v} }{ \dot{r}} 
\cr && \cr
&=& \int_{r_{\rm min}}^{ r_{\rm max}} \frac{dr}{f(r)} \left( \frac{L}{r} \right)^{z+1} \Big[
1 - \frac{P_v}{ \sqrt{P_v^2 + \frac{r^{ 2 (d+z - \theta_e (d+1) )}}{L^{2 (d+z)}} r_F^{2 \theta_e (d+1) } a(r)^2 f(r) }}
\Big],
\nonumber
\\
\label{t-R}
\eea 
where $r^\ast_\infty = r^\ast (\infty)$. Next, by multiplying eq. \eqref{t-R}
%%%%%%%%%%%%%%%%%%%%%%%%%%%%%%%%
\footnote{We change the upper limit of the integral in eq. \eqref{t-R} from $\infty$ to $r_{\rm max}$.}
%%%%%%%%%%%%%%%%%%%%%%%%%%%%%%%%
by $P_v$ and adding it to eq. \eqref{C-gen-HV-1}, one can rewrite the latter as follows
\bea
\mathcal{C}_{\rm gen} &=& \frac{2 V_d}{G_N L} \Bigg\{
\int_{r_{\rm min}}^{ r_{\rm max}} \frac{dr}{f(r)} \left( \frac{L}{r} \right)^{z+1}
\Bigg[
\sqrt{P_v^2 + \frac{r^{ 2 (d+z - \theta_e (d+1) )}}{L^{2 (d+z)}} r_F^{2 \theta_e (d+1) } a(r)^2 f(r) } - P_v 
\Bigg] 
\cr && \cr
&&
\;\;\;\;\;\;\;\;\;\;\;\;\;\;\;\;\;\;\;\;\;\;\;\;\;\;\;\;\;\;\;\;\;\;\;\;\;\;\;\;\;\;\;\; + P_v \big( t_R + r^\ast_{ \infty} - r^\ast (r_{ \rm min}) \big) 
\Bigg\}.
\label{C-gen-HV-2}
\eea 
Then, by taking the derivative of the above expression with respect to the time $t_R$, one obtains
\bea
\frac{d \mathcal{C}_{ \rm gen} }{d t_R}  &=& \frac{2 V_d}{ G_N L} \Bigg\{
\frac{d P_v}{ d t_R} \int_{r_{\rm min}}^{ r_{\rm max}} dr \left( \frac{L}{r} \right)^{z+1} \frac{1}{f(r)}
\Bigg[
\frac{P_v}{ \sqrt{P_v^2 + \frac{r^{ 2 (d+z - \theta_e (d+1) )}}{L^{2 (d+z)}} r_F^{2 \theta_e (d+1) } a(r)^2 f(r) } } - 1
\Bigg]
\cr && \cr 
&&
\;\;\;\;\;\;\;\;\;\;\;\;\; + \frac{d P_v}{ d t_R} \big( t_R + r^\ast_{ \infty} - r^\ast (r_{ \rm min}) \big) + P_v
\Bigg\} 
\cr && \cr 
&=& P_v,
\label{dC-dtR}
\eea 
where in the last line we applied eq. \eqref{t-R}. Since $\tau= 2 t_R$, one can rewrite the above equation as follows
%The time derivative of the generalized complexity with respect to the boundary time $\tau$ is given by
\bea
\frac{d \mathcal{C}_{ \rm gen} }{d \tau} =
%= \frac{1}{2} P_t \mid_{ \partial \Sigma (t)} = 
\frac{V_d}{ G_N L}P_v ( \tau) = \frac{V_d}{ G_N L} \sqrt{ \hat{U} (r_{\rm min}) },
\label{dC-dtau-1}
\eea 
where we used eq. \eqref{Pv-U-rmin}. At late times, $r_{\rm min} \rightarrow r_f$, and hence one has
\bea
\lim_{\tau \to \infty}\frac{d \mathcal{C}_{ \rm gen} }{d \tau} = \frac{V_d}{ G_N L} P_{\infty} = \frac{V_d}{ G_N L} \sqrt{ \hat{U} (r_f)}.
\label{dC-dtau-2}
\eea
Therefore, the growth rate of $\mathcal{C}_{\rm gen}$ at late times, is {\it linear} and determined by the value of the effective potential at $r=r_f$. Notice that $r_f$ is a maximum of $\hat{U}(r)$ and is located inside the horizon. Therefore, the maximum of $\hat{U}(r)$ inside the horizon determines the growth rate of generalized volume-complexity at late times. In Section \ref{Sec: Maxima of the Effective Potential For lambda-neq-0}, we investigate such maxima.
\\Now, we want to prove that the generalized volume-complexity approaches its late time value in eq. \eqref{dC-dtau-2} from below. To do so, we expand $\hat{U}(r_{\rm min})$ in eq. \eqref{dC-dtau-1} around $r= r_f$ as follows
\bea
\frac{d \mathcal{C}_{ \rm gen} }{d \tau} =  \frac{V_d}{ G_N L} \Bigg[
\sqrt{ \hat{U} (r_f) } + \frac{\hat{U}''(r_f) (r- r_f)^2}{4 \sqrt{U(r_f)}} + \mathcal{O} \left( (r- r_f)^3 \right)
\Bigg],
\label{dC-dtau-expansion}
\eea 
where we used the fact that $\hat{U}'(r_f) =0$. 
%As mentioned below eq. \eqref{U-r}, $\hat{U}(r)$ is zero at $r=0$ and $r=r_h$. Moreover, $\hat{U}(r)$ is positive inside the horizon. Therefore, $r_f$ has to be a local maximum of $\hat{U}(r)$, and hence $\hat{U}''(r_f) <0$. Consequently, 
Since $r_f$ is a maximum of $\hat{U}(r)$, the second term in eq. \eqref{dC-dtau-expansion} is negative and hence $\frac{d \mathcal{C}_{\rm gen} }{d \tau}$ approaches its late time value from below. This behavior can be easily seen in Figures \ref{fig: dCV-dtau-lambda=0-d}, \ref{fig: dCV-dtau-lambda=0-z}, \ref{fig: dCV-dtau-lambda=0-theta},  \ref{fig: dCgen-dtau-lambda-neq-0-d}, \ref{fig: dCgen-dtau-lambda-neq-0-z}, \ref{fig: dCgen-dtau-lambda-neq-0-theta} and \ref{fig: dCgen-dtau-lambda-neq-0-lambda}.
%\\ In Figure \ref{fig: dC-dtau-theta=0-d=2-z}, we plotted the growth rate $ \lim_{\tau \to \infty} \frac{d \mathcal{C}_{\rm gen}}{d \tau}$ at late times for different values of $z$ and $\lambda$. It is observed that $ \lim_{\tau \to \infty} \frac{d \mathcal{C}_{\rm gen}}{d \tau} $ is an increasing function of $\lambda$. Moreover, for small values of $\lambda$, it is lower than that for the case $\lambda=0$. However, by increasing $\lambda$, it eventually exceeds the growth rate for the case $\lambda=0$.

%%%%%%%%%%%%%%%%%%%%%%%%%%
\subsection{$\lambda =0 $: Volume-Complexity}
\label{Sec: lambda-0: Volume-Complexity }
%%%%%%%%%%%%%%%%%%%%%%%%%%

As mentioned before, for the case $\lambda =0 $, one has $a(r)=1$ and hence $\mathcal{C}_{ \rm gen}$ reduces to the volume-complexity $\mathcal{C}_{ \rm V}$. Moreover, it is straightforward to check that the local maximum of $\hat{U}(r)$ is given by
\bea
r_f = r_h w_f^{ \frac{1}{ (d_e +z) } },
\label{rf-wf-lambda-0}
\eea 
where 
\bea
w_f = \frac{d_e + z - 2 \theta_e }{ 2 d_e +2 z - 2 \theta_e}.
\label{wf-lambda-0}
\eea 
Since $d_e > 0$ and $z \geq 1$, one has $d_e +z >0$. Therefore, $w_f < 1$ and this maximum is located inside the horizon. Next, by plugging eq. \eqref{wf-lambda-0} into eq. \eqref{dC-dtau-2}, one obtains the late time growth rate of $\mathcal{C}_V$ as follows 
\bea
\lim_{\tau \to \infty} \frac{d \mathcal{C}_{ \rm V} }{d \tau} = \frac{V_d}{ 2 G_N } \frac{r_F^{ \theta_e (d+1)} r_h^{d_e +z - \theta_e } } {L^{d+z +1 }} \Upsilon,
\label{dC-dtau-lambda-0-1}
\eea
where
\bea 
\Upsilon = \frac{ \sqrt{ (d_e +z) \big( (d + z) - (d+2) \theta_e \big)}}{ \big( (d+z) - (d+1) \theta_e \big)}
\times \left( \frac{ (d+z) - (d+2) \theta_e }{ 2 (d+z) -2 ( d+1) \theta_e }\right)^{ - \frac{ \theta_e }{ ( d_e +z)}}.
\label{Upsilon}
\eea
%\bea
%\frac{d \mathcal{C}_{ \rm V} }{d \tau} \mid_{\tau \rightarrow \infty} \!\!\!\! &=& \!\!\!\! \frac{V_d}{ 2 G_N } \frac{r_F^{ \theta_e (d+1)} r_h^{d_e +z - \theta_e } } {L^{d+z +1 }} \frac{ \sqrt{ (d_e +z) \big( (d + z) - (d+2) \theta_e \big)}}{ \big( (d+z) - (d+1) \theta_e \big)}
%\cr && \cr 
%&&
%\times \left( \frac{ (d+z) - (d+2) \theta_e }{ 2 (d+z) -2 ( d+1) \theta_e }\right)^{ - \frac{ \theta_e }{ ( d_e +z)}},
%\label{dC-dtau-lambda-0}
%\eea
Moreover, one can rewrite eq. \eqref{dC-dtau-lambda-0-1} as follows
\bea
\lim_{\tau \to \infty} \frac{d \mathcal{C}_{ \rm V} }{d \tau} &=& \frac{8 \pi M^{1- \frac{ \theta_e}{d_e+z}}}{d_e} r_F^{ \theta_e \left( 1 + \frac{\theta }{d_e +z} \right)} \left( \frac{16 \pi G_N L^{d+z+1}}{V_d d_e} \right)^{- \frac{\theta_e}{d_e +z}} \Upsilon
\cr && \cr 
&=& \frac{8 \pi M}{d_e} r_F^{\theta_e} \left( \frac{4 \pi L^{z+1}}{ d_e+z}\right)^{- \frac{\theta_e}{z}} \Upsilon \; T^{- \frac{\theta_e}{z}} 
\cr && \cr 
&=& M T^{- \frac{\theta_e}{z}}.
\label{dC-dtau-lambda-0-2}
\eea 
It should be pointed out that for another kind of HV black branes it was shown in ref. \cite{Swingle:2017zcd} that $\lim_{\tau \to \infty} \frac{d \mathcal{C}_{ \rm V} }{d \tau} \propto M T^{- \frac{\theta_e}{z}}$. On the other hand, for $\theta=0$ and $z=1$, the black brane solution becomes a planar AdS-Schwarzschild black hole whose mass is given by
%%%%%%%%%%%%%%%%%%%%%%%%%%%%
\footnote{
%In the literature the metric of the transverse directions in eq. \eqref{metric-BB} is written as $d \Omega_d^2 = \frac{1}{L^2} \sum_{i=1}^{d} dx_i^2$. Notice that $\Omega_d = \frac{V_d}{L^d}$, where $\Omega_d$ and $V_d$ are the volumes of the metrics $d \Omega_d^2$ and $\sum_{i=1}^d d x_i^2$, respectively. 
As mentioned in footnote \ref{footnote-7}, one has $\Omega_d = \frac{V_d}{L^d}$. Therefore, the mass of a planar AdS-Schwarzschild black hole is also written as $M= \frac{d \; \Omega_d r_h^{d+1} }{16 \pi G_N L^{2}}$.}
%%%%%%%%%%%%%%%%%%%%%%%%%%%%
\bea
\tilde{M} = \frac{d \; V_d r_h^{d+1} }{16 \pi G_N L^{d+2}}.
\label{M-AdS-BH}
\eea 
In this case, $\Upsilon =1$ and hence the late time growth rate of $\mathcal{C}_{\rm V}$ in eq. \eqref{dC-dtau-lambda-0-2} reduces to
\bea
\lim_{\tau \to \infty}  \frac{d \mathcal{C}_{ \rm V} }{d \tau} = \frac{V_d r_h^{d+1}}{2 G_N L^{d+2} } = \frac{8 \pi \tilde{M}}{d},
\label{dC-dtau-lambda-0-theta-0-z-1}
\eea 
%where $M= \frac{d \; V_d r_h^{d+1} }{16 \pi G_N L^{d+1}}$ is the mass of the black hole. The above expression 
which is the same as that for a planar AdS-black hole in ref. \cite{Carmi:2017jqz}.

%%%%%%%%%%%%%%%%%%%%%%%%%%%
\subsubsection{General Time Dependence}
\label{Sec: General Time Dependence-lambda-0}
%%%%%%%%%%%%%%%%%%%%%%%%%%%

\begin{figure}
	\begin{center}
		\includegraphics[scale=0.31]{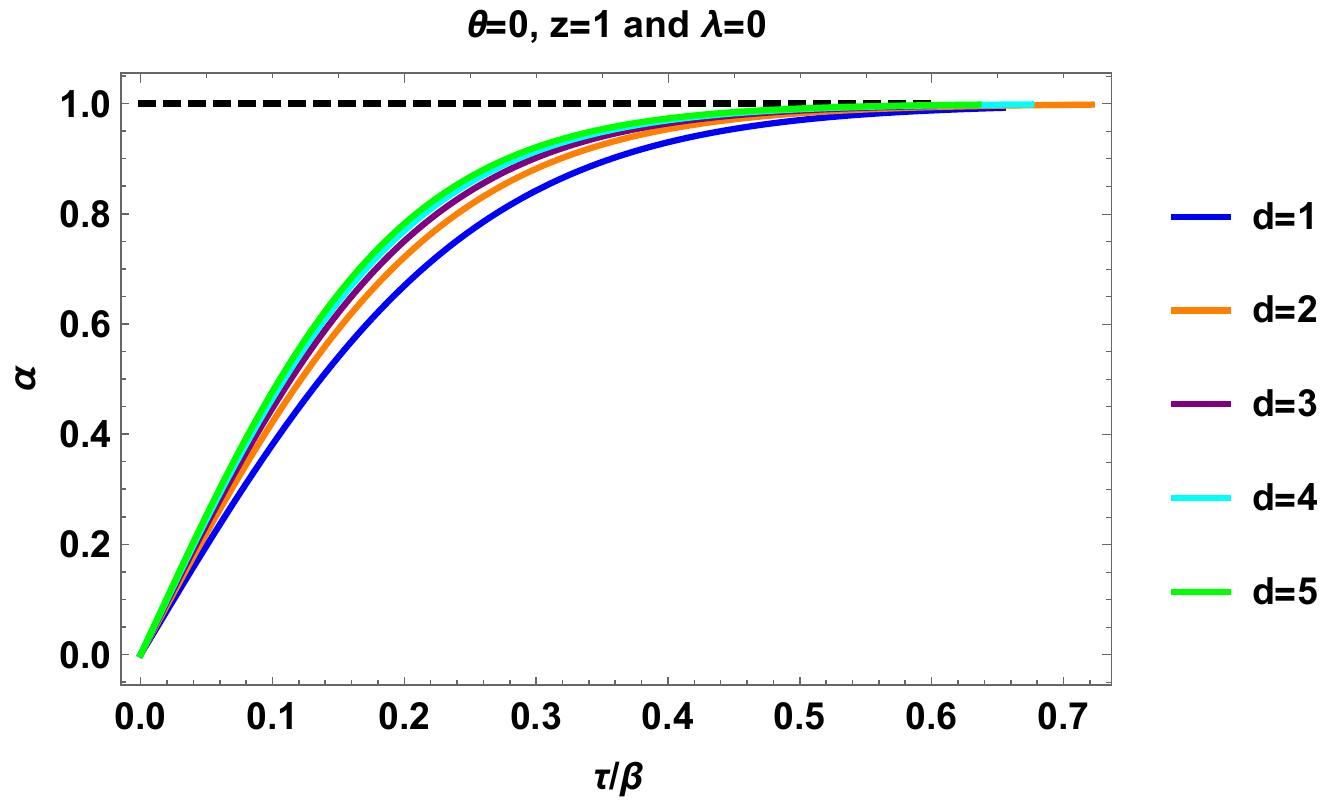}
		\hspace{0.5cm}
%		\includegraphics[scale=0.29]{dCV-dtau-lambda-0-theta-1-z-1-d.pdf}
%		\\
	    \includegraphics[scale=0.31]{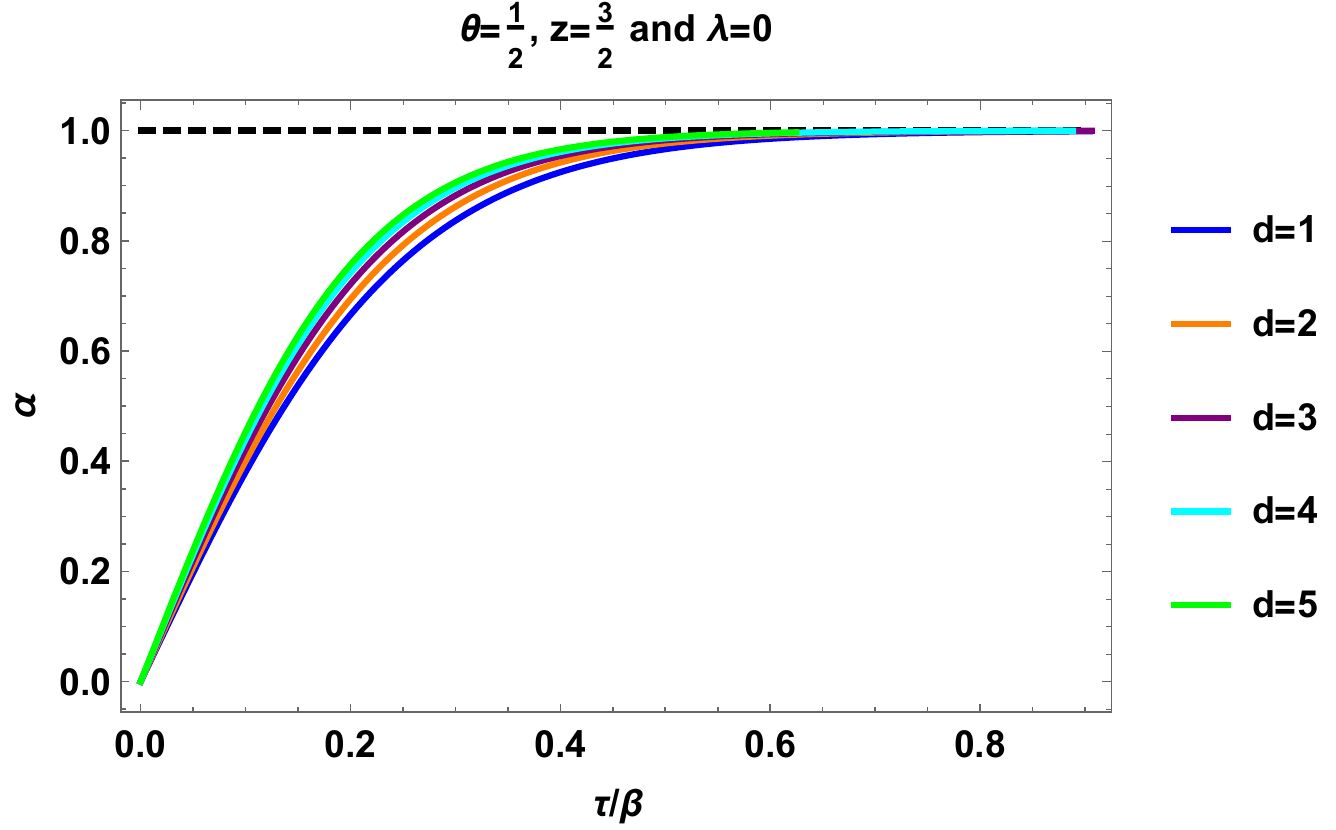}
%		\hspace{0.5cm}
%		\includegraphics[scale=0.29]{dCV-dtau-lambda-0-d-4-z-2-theta.pdf}
	\end{center}
	\caption{ The normalized growth rate $\alpha = \frac{ \frac{d \mathcal{C}_V}{d \tau}}{ \lim_{\tau \to \infty} \frac{d \mathcal{C}_V}{d \tau} }$ 
		%in eq. \eqref{alpha}, 
		as a function of $\tau / \beta$ for $\lambda=0$ and different values of $d$, $\theta$ and $z$:
		{\it Left}) $\theta=0$ and $z=1$. This figure is the same as Figure 7 in ref. \cite{Belin:2021bga}, since for these values of $\theta$ and $z$, the black brane becomes a planar AdS-Schwarzschild black hole. 
     	{\it Right}) $\theta = \frac{1}{2}$ and $z= \frac{3}{2}$.
     	 %$\theta=1$ and $z=1$.
%     	 {\it Down Left}) $d=3$ and $\theta=2$.
%     	  {\it Down Right}) $d=4$ and $z=2$.
		We set $ L= r_h = r_F= 1$.
	}
	\label{fig: dCV-dtau-lambda=0-d}
\end{figure}

\begin{figure}
	\begin{center}
		\includegraphics[scale=0.31]{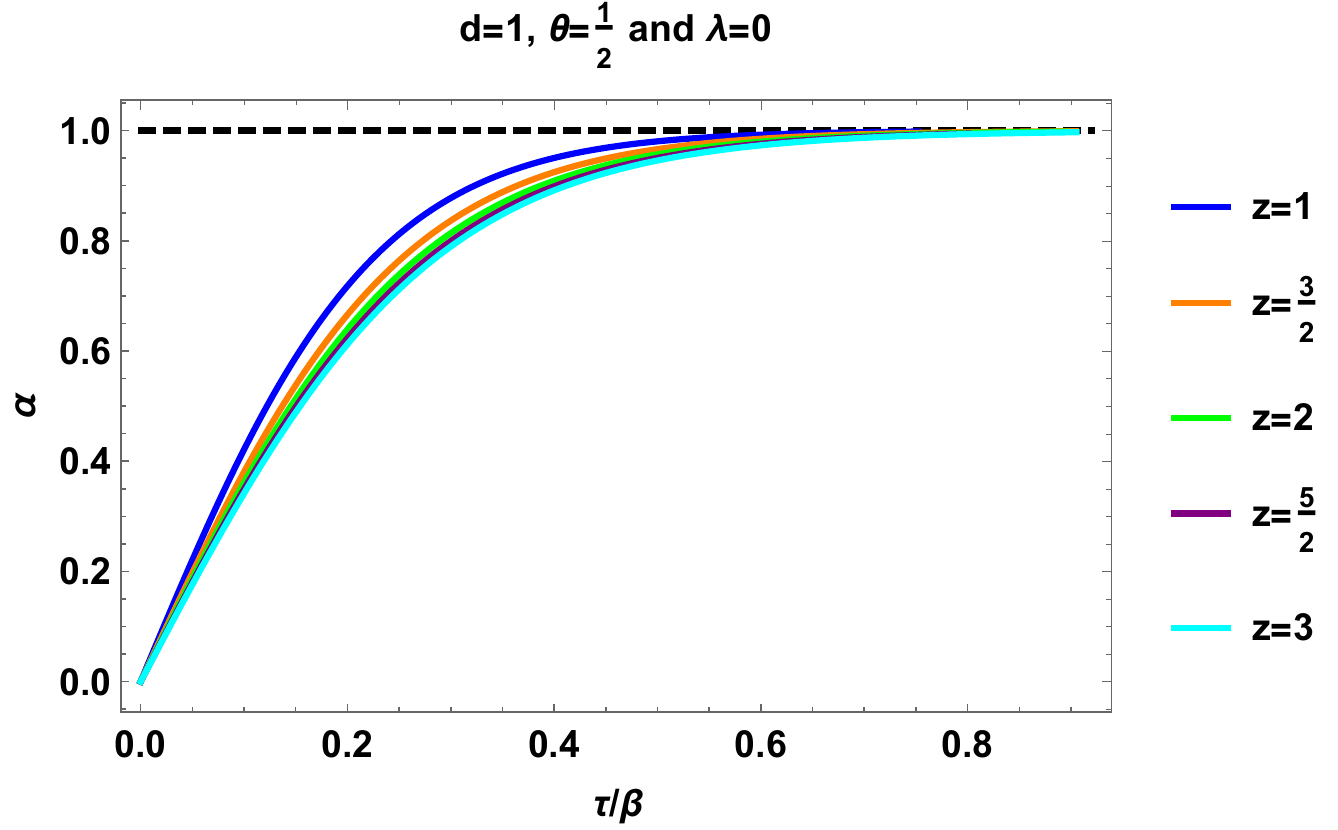}
		\hspace{0.5cm}
		\includegraphics[scale=0.31]{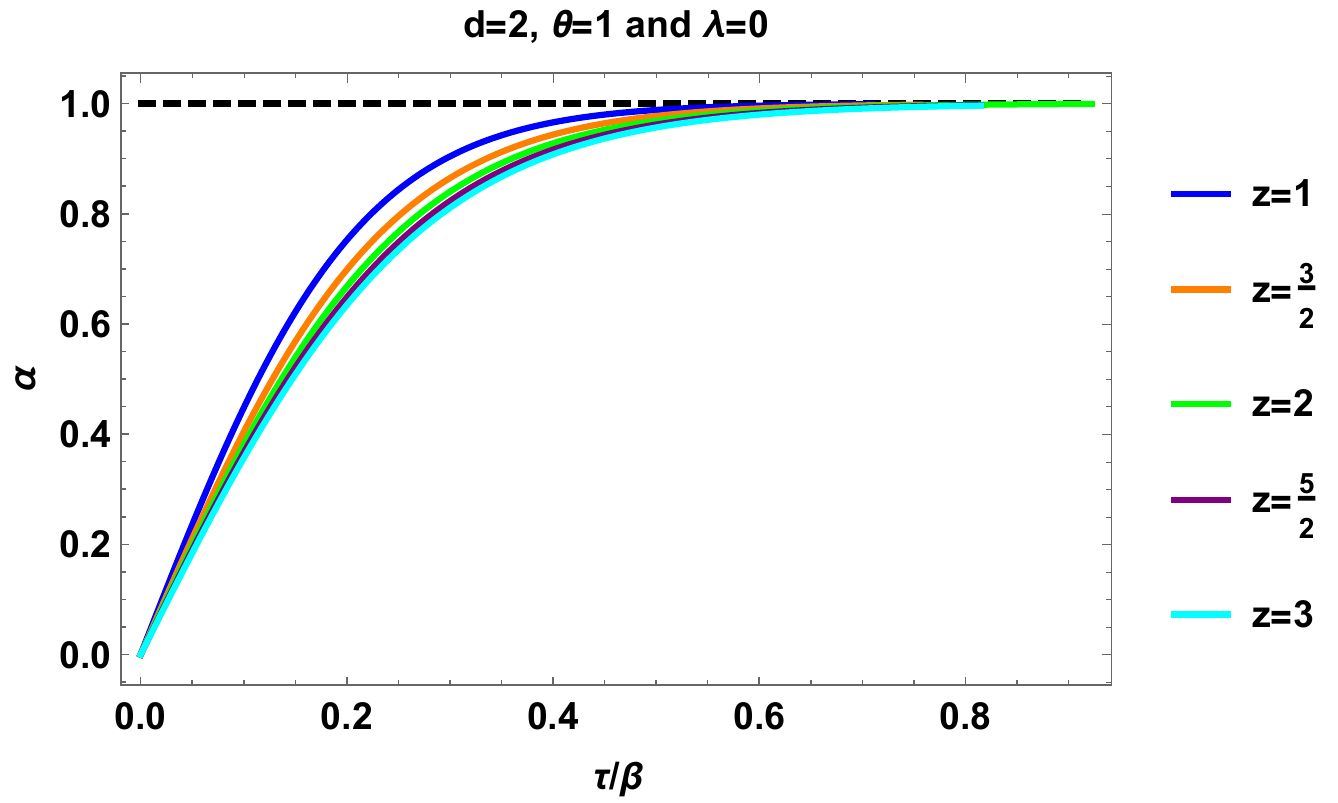}
		\\
		\includegraphics[scale=0.31]{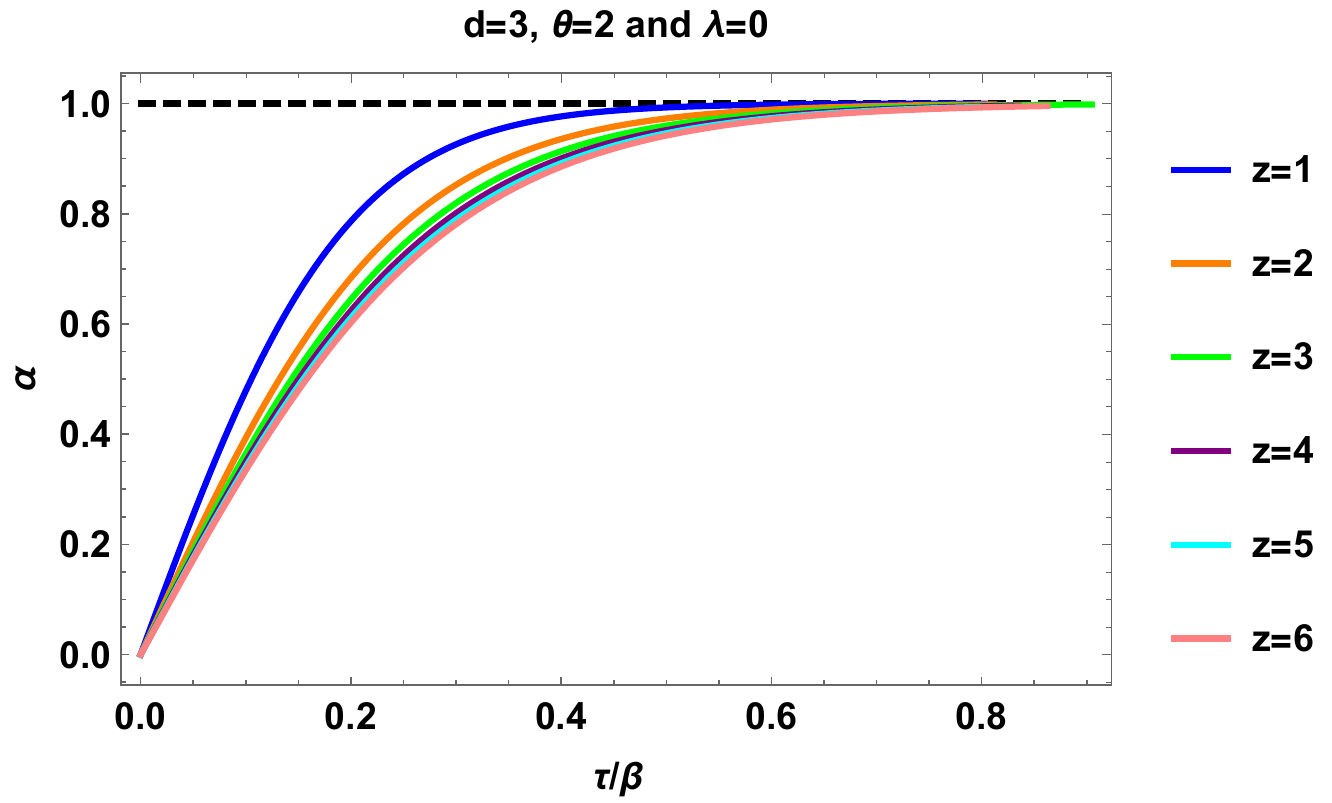}
		\hspace{0.5cm}
		\includegraphics[scale=0.31]{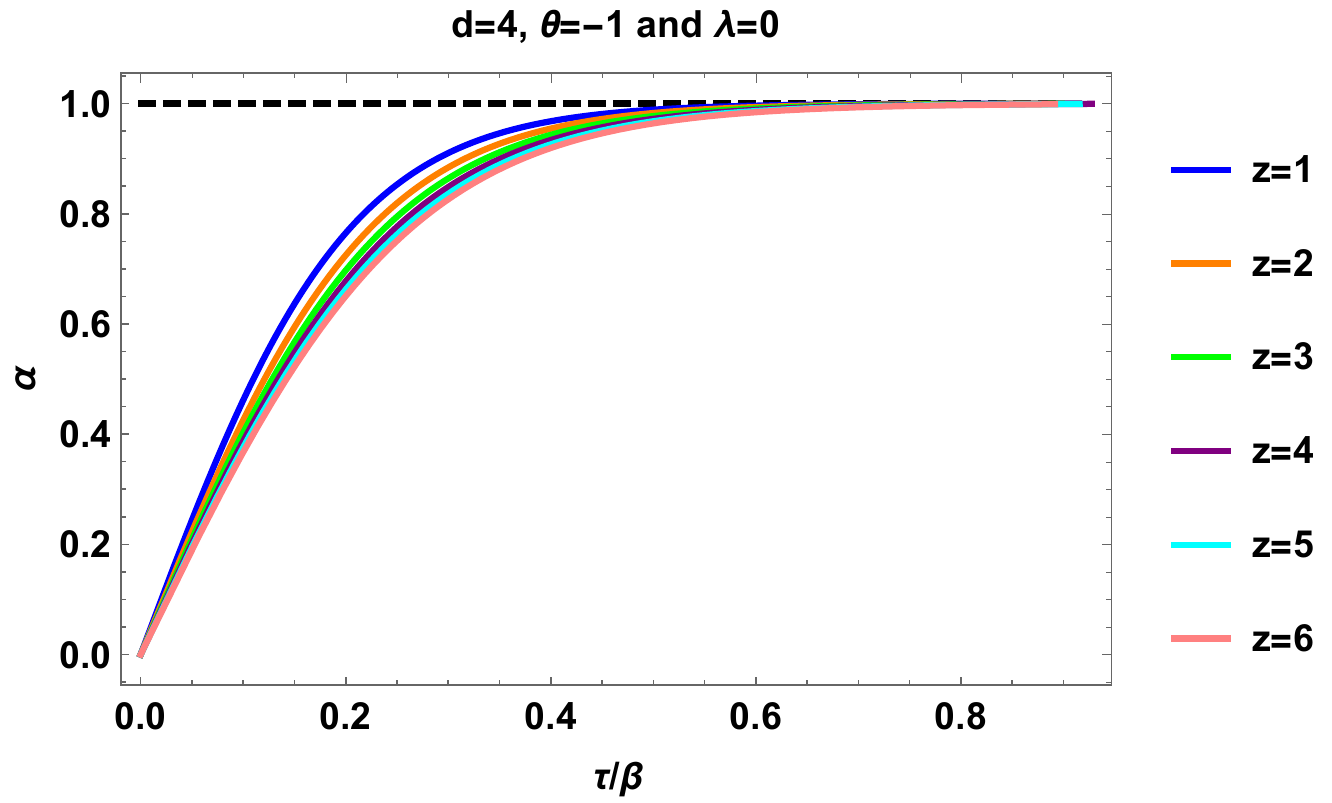}
	\end{center}
	\caption{ The normalized growth rate $\alpha$
	%in eq. \eqref{alpha}, 
	as a function of $\tau / \beta$ for $\lambda=0$ and different values of $d$, $\theta$ and $z$:
	{\it Top Left}) $d=1$ and $\theta= \frac{1}{2}$,
	{\it Top Right}) $d=2$ and $\theta= 1$,
	{\it Down Left}) $d=3$ and $\theta=2$,
	{\it Down Right}) $d=4$ and $\theta=-1$.
	We set $ L= r_h = r_F= 1$.
	}
	\label{fig: dCV-dtau-lambda=0-z}
\end{figure}

\begin{figure}
	\begin{center}
		\includegraphics[scale=0.31]{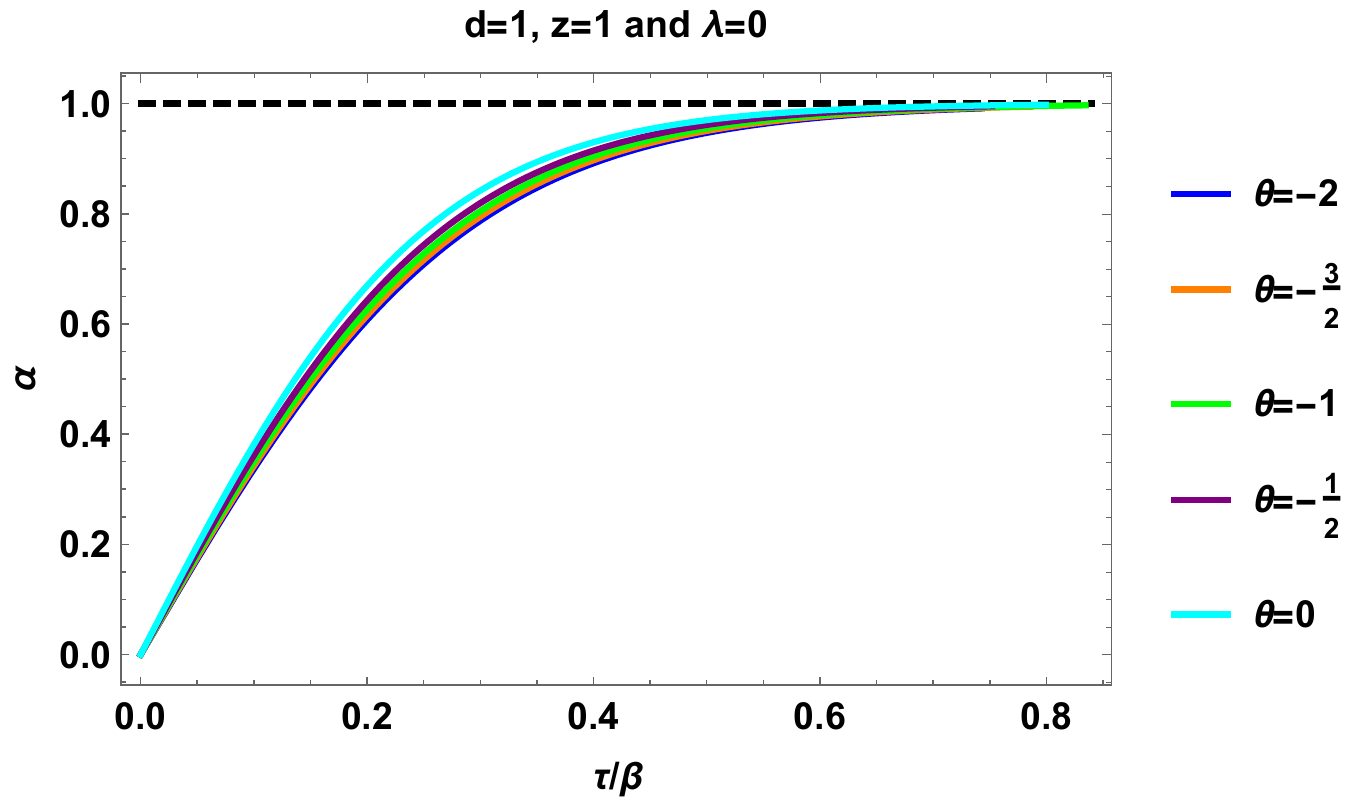}
		\hspace{0.5cm}
		\includegraphics[scale=0.31]{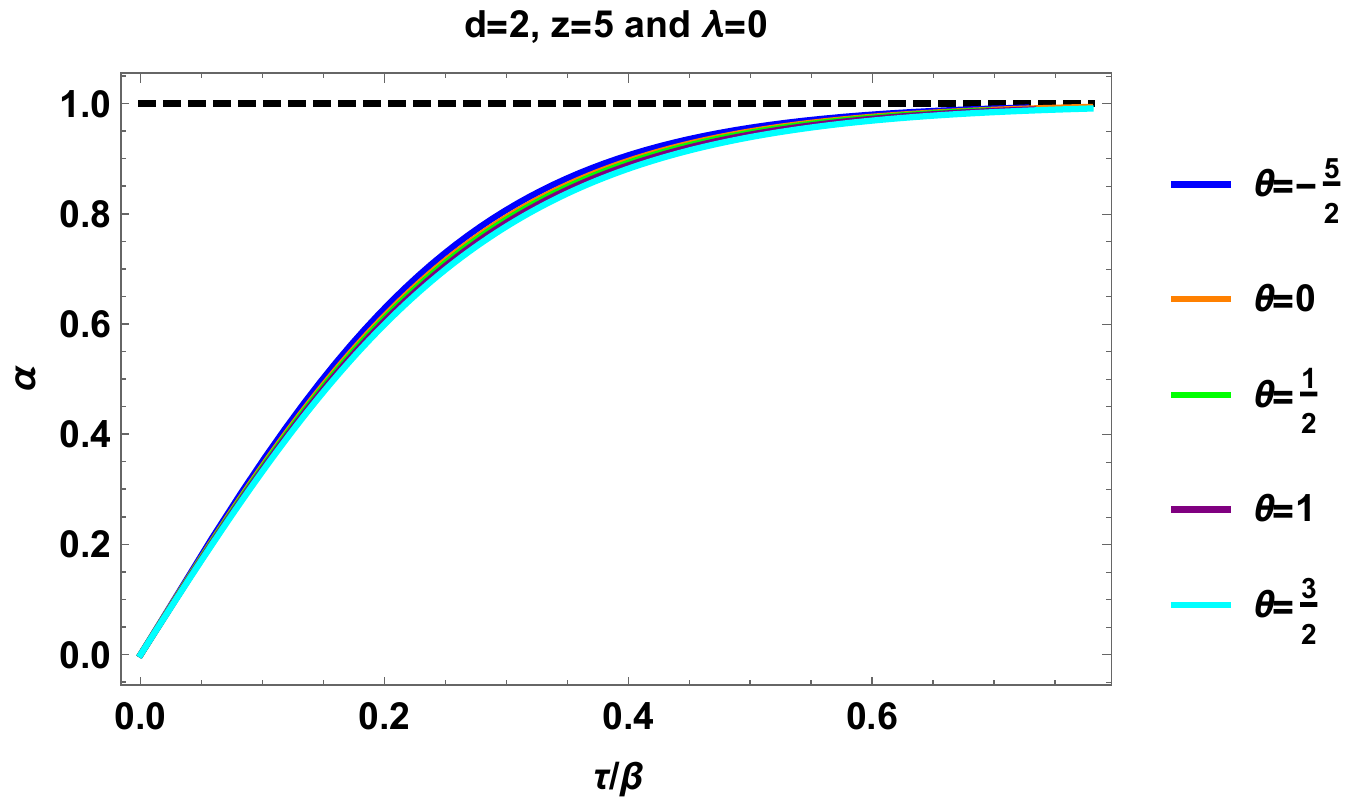}
		\\
		\includegraphics[scale=0.31]{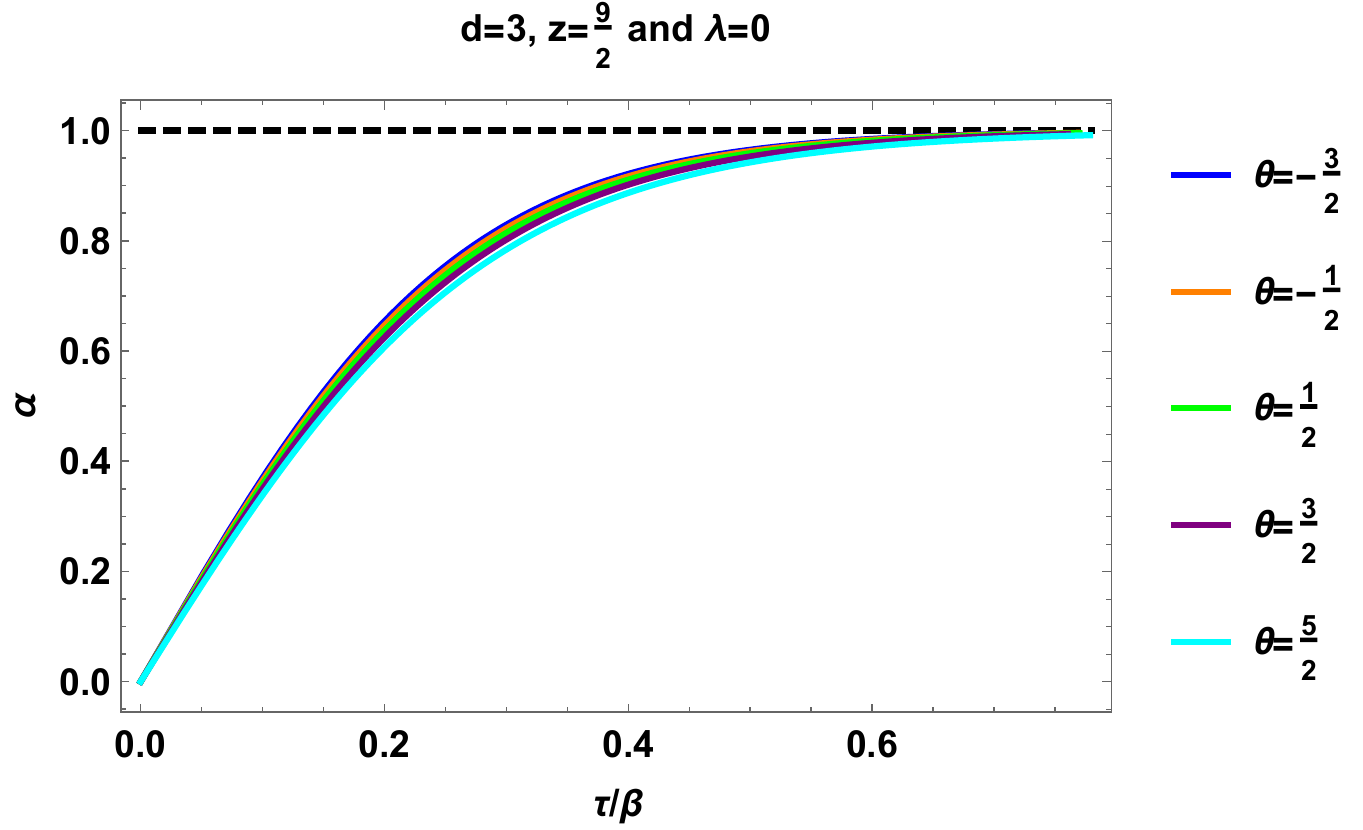}
		\hspace{0.5cm}
		\includegraphics[scale=0.31]{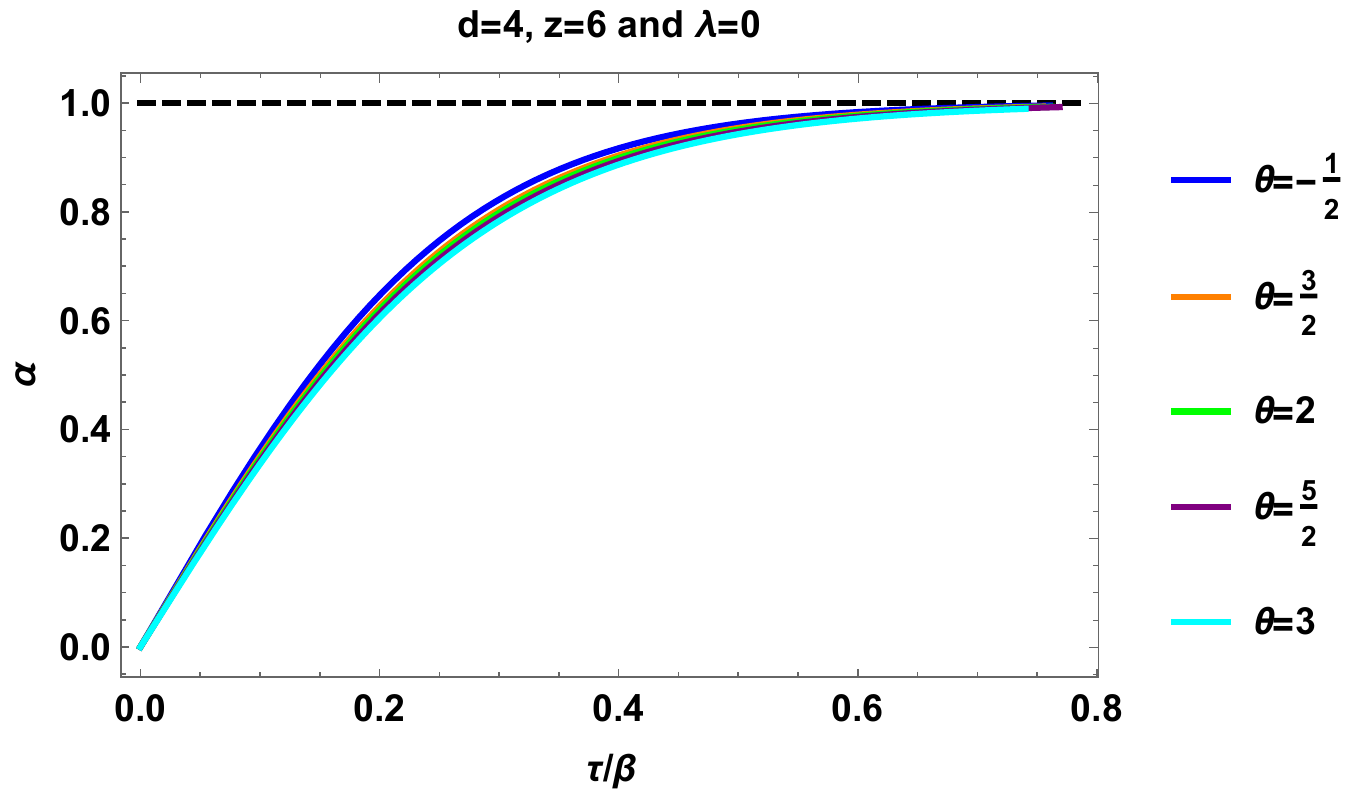}
	\end{center}
	\caption{ The normalized growth rate $\alpha$
	%$\alpha = \frac{ \frac{d \mathcal{C}_V}{d \tau}}{ \lim_{\tau \to \infty} \frac{d \mathcal{C}_V}{d \tau} }$
	% in eq. \eqref{alpha}, 
	 as a function of $\tau / \beta$ for $\lambda=0$ and different values of $d$, $\theta$ and $z$:
	{\it Top Left}) $d=1$ and $z=1$,
	{\it Top Right}) $d=2$ and $z=5$,
	{\it Down Left}) $d=3$ and $z= \frac{9}{2}$,
	{\it Down Right}) $d=4$ and $z=6$.
	We set $ L= r_h = r_F= 1$.
	}
	\label{fig: dCV-dtau-lambda=0-theta}
\end{figure}

In this section, we want to numerically calculate the time dependence of the volume-complexity for an arbitrary time. To this end, we first define (See also \cite{Carmi:2017jqz})
\bea
s= \frac{r}{r_h}, \;\;\;\;\;\;\;\;\;\;\;\;\;\;\;\;\;\;\;\;\;\;\;\;\;\;\;\; s_{\rm min} = \frac{r_{\rm min}}{r_h}, 
\label{s-r}
\eea
and the normalized growth rate
\bea
\alpha = \frac{ \frac{d \mathcal{C}_V}{d \tau}}{ \lim_{\tau \to \infty}  \frac{d \mathcal{C}_V}{d \tau} },
\label{alpha}
\eea 
where the late time growth rate is given in eq. \eqref{dC-dtau-lambda-0-2}. Then by applying eqs. \eqref{U-r} and \eqref{dC-dtau-lambda-0-1}, one can rewrite eq. \eqref{dC-dtau-1} as follows
\bea
\alpha = \frac{2}{\Upsilon} s_{\rm min}^{ \frac{d_e+z}{2} - 2 \theta_e} \sqrt{1 - s_{\rm min}^{d_e +z} }.
\label{alpha-lambda=0}
\eea 
On the other hand, by applying eqs. \eqref{r-star} and \eqref{Pv-U-rmin}, one can recast eq. \eqref{t-R} to the following form
%%%%%%%%%%%%%%%%%%%%%%%%%%%%%
\footnote{Notice that $0 < s_{\rm min} <1 $ and $s > s_{\rm min}$. Moreover, depending on the values of $d$, $\theta$ and $z$, for some values of $s_{\rm min}$, $\tau$ becomes imaginary and one has to discard them. Furthermore, the integrand is singular at $s=1$. To avoid the singularity in the numerical calculations, we changed the limit of the integral in eqs. \eqref{tau-lambda=0} and \eqref{tau-lambda neq 0} as follows $\int_{s_{\rm min}}^\infty ds F(s)= \int_{s_{\rm min}}^{1 - \epsilon} ds F(s)+ \int_{1+ \epsilon}^\infty ds F(s)$, where $\epsilon=2 \times 10^{-8}$. 
%We would like to thank Ghadir Jafari for his helpful comments on this point.
}
%%%%%%%%%%%%%%%%%%%%%%%%%%%%%
\bea
\tau = \frac{(d_e +z) \beta \alpha L}{4 \pi} \int_{s_{\rm min}}^{\infty} ds \frac{s^{d_e -1}}{ \big(1- s^{d_e+z} \big) \sqrt{ s_{\rm min}^{d_e+z- 2 \theta_e} \big(1- s_{\rm min}^{d_e+z}  \big) - s^{d_e +z - 2 \theta_e} \big( 1- s^{d_e +z} \big) } }, \;\;\;\;\;\;
%\nonumber
%\\
\label{tau-lambda=0}
\eea 
where $\beta$ is the inverse temperature. In Figures \ref{fig: dCV-dtau-lambda=0-d}, \ref{fig: dCV-dtau-lambda=0-z} and \ref{fig: dCV-dtau-lambda=0-theta}, we plotted the normalized growth rate $\alpha$ as a function of $\frac{\tau}{\beta}$ for different values of $d$, $\theta$ and $z$. It is observed that, for all values of $d$, $\theta$ and $z$, $C_V$ approaches its late time value given in eq. \eqref{dC-dtau-lambda-0-2} from {\it below} which verifies the discussion below eq. \eqref{dC-dtau-expansion}. Moreover, from Figure \ref{fig: dCV-dtau-lambda=0-d}, it is observed that $\alpha$ is an increasing function of $d$. In Figure \ref{fig: dCV-dtau-lambda=0-z}, $\alpha$ is a decreasing function of $z$. On the other hand, in Figure \ref{fig: dCV-dtau-lambda=0-theta}, it is observed that for $ \{ d=1, z=1 \}$, $\alpha$ is an increasing function of $\theta$. However, for $\{d=2, z=5 \}$, $\{ d=3, z= \frac{9}{2}\}$ and $\{ d=4, z=6\}$ it is a decreasing function of $\theta$. 
%Furthermore, it is an increasing function of $d$ and $\theta$. Moreover, $\alpha$ is a decreasing function of $z$. 
It should be pointed out that for $\theta=0$ and $z=1$, the behavior of the growth rate is again the same as that in ref. \cite{Carmi:2017jqz}.
%%%%%%%%%%%%%%%%%%%%%%%%%%%%%%%%%
%%%%%%%%%%%%%%%%%%%%%%%%%
%%%%%%%%%%%%%%%%%%%%%%%%%%%%%%%%%
%%%%%%%%%%%%%%%%%%%%%%%%%
%%%%%%%%%%%%%%%%%%%%%%%%%%%%%%%%%
%%%%%%%%%%%%%%%%%%%%%%%%%

%%%%%%%%%%%%%%%%%%%%%%%%%%
\subsection{Maxima of the Effective Potential For $\lambda \neq 0$}
\label{Sec: Maxima of the Effective Potential For lambda-neq-0}
%%%%%%%%%%%%%%%%%%%%%%%%%%

In this section, we investigate the local maxima $w_f$ of the effective potential $\hat{U}(w)$
%%%%%%%%%%%%%%%%%%%%%%%%%%%%%
\footnote{For convenience, we work in the $w$ coordinate instead of $r$ in this section. The two coordinates are related to each other by eq. \eqref{w-r-coordinate}.}
%%%%%%%%%%%%%%%%%%%%%%%%%%%%%
which are located inside the horizon, i.e.
\bea
\hat{U}''(w_f) <0, \;\;\;\;\;\;\;\;\;\;\;\;\;\;\;\;\;\;\;\;\;\; 0 < w_f < 1.
\label{w-maxima-inside}
\eea 
As mentioned below  eq. \eqref{dC-dtau-2}, these maxima determine the growth rate of $\mathcal{C}_{\rm gen}$ at late times. For the case $\lambda=0$, we found the local maximum of $\hat{U}(r)$ in eq. \eqref{rf-wf-lambda-0}. In the following, we consider the case $\lambda \neq 0$.
%for the case $\lambda \neq 0$. It was argued in ref. \cite{Belin:2021bga} that, these maxima are the constant-$r$ hypersurface $r=r_f$ where $\Sigma (\tau) $ approaches at late times, i.e. $\tau \rightarrow \infty$. Therefore, the presence of these maxima make $\mathcal{C}_{\rm gen}$ to have a linear growth at late times. Moreover, the value of the potential at $r=r_f$ gives this rate (See eq. \eqref{dC-dtau-2}).
In this case, by taking the derivative of $\hat{U}(w)$ with respect to $w$, one has
\bea
\frac{ d \hat{U} (w)}{ d w} = \frac{A(\lambda, w) B ( \lambda , w)}{w^{1+ \frac{2 \theta_e}{d_e+z}} },
\eea 
where 
\bea
A( \lambda, w) &=& d+1 + (d-1) \left( \frac{r_h}{r_F} \right)^{ 4 \theta_e} w^{-2 + \frac{4 \theta_e}{d_e +z}} \big( 2 w z (z-1) - d_e (d_e -z +2 \big)^2 \lambda,
\label{A}
\\
B(\lambda, w)  &=& B_0 ( w) + \lambda B_1 (w),
\label{B}
\eea 
and the functions $B_{0,1} ( w)$ are  defined as follows
\bea
B_0(w) &=& - w (d+1) \big( d(2w -1) (d+z) + (d +2 -2 (d+1) w ) \theta \big),
\cr && \cr
B_1(w) &=& \left( \frac{r_h}{r_F} \right)^{4 \theta_e} \frac{(d-1) }{w^{1- \frac{4 \theta_e}{d_e+ z}}}
\big( d^2 - 2 w z (z-1) + \theta (\theta+z-2) -d (z+2 ( \theta -1)  \big)
\cr && \cr
&& \times 
\Bigg[
4(w-1) \Big(  \theta \big( (d+ 4w) z^2 - d (4+3d (d+2)) -4 z w \big)
\cr && \cr 
&& 
\;\;\;\;\;\;  + d^2 (d-z+2) (d+z) + \big( 4+ 3d (d+2) -2z \big) \theta^2 - (d+2) \theta^3 \Big)
\cr && \cr 
&&
\;\;\;\;\;\;   + \Big( d (d_e +z) (1-2w)  +2 (w-1)  \theta \Big)
\cr && \cr
&& 
\;\;\;\;\;\;   \times
\Big( d^2 -2w z(z-1) + \theta (z +\theta -2) -d (z+2 ( \theta -1) ) \Big)
\Bigg].
\nonumber
\\
\label{B0-B1}
\eea
By setting $A ( \lambda, w)= 0$ or $B( \lambda, w) =0$, one can find $\lambda$ in terms of $w$ as follows
%\bea
%A ( \lambda, w) \!\!\!\! &=&  \!\!\!\! 0,
%\\
%B( \lambda, w) \!\!\!\! &=&  \!\!\!\! 0,
%\label{A=0-B=0}
%\eea
\bea
\lambda &=& - \left( \frac{r_F}{r_h} \right)^{4 \theta_e} \frac{(d+1) w^{2 - \frac{4 \theta_e}{d_e +z}}}{ (d-1) \big( 2 z (z-1)  w  - d_e(d_e -z +2) \big)^2},
\label{lambda-A}
\\
\lambda &=& - \frac{B_0(w)}{B_1(w)},
\label{lambda-B-1}
\eea 
%\bea
%\lambda = - \frac{(d+1) w^2}{ (d-1) (d^2 - d (z-2) - 2 z (z-1)  w)}.
%\label{lambda-A}
%\eea 
respectively. Next, from the above equations, one can find the extrema $w$ of $\hat{U}(w)$. However, it is not possible to solve eqs. \eqref{lambda-A} and \eqref{lambda-B-1} analytically for arbitrary values of $d$, $\theta$ and $z$. Therefore, in the following we consider some specific values of these parameters for which there are analytic solutions.
%we set $\theta =0$, in the following. 

%%%%%%%%%%%%%%%%%%%%%%%%%%
\subsubsection{$\theta=0$}
\label{Sec: theta=0}
%%%%%%%%%%%%%%%%%%%%%%%%%%

\begin{figure}
	\begin{center}
		\includegraphics[scale=0.26]{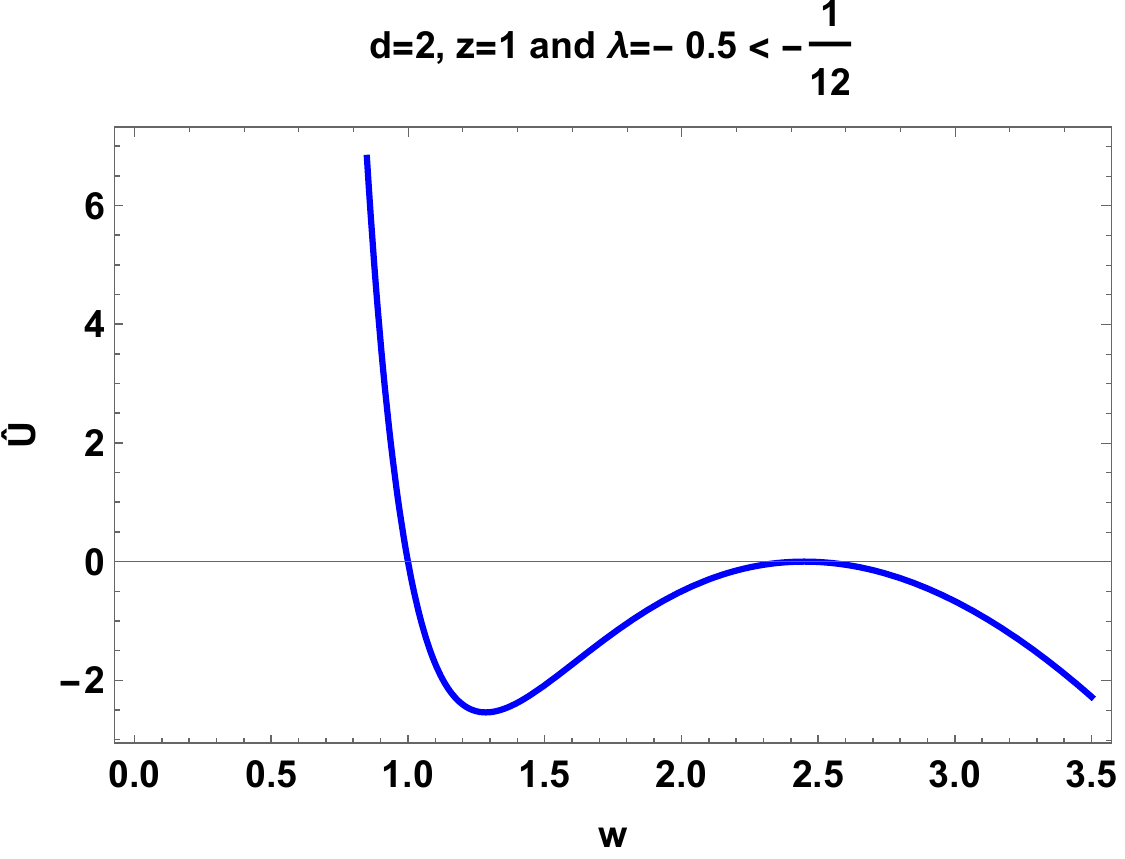}
		\hspace{0.1cm}
		\includegraphics[scale=0.26]{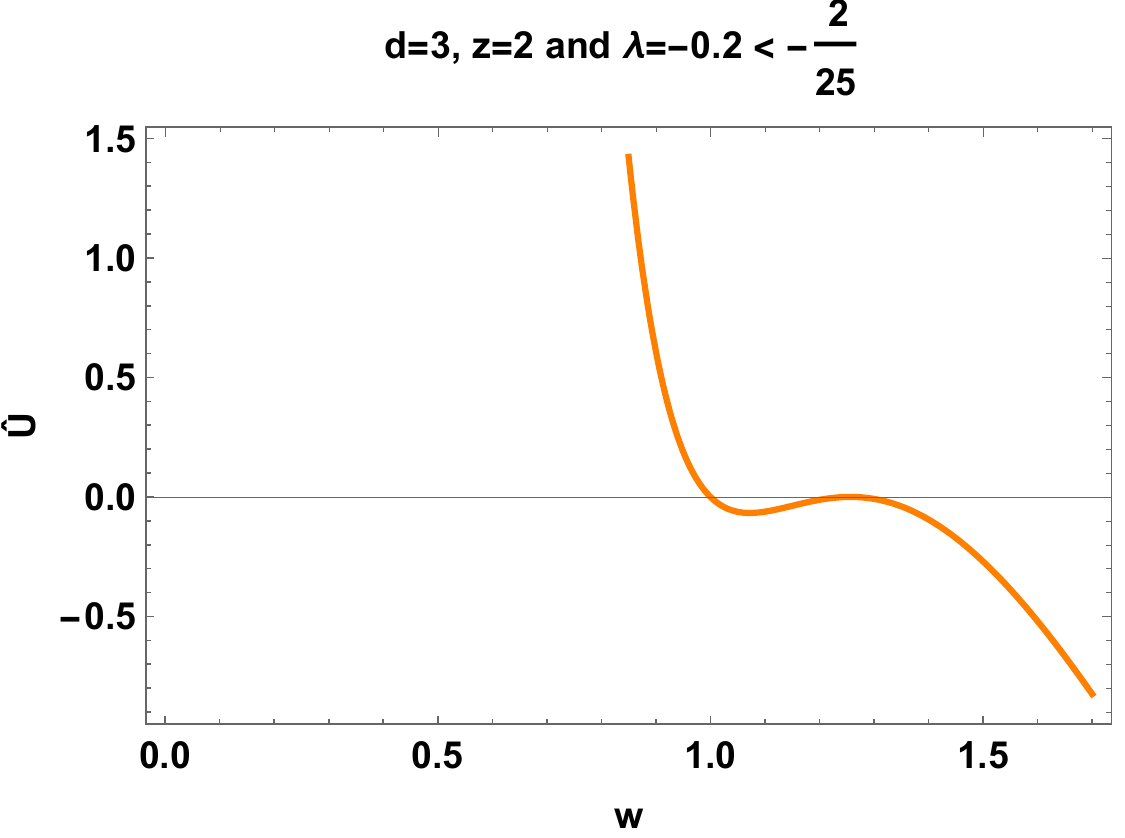}
        \hspace{0.1cm}
		\includegraphics[scale=0.26]{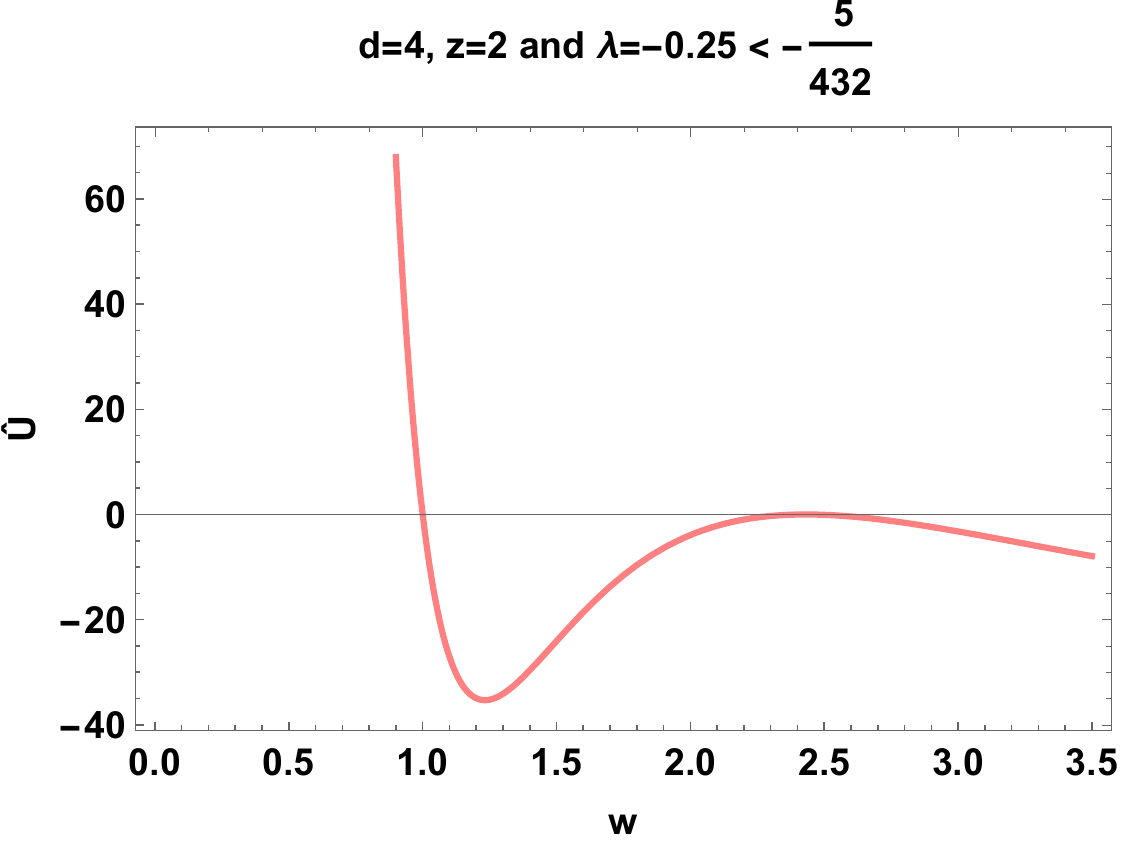}
	\end{center}
	\caption{ The effective potential $\hat{U}(w)$ as a function of $w$ for $\theta=0$ and different values of $d$, $z$ and $\lambda$. 
	%	In the first row $d=2$, second row $d=3$, third row $d=4$ and the last one $d=5$.
		%{\it Left)} $z=1$ {\it Middle)} $z=5$ {\it Right)} $z=6$. 
		For the values of  $\lambda$ given in Table \ref{Table: wfA1-outside}, there is a maximum $w_{fA1}$ outside the horizon, i.e. $w_{fA1} > 1$. We set $L=r_h=r_F=1$.
	}
	\label{fig: theta=0-z-wfA1-outside}
\end{figure}

\begin{figure}
	%\vspace{-1cm}
	\begin{center}
		\includegraphics[scale=0.26]{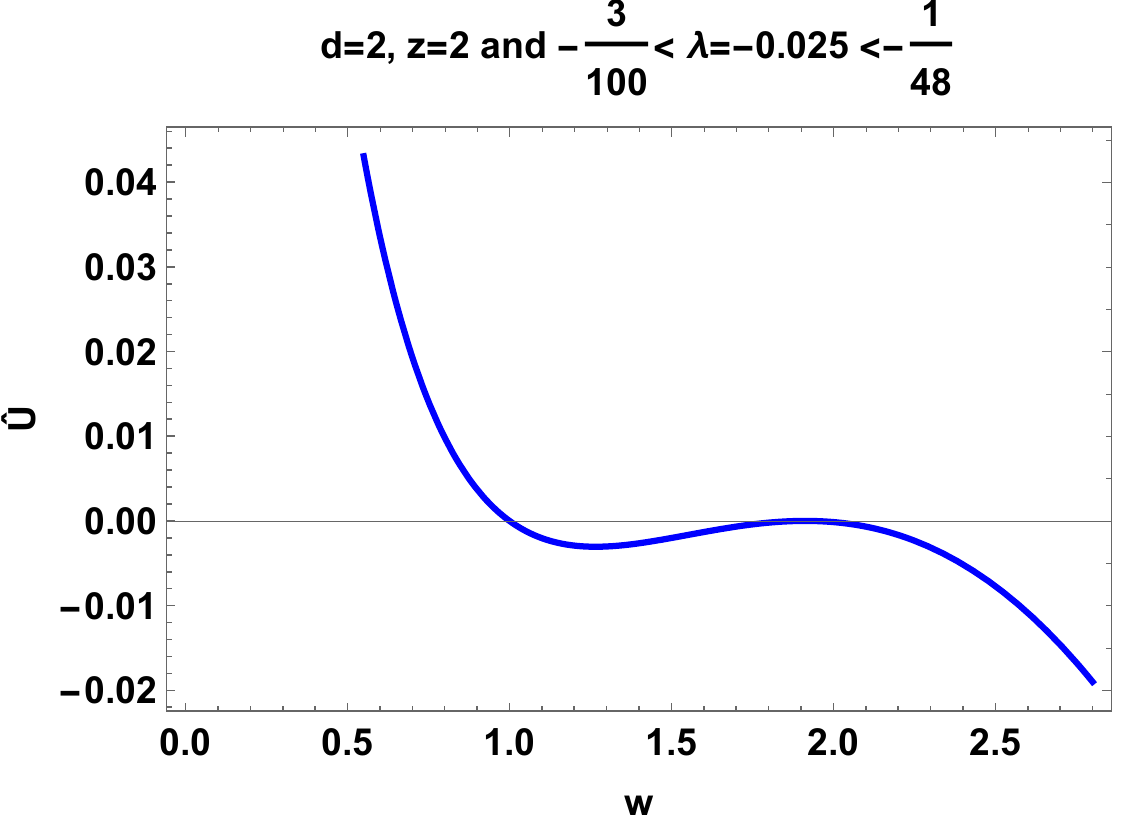}
		\hspace{0.1cm}
		\includegraphics[scale=0.26]{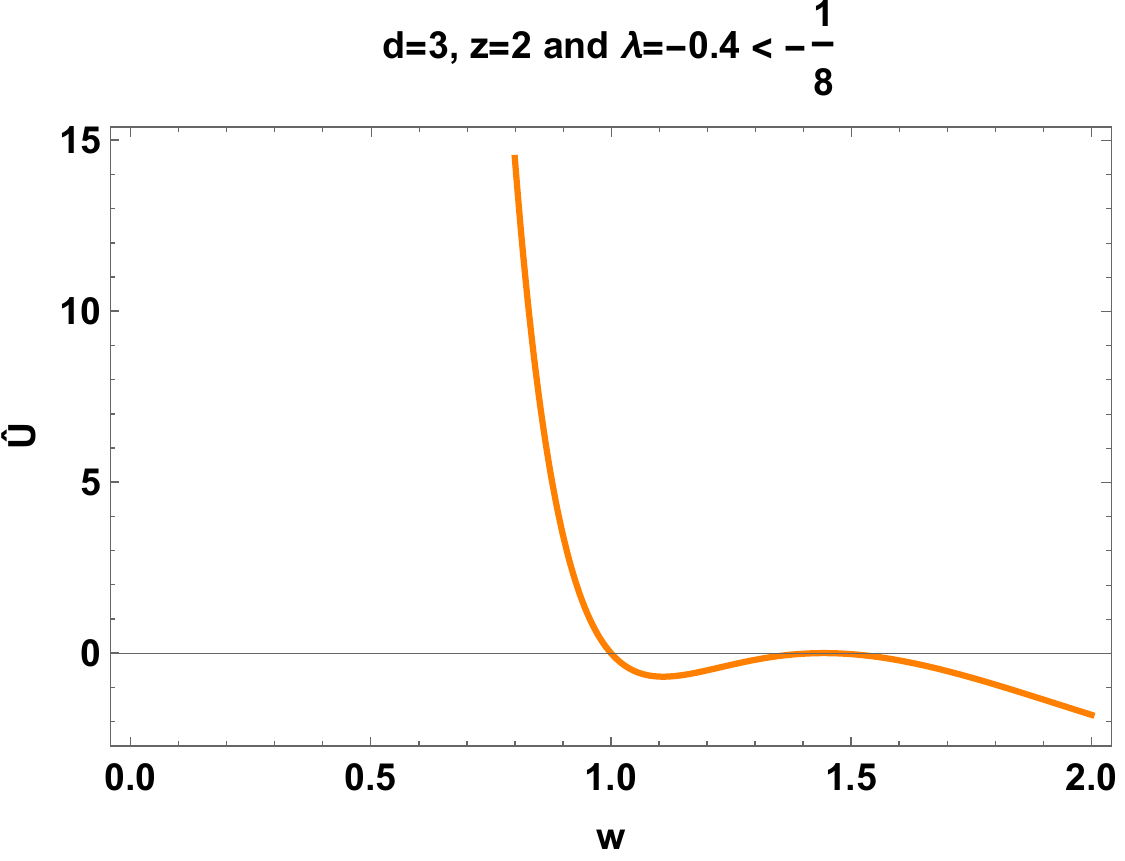}
        \hspace{0.1cm}
		\includegraphics[scale=0.26]{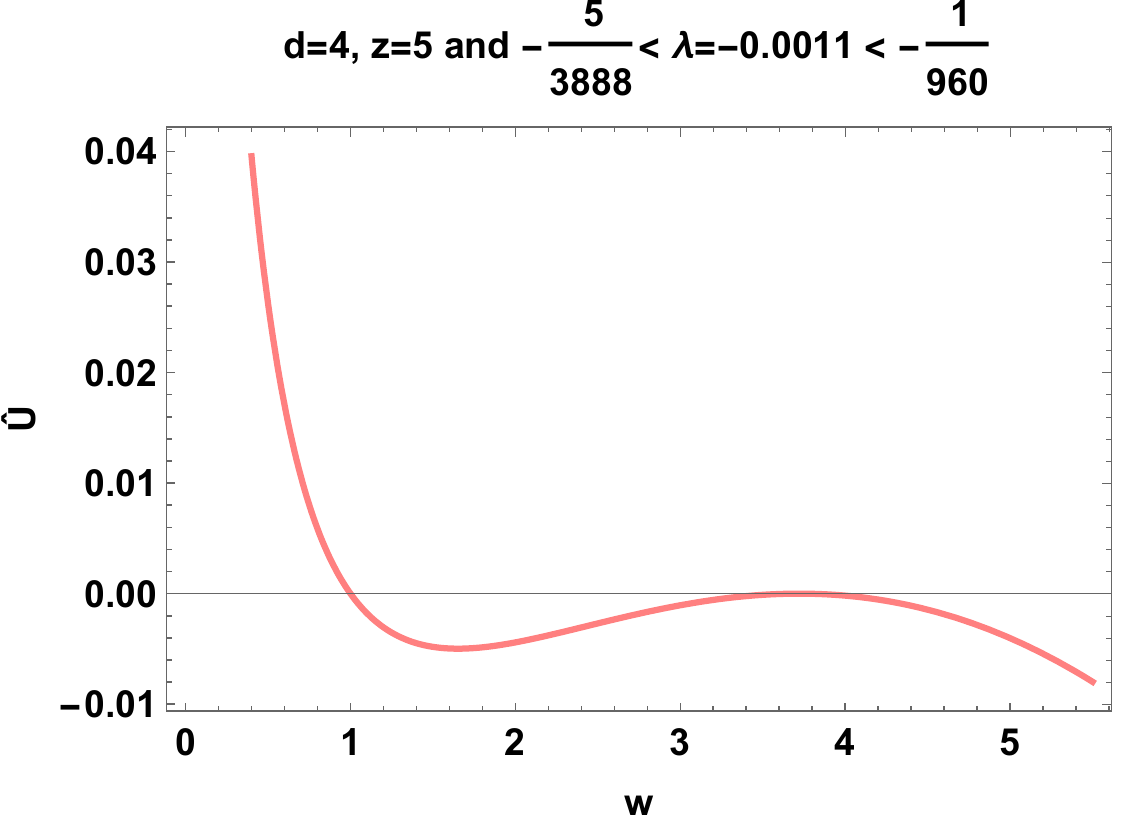}
	\end{center}
	\caption{ The effective potential $\hat{U}(w)$ as a function of $w$ for $\theta=0$ and different values of $d$, $z$ and $\lambda$. 
		%	In the first row $d=2$, second row $d=3$, third row $d=4$ and the last one $d=5$.
		%{\it Left)} $z=1$ {\it Middle)} $z=5$ {\it Right)} $z=6$. 
		For the values of  $\lambda$ given in Table \ref{Table: wfA2-outside}, there is a maximum $w_{fA2}$ outside the horizon. 
		%We set $L=r_h=r_F=1$.
	}
	\label{fig: theta=0-z-wfA2-outside}
\end{figure}

\begin{figure}
%	\vspace{-0.5cm}
	\begin{center}
		\includegraphics[scale=0.26]{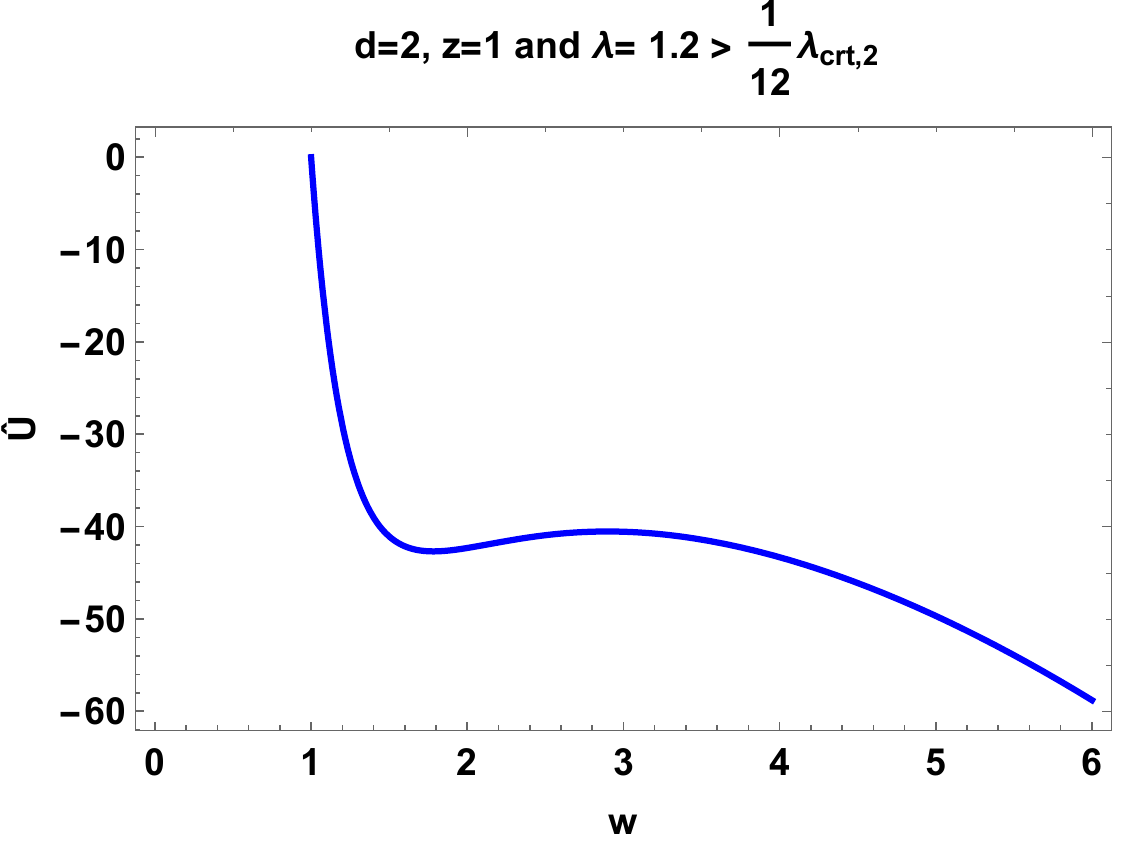}
		\hspace{0.1cm}
		\includegraphics[scale=0.26]{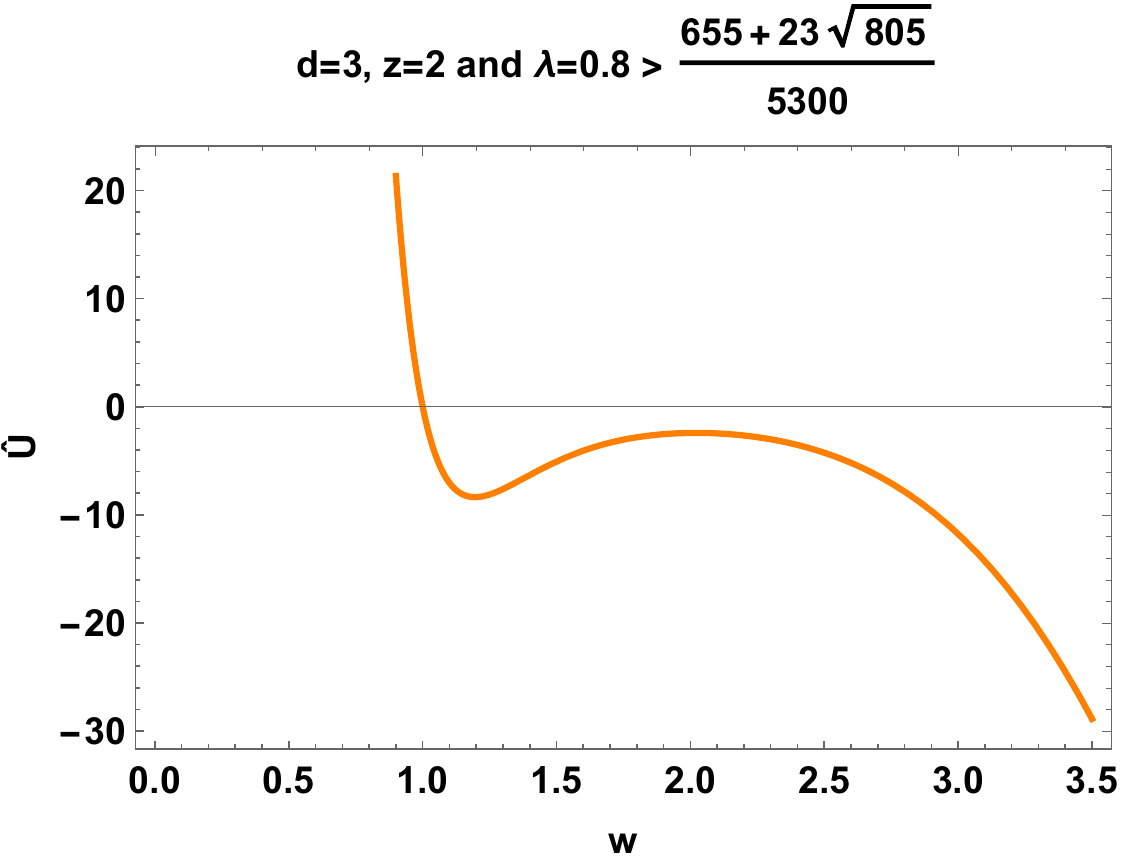}
		\hspace{0.1cm}
		\includegraphics[scale=0.26]{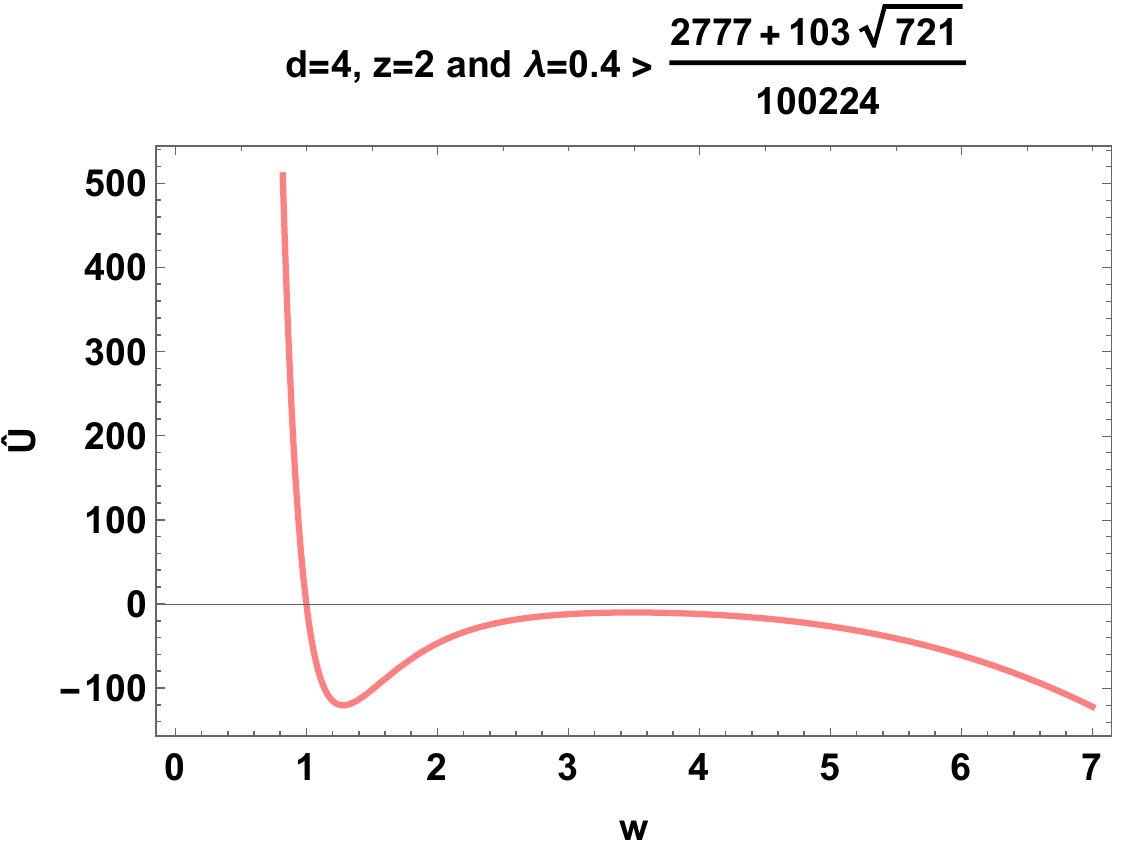}
	\end{center}
	\caption{ The effective potential $\hat{U}(w)$ as a function of $w$ for $\theta=0$ and different values of $d$, $z$ and $\lambda$.
		For the values of $\lambda$ given in Table \ref{Table: wfB-outside} there is a maximum $w_{fB}$ outside the horizon.
		%, i.e. $w_{fB} > 1$. 
		%The appropriate range of $\lambda$ is given in Table \ref{Table: wfB-outside}. 
		%We set $L=r_h=r_F=1$.
	}
	\label{fig: theta=0-z-wfB-outside}
\end{figure}

\begin{figure}
%	\vspace{-2cm}
	\begin{center}
		\includegraphics[scale=0.24]{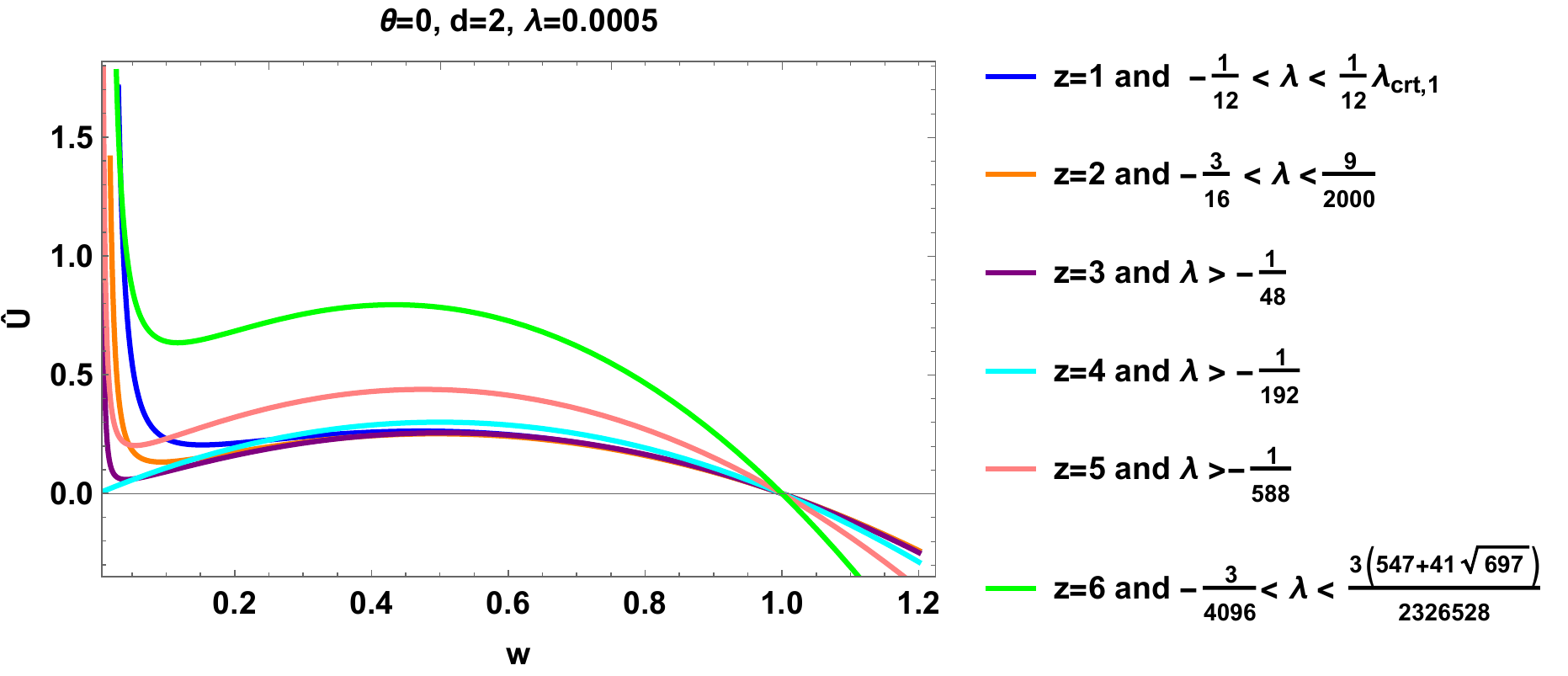}
		\includegraphics[scale=0.24]{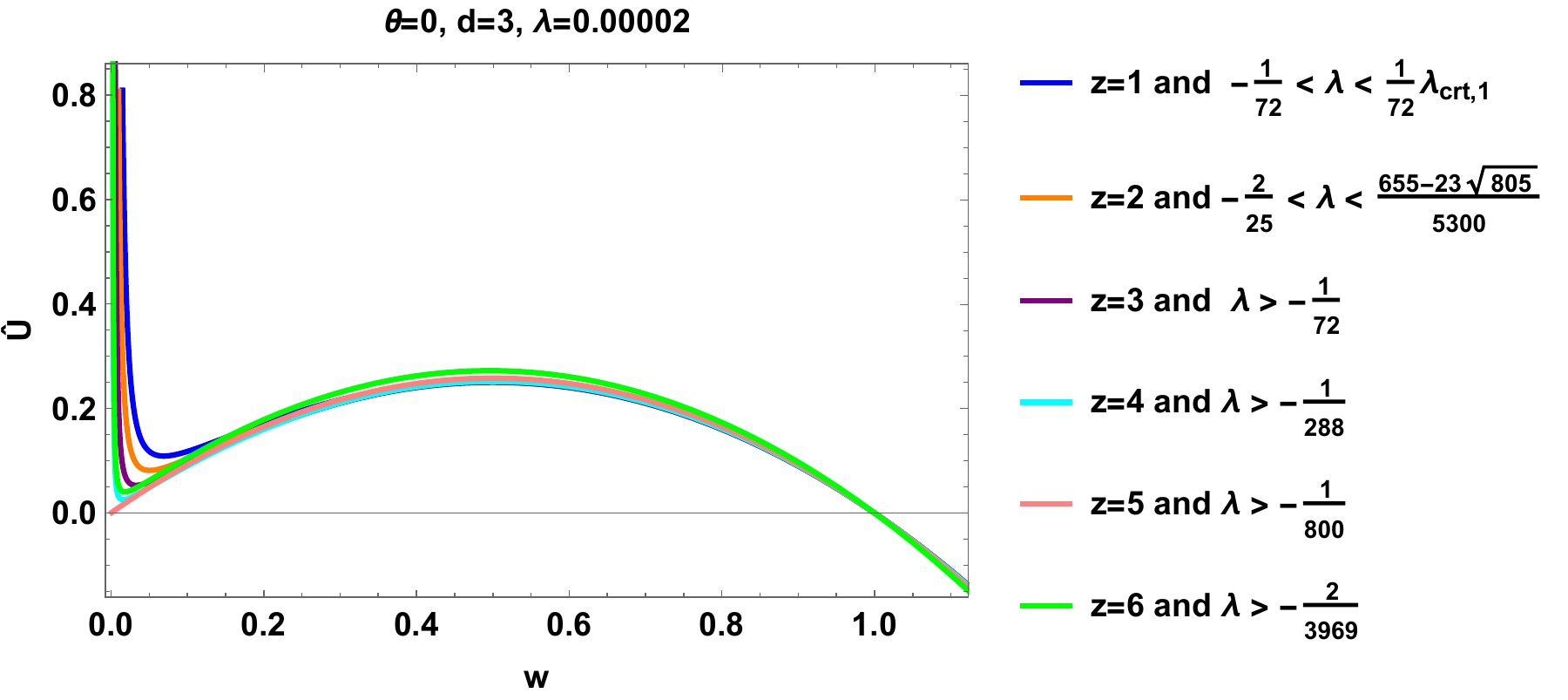}
		\\
		\includegraphics[scale=0.24]{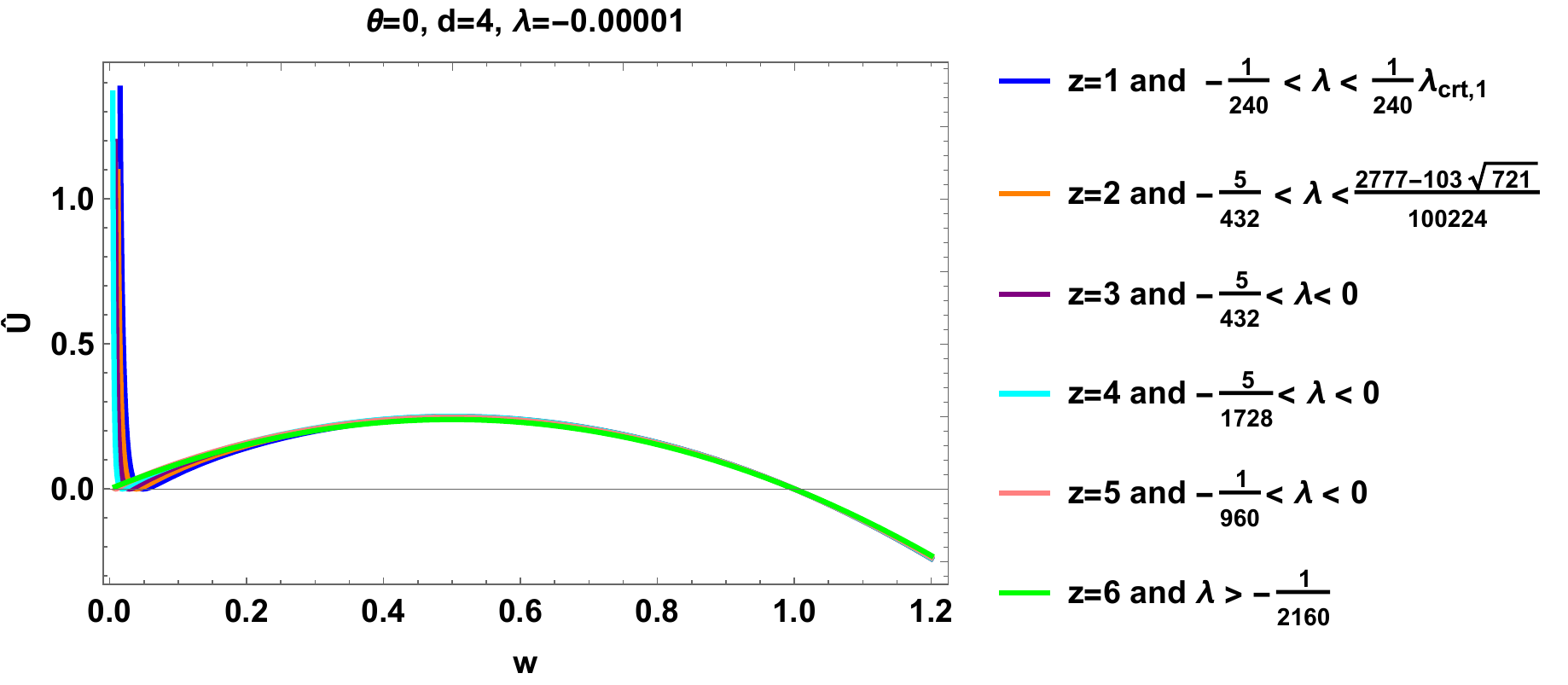}
		\includegraphics[scale=0.24]{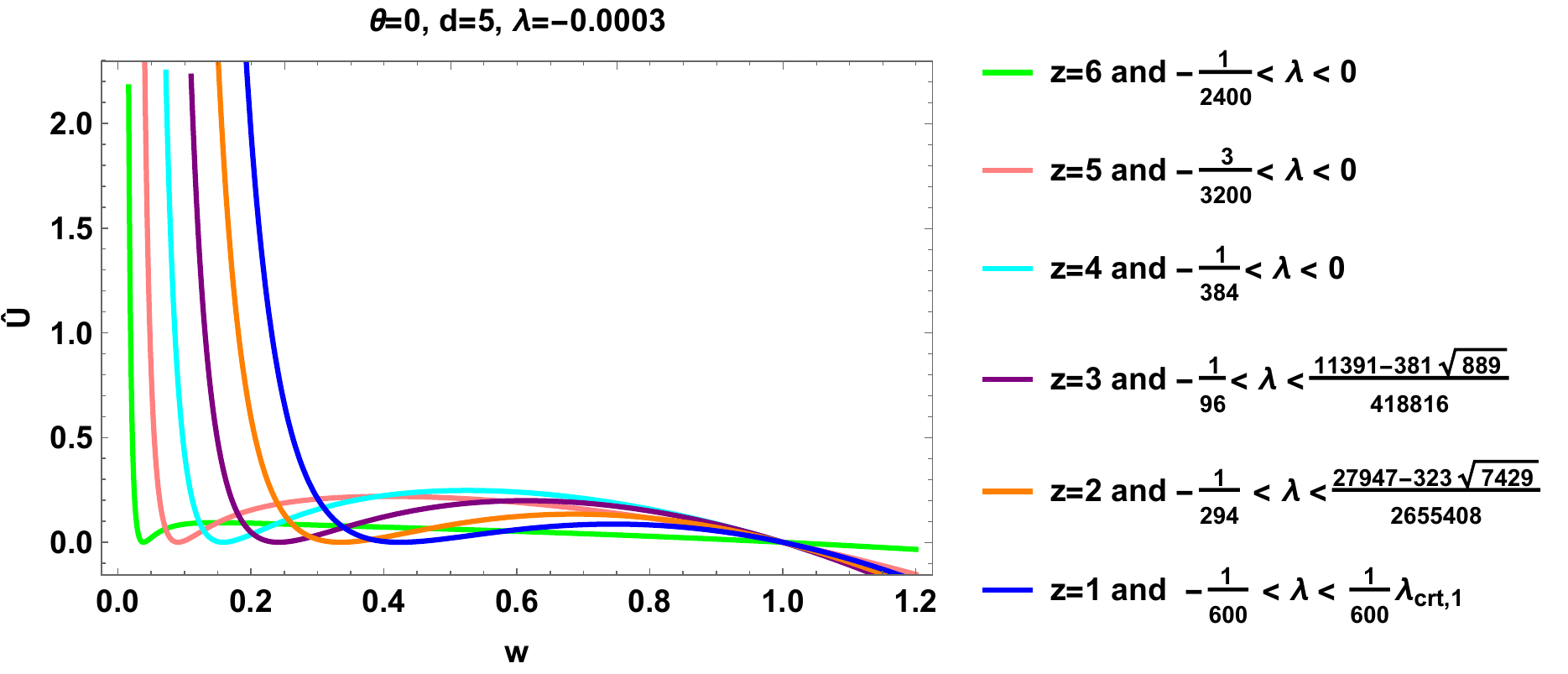}
	\end{center}
	\caption{The effective potential $\hat{U}(w)$ as a function of $w$ for $\theta=0$ and different values of $d$ and $z$. For all values of $z$, there is a maximum $w_{fB}$ inside the horizon.
	%, i.e. $w_{fB} < 1$. 
	The appropriate ranges of $\lambda$ are given in Table \ref{Table: wfB-inside}. 
	%We set $L=r_h=r_F=1$.
	}
	\label{fig: theta=0-z-wfB-inside}
\end{figure}

\begin{figure}
	\begin{center}
	\includegraphics[scale=0.31]{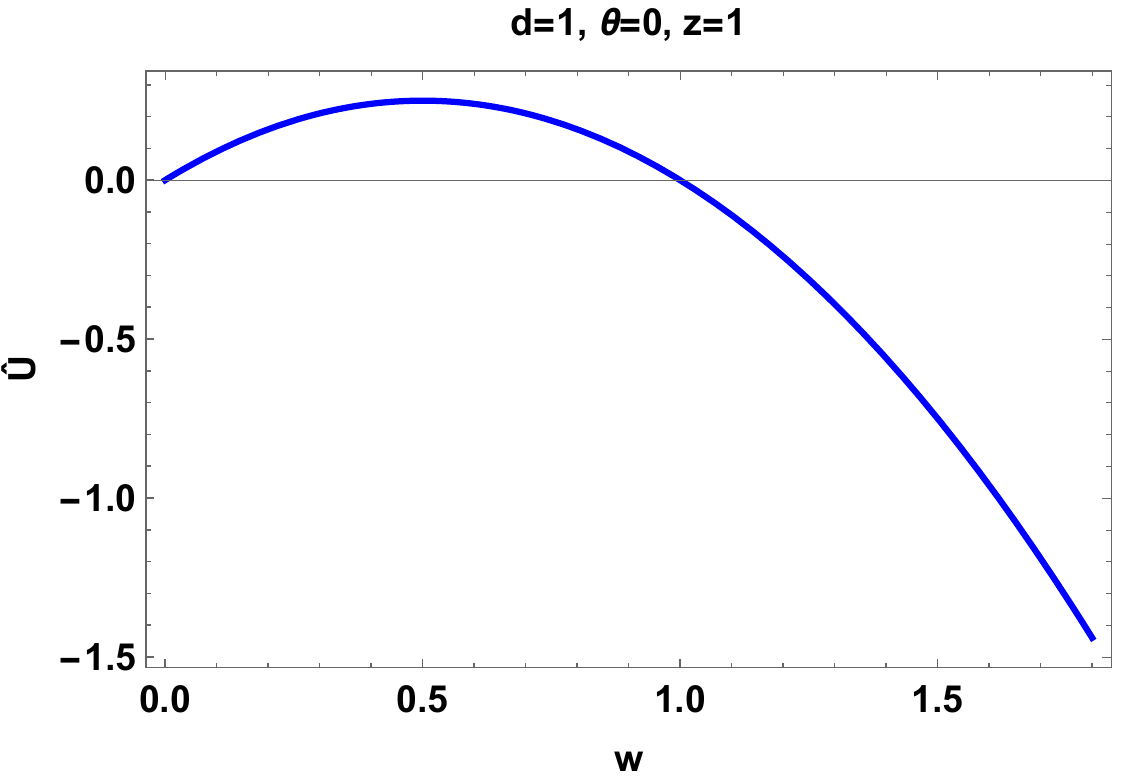}
	\end{center}
	\caption{The effective potential $\hat{U}(w)$ as a function of $w$ for $d=1$, $\theta=0$ and $z=1$. In this case, $\hat{U}(w)$ is independent of $\lambda$, and there is a maximum inside the horizon at $w_{fB} = \frac{1}{2}$.
	}
	\label{fig: theta=0-z=1-d=1}
\end{figure}

In this case, the background becomes a Lifshitz black brane. Moreover, it is straightforward to show that eq. \eqref{lambda-A} has the following solutions when $d \neq 1$
\bea
w_{fA1} &=& \frac{d \left( 2 z  (d-1) (d - z +2) (z-1) \lambda + \sqrt{ \lambda ( 1- d^2)} | d - z +2 | \right)}{ d+1 + 4 (d-1) (z-1)^2 z^2 \lambda},
\\
w_{fA2} &=& \frac{d \left( 2 z  (d-1) (d - z +2) (z-1) \lambda - \sqrt{ \lambda ( 1- d^2)} | d - z +2 | \right)}{ d+1 + 4 (d-1) (z-1)^2 z^2 \lambda}.
\label{w-fA1}
\eea 
It is clear that to have a real value for $w_{fA1,2}$, one needs $\lambda < 0$. Moreover, asking for $w_{fA1,2}$ to be a local maximum, i.e. $\hat{U}''(w_{fA1,2}) <0$, imposes a constraint on the value of $\lambda$. These values are reported in Tables \ref{Table: wfA1-outside} and \ref{Table: wfA2-outside}, respectively. It is straightforward to verify that for these values of $\lambda$, $w_{fA1,2}$ are always located outside the horizon. Therefore, they have no contributions to the late time growth rate of $\mathcal{C}_{\rm gen}$, and we are not interested to them here. In Figures \ref{fig: theta=0-z-wfA1-outside} and \ref{fig: theta=0-z-wfA2-outside}, we plotted $\hat{U}(w)$ for $\theta =0$ and different values of $d$ and $z$. It is observed that $w_{fA1,2}$ are maxima of $\hat{U}(w)$ and located {\it outside} the horizon, i.e. $w_{fA1,2} > 1$. On the other hand, for $d=1$, from eq. \eqref{A}, one has $A(\lambda,w) = d+1$. Therefore, equation $A(\lambda, w)= 0$, has no solutions.
%It is straightforward to check that $w_{fA}$ is a local maximum outside the horizon, i.e. $w_{fA} > 1$, for some values of $\lambda$ which are shown in Table \ref{Table: wfA1-outside}.
%%%%%%%%%%%%%%%%%%%%%%%%%%%
\begin{table}
\begin{center}
	\scalebox{0.9}{
	\begin{tabular}{| l | l | l | l | l | l |}
		\hline
		 & $d=2$ & $d=3$ & $d=4$ & $d=5$ 
		 \\ \hline
		$z=1$ & $ \lambda < - \frac{1}{12}$ & $ \lambda < - \frac{1}{72}$ & $ \lambda < - \frac{1}{240}$ & $ \lambda < - \frac{1}{600}$
	    \\ \hline
		$ z=2$ & $\hat{U}'' (w_{fA}) > 0$	& $\lambda < - \frac{2}{25} $ & $ \lambda < - \frac{5}{432}$ & $ \lambda < - \frac{1}{294}$
		\\ \hline
		$ z=3$ & $\hat{U}'' (w_{fA}) > 0$ & $\hat{U}'' (w_{fA}) > 0$ & $\hat{U}'' (w_{fA}) > 0$ & $ \lambda < - \frac{3}{128}$
		\\ \hline
		$ z=4$ & $w_{fA} =0$ & $\hat{U}'' (w_{fA}) > 0$  & $\hat{U}'' (w_{fA}) > 0$ & $\hat{U}'' (w_{fA}) > 0$
		\\ \hline
		$ z=5$ & $ - \frac{3}{1600} < \lambda < - \frac{1}{588} $ & $w_{fA} =0$ & $\hat{U}'' (w_{fA}) > 0$ & $\hat{U}'' (w_{fA}) > 0$
		\\ \hline
		$ z=6$ & $ - \frac{1}{1200} < \lambda < - \frac{3}{4096} $ & $ - \frac{1}{1800} < \lambda < - \frac{2}{3969}$ & $w_{fA} =0$ & $\hat{U}'' (w_{fA}) > 0$
		\\ \hline
	\end{tabular} }
	\caption{The appropriate ranges of the coupling constant $\lambda$ for which $w_{fA1}$ is a maximum outside the horizon. For some values of $d$ and $z$, one has $\hat{U}'' (w_{fA1}) > 0$. In these cases, $w_{fA1}$ is a minimum. Moreover, for some cases, one has $w_{fA1}=0$, which should be discarded.}
\label{Table: wfA1-outside}
\end{center}
\end{table}
%%%%%%%%%%%%%%%%%%%%%%%%%%%
%%%%%%%%%%%%%%%%%%%%%%%%%%%
\begin{table}
	\begin{center}
		\scalebox{0.9}{
		\begin{tabular}{| l | l | l | l | l | l |}
			\hline
			& $d=2$ & $d=3$ & $d=4$ & $d=5$ 
			\\ \hline
			$z=1$ & There is no $ \lambda$ &There is no $ \lambda$ & There is no $ \lambda$ & There is no $ \lambda$
			\\ \hline
			$ z=2$ & $ \lambda < - \frac{3}{16}$ & $\lambda < - \frac{1}{8} $ & $ \lambda < - \frac{5}{48}$ & $ \lambda < - \frac{3}{32}$
			\\ \hline
			$ z=3$ & $ - \frac{3}{100} < \lambda < - \frac{1}{48}$ & $- \frac{1}{18} < \lambda < - \frac{1}{72}$ & $  \lambda < - \frac{5}{432}$ & $ \lambda < - \frac{1}{96}$
			\\ \hline
			$ z=4$ & There is no $\lambda$ & $ - \frac{2}{441} < \lambda < - \frac{1}{288}$  & $ - \frac{5}{768} < \lambda < - \frac{5}{1728}$ & $ - \frac{1}{54} < \lambda < - \frac{1}{384}$
			\\ \hline
			$ z=5$ & There is no $\lambda$ & There is no $ \lambda$ & $ - \frac{5}{3888} < \lambda < - \frac{1}{960}$ & $ - \frac{1}{600} < \lambda < - \frac{3}{3200}$
			\\ \hline
			$ z=6$ & There is no $\lambda$ & There is no $ \lambda$ & There is no $ \lambda$ & $ - \frac{3}{6050} < \lambda < - \frac{1}{2400}$
			\\ \hline
		\end{tabular} }
		\caption{The appropriate ranges of the coupling constant $\lambda$ for which $w_{fA2}$ is a maximum outside the horizon. For some cases, there is no $\lambda$.}
		\label{Table: wfA2-outside}
	\end{center}
\end{table}
%%%%%%%%%%%%%%%%%%%%%%%%%%%
%%%%%%%%%%%%%%%%%%%%%%%%%%%%%%%%%
%%%%%%%%%%%%%%%%%%%%%%%%%
%%%%%%%%%%%%%%%%%%%%%%%%%%%%%%%%%
%%%%%%%%%%%%%%%%%%%%%%%%%
%It should be pointed out that to have a real value for $w_{fA}$, one should has $\lambda <0$. 
%\\Moreover, for $z=1$ and 
%%$ \lambda <  - \frac{1}{(d-1) d^2 (d+1)}$
%%$ - \frac{1}{12} <\lambda <0$
%\bea
%\tilde{\lambda} < - 1,
%\label{z=1-lambda-wfA-max-outside}
%\eea 
%where the effective coupling constant $\tilde{\lambda}$ is defined as follows
%\bea 
%\tilde{\lambda} = d^2 (d^2 -1) \lambda,
%%(d-1) d^2 (d+1) \lambda
%\label{lambda-t-z=1}
%\eea
%$w_{fA}$ is a local maximum located outside the horizon (See the orange curves in the right panels of figure \ref{fig: theta=0-z=1}). Furthermore, form Table \ref{Table: wfA1-outside}, it is clear that for $d- z = -1, 0$, one has $\hat{U}''(w_{fA}) >0 $, and hence $w_{fA}$ is a minimum instead of a maximum. Moreover, for $d \geq 4$ and $d-z = +1$, one has $\hat{U}''(w_{fA}) >0 $. On the other hand, for $d- z = -2$, one has $w_{fA} = 0$.
% and for $ \lambda < - \frac{1}{12}$, it is a maximum located outside the horizon (See the orange curve in the right panel of figure \ref{fig: z=1}).
%\\In Figure \ref{fig: theta=0-z=1}, we restricted ourself to the case $z=1$. In the left panels of Figure \ref{fig: theta=0-z=1}, the maxima are located inside the horizon. However, in the right panels of Figure \ref{fig: theta=0-z=1}, either there is no maximum or it is located outside the horizon. 
\\Next, we find the maxima given by eq. \eqref{lambda-B-1}. By setting $\theta =0$, it simplifies to
\bea
\lambda \! = \! \frac{ (d+1) w^2 ( 2 w -1) }{( d -1 ) \! \left( d^2 \! - \! d (z-2) -2 w z (z-1) \right) \! \left( d^2 ( 2 w \! - \! 3) - d (2 w \! - \! 3) (z -2) \! + \! 2 z (z-1 )w (2 w \! - \! 1)  \right)},
\nonumber
\\
\label{lambda-B-2}
\eea 
from which one can find another extremum $w_{fB}$. 
%%%%%%%%%%%%%%%%%%%%%%%%%%%%%
\footnote{It should be emphasized that eq. \eqref{lambda-B-2} gives three extrema $w_{fB1}$, $w_{fB2}$ and $w_{fB3}$.
	%% depending on the value of $\lambda$. 
The extrema $w_{fB2}$ and $w_{fB3}$ generally have imaginary parts. However, for some values of $\lambda$, the ratio of their imaginary parts to their real parts can be as small as $10^{-15}$ or $10^{-16}$. In the following, we discard $w_{fB2,3}$ and denote $w_{fB1}$ by $w_{fB}$.}
%\footnote{It should be emphasized that eq. \eqref{lambda-B-2} gives two other extrema
%% depending on the value of $\lambda$. 
%which have imaginary parts. However, for some values of $\lambda$, the ratio of their imaginary parts to their real parts can be as small as $10^{-15}$ or $10^{-16}$. In the following, we discard these extrema.}
%%%%%%%%%%%%%%%%%%%%%%%%%%%%%
Since the corresponding expression is very complicated, we do not write it here. However, for $\lambda \ll 1$, one can expand it in powers of $\lambda$ as follows
\bea
w_{fB} &=& \frac{1}{2} - \frac{4d (d-1) (d-z +2) \big( d(d+2) +z (1- d -z \big) }{(d+1) } \lambda
\cr && \cr 
\!\!\!\!\!\!\!\!\!\!\!\!  &&
- \frac{16 (d-1)^2}{ (d+1)^2} d (d-z+2) \big( d (d+2) + z (1 - d -z ) \big) 
\!\! \times \!\!
\Bigg( \!\! 5d^2 (d+2)^2 -2 d (d+2) (5d -1) z 
\cr && \cr
\!\!\!\!\!\!\!\!\!\!\!\! &&
+ \big(-1 +3 d (d-2) \big) z^2 +2 (d+1) z^3 - z^4 
\Bigg) \lambda^2 + \mathcal{O} \left( \lambda^3 \right) + \cdots . \!\!\!\!\!\!\!
\label{wfB-lambda-neq-0-theta=0-expansion}
\eea 
%Next, one has to find the appropriate range of $\lambda$ for which $\hat{U}''(w_{fB}) < 0$ and $0 < w_{fB} <1 $. These values are reported in Table \ref{Table: wfB-inside}. 
It should be pointed out that $w_{fB}$ can be a local maximum either {\it outside} or {\it inside} the horizon depending on the value of the coupling $\lambda$. The corresponding values of $\lambda$ are shown in Tables \ref{Table: wfB-outside} and \ref{Table: wfB-inside}, respectively. 
%%%%%%%%%%%%%%%%%%%%%%%%%%%
\begin{table}
	\begin{center}
	\scalebox{0.9}{
		\begin{tabular}{| l | l | l | l | l | l |}
			\hline
			& $d=2$ & $d=3$ & $d=4$ & $d=5$ 
			\\ \hline
			$z=1$ & $  \lambda >  \frac{1}{12}  \lambda_{crt,2} $ & $  \lambda >  \frac{1}{72} \lambda_{crt,2} $ & $   \lambda >  \frac{1}{240} \lambda_{crt,2} $ & $  \lambda >  \frac{1}{600} \lambda_{crt,2}$
			\\ \hline
			$ z=2$ & There is no $\lambda$ & $\lambda > \frac{655 + 23 \sqrt{805}}{ 5300} $ & $ \lambda >  \frac{2777 + 103 \sqrt{721}}{100224}$ & $ \lambda > \frac{27947 + 323 \sqrt{742s9}}{2655408}$
			\\ \hline
			$ z=3$ & There is no $\lambda$ & There is no $\lambda$ & There is no $\lambda$ & $ \lambda > \frac{11391 + 381 \sqrt{889}}{418816}$
			\\ \hline
			$ z=4$ & There is no $\lambda$ & There is no $\lambda$ & There is no $\lambda$ & There is no $\lambda$
			\\ \hline
			$ z=5$ & There is no $\lambda$ & There is no $\lambda$ & There is no $\lambda$ & There is no $\lambda$
			\\ \hline
			$ z=6$ & There is no $\lambda$ & There is no $\lambda$ & There is no $\lambda$ & There is no $\lambda$
			\\ \hline
		\end{tabular} }
		\caption{The appropriate ranges of the coupling constant $\lambda$ for which $w_{fB}$ is a maximum outside the horizon. Here, $\lambda_{crt,2}$ is defined by eq. \eqref{lambda-crt-z=1}.}
		\label{Table: wfB-outside}
	\end{center}
\end{table}
%%%%%%%%%%%%%%%%%%%%%%%%%%%
%\hspace{-2cm}
%%%%%%%%%%%%%%%%%%%%%%%%%%% 
\small{
\begin{table}
	\begin{center}
\scalebox{0.8}{
	\begin{tabular}{| l | l | l | l | l | l  |}
		\hline
		&  $d=2$ & $d=3$ & $d=4$ & $d=5$ 
		\\ \hline
		$z=1$ & 
		$ - \frac{1}{12} < \lambda <  \frac{ 1}{12} \lambda_{crt,1} $ & $ - \frac{1}{72} < \lambda <  \frac{1}{72} \lambda_{crt,1}$ & $  - \frac{1}{240} < \lambda <  \frac{1}{240} \lambda_{crt,1} $ & $ - \frac{1}{600} < \lambda < \frac{1}{600} \lambda_{crt,1}$
		\\ \hline
		$ z=2$ & 
		$ - \frac{3}{16} < \lambda < \frac{9}{2000}$ & $ - \frac{2}{25} < \lambda < \frac{655 - 23 \sqrt{805}}{5300} $ & $ - \frac{5}{432}< \lambda <  \frac{2777 - 103 \sqrt{721}}{100224}$ & $ - \frac{1}{294} < \lambda < \frac{27947 - 323 \sqrt{7429}}{2655408}$
		\\ \hline
		$ z=3$ & 
		$ \lambda > - \frac{1}{48}$ & $ \lambda > - \frac{1}{72}$ & $ - \frac{5}{432}< \lambda < 0$ & $ - \frac{1}{96} < \lambda < \frac{11391 - 381 \sqrt{889}}{418816}$
		\\ \hline
		$ z=4$ & 
	   $ \lambda > - \frac{1}{192}$ & $ \lambda > - \frac{1}{288}$ & $ - \frac{5}{1728} < \lambda < 0$ & $ - \frac{1}{384} < \lambda < 0$
		\\ \hline
		$ z=5$ &  
	    $ \lambda > - \frac{1}{588}$ & $ \lambda > - \frac{1}{800}$ & $ - \frac{1}{960} < \lambda < 0$ & $ - \frac{3}{3200} < \lambda < 0$
		\\ \hline
		$ z=6$ & 
	   $ - \frac{3}{4096} < \lambda < \frac{3 ( 547 + 41 \sqrt{697})}{2326528} $ & $ \lambda > - \frac{2}{3969}$ & $w_{fB} = \frac{1}{2}$ and $ \lambda > - \frac{1}{2160}$ & 
		$ - \frac{1}{2400} < \lambda < 0$
		\\ \hline
	\end{tabular}
}
	\caption{The appropriate ranges of the coupling constant $\lambda$ for which $w_{fB}$ is a maximum inside the horizon. Here, $\lambda_{crt,1}$ is defined by eq. \eqref{lambda-crt-z=1}.}
	\label{Table: wfB-inside}
		\end{center}
\end{table}
}
%%%%%%%%%%%%%%%%%%%%%%%%%%%
According to Table \ref{Table: wfB-outside}, it is easy to see that for $z=4,5,6$, $w_{fB}$ cannot be a maximum outside the horizon when $d=2,3,4,5$. On the other hand, from Table \ref{Table: wfB-inside}, for every values of $d$ and $z$ there is always a range of $\lambda$ for which $w_{fB}$ is a maximum inside the horizon. 
%However, for $d=1$, $w_{fB}$ is zero, and it coincides with the singularity at $r=0$. Therefore, for $d=1$, there is no maximum inside the horizon. 
In Figures \ref{fig: theta=0-z-wfB-outside} and \ref{fig: theta=0-z-wfB-inside}, we plotted $\hat{U}(w)$ as a function of $w$ for the case $\theta=0$ and different values of $d$ and $z$ to show its maxima. Figure \ref{fig: theta=0-z-wfB-outside} indicates that for the values of $\lambda$ in Table \ref{Table: wfB-outside}, $w_{fB}$ is a maximum
%of $\hat{U}(w)$ 
outside the horizon. Moreover, Figure \ref{fig: theta=0-z-wfB-inside} shows that for the values of $\lambda$ given in Table \ref{Table: wfB-inside}, $w_{fB}$ is a maximum
%of $\hat{U}(w)$ 
inside the horizon. Therefore, for the case $\theta=0$, the only ranges of $\lambda$ that we are interested in are given by Table \ref{Table: wfB-inside}.
%%%%%%%%%%%%%%%%%%%%%%%%%%%%%%%%%
%%%%%%%%%%%%%%%%%%%%%%%%%
%%%%%%%%%%%%%%%%%%%%%%%%%%%%%%%%%
%%%%%%%%%%%%%%%%%%%%%%%%%
On the other hand, for the case $d=1$, from eq. \eqref{B}, one has
\bea
B(\lambda, w) = 4 w (1- 2 w),
\label{B-1}
\eea 
which is independent of $\lambda$. In this case, equation $B(\lambda, w) =0$ gives two extrema $w= \{ 0, \frac{1}{2} \}$. The first one coincides with the singularity and should be discarded. The second one is a maximum inside the horizon. Therefore, for $d=1$, $w_{fB} = \frac{1}{2}$ is always a maximum inside the horizon (See Figure \ref{fig: theta=0-z=1-d=1}) and in this case there is no constraint on the value of $\lambda$. 
%%%%%%%%%%%%%%%%%%%%%%%%%%%%%%%%%
%%%%%%%%%%%%%%%%%%%%%%%%%
\\Eventually, by plugging eq. \eqref{wfB-lambda-neq-0-theta=0-expansion} into eq. \eqref{dC-dtau-2}, one obtains the late time growth rate as follows
\bea
\lim_{ \tau \rightarrow \infty} \frac{d \mathcal{C}_{\rm gen}}{d \tau}  &=& \frac{8 \pi M}{d} \Bigg[
1 + \frac{4 (d-1) \big( d (d+2) + z (1- d-z )\big)^2}{d+1} \lambda
\cr && \cr
&&
+ 2 \left( \frac{ 4 d (d-1) (d-z+2) \big( d( d+2) + z (1-d-z ) \big)}{d+1 } \right)^2 \lambda^2 + \mathcal{O}\left( \lambda^3 \right) + \cdots
\Bigg], \;\; 
\label{dC-dtau-lambda-neq-0-theta=0-expansion}
\eea 
where for $\lambda= \theta=0$, it is consistent with eq. \eqref{dC-dtau-lambda-0-2}. Moreover, for $d=1$, one arrives at
\bea
\lim_{ \tau \rightarrow \infty} \frac{d \mathcal{C}_{\rm gen}}{d \tau}  &=& 8 \pi M,
\label{dC-dtau-lambda-neq-0-theta=0-d=1}
\eea 
which is independent of $\lambda$.

%%%%%%%%%%%%%%%%%%%%%%%%%%
\subsubsection{$\theta=0$ and $z=1$}
\label{Sec: theta=0 and z=1}
%%%%%%%%%%%%%%%%%%%%%%%%%%

\begin{figure}
	\begin{center}
		   \includegraphics[scale=0.26]{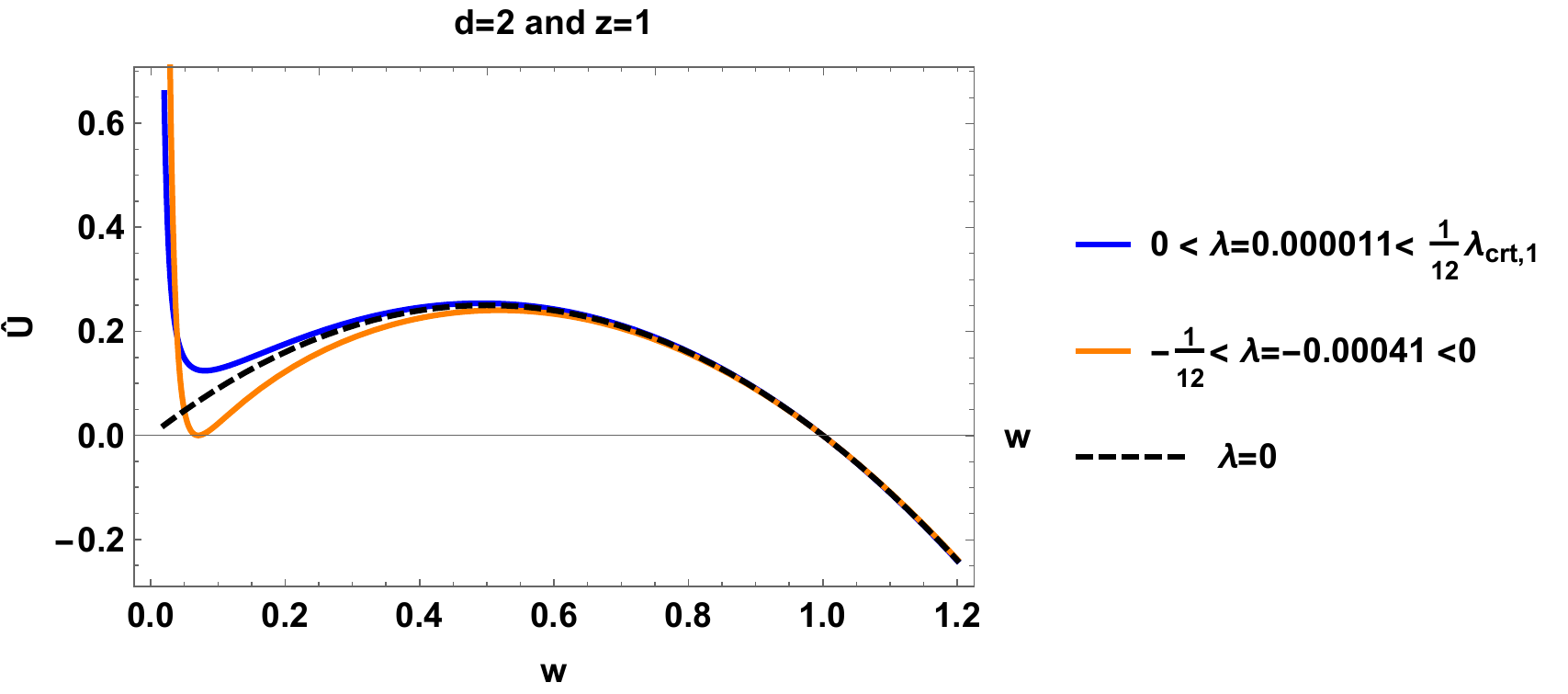}
			\hspace{0.1cm}
		   \includegraphics[scale=0.26]{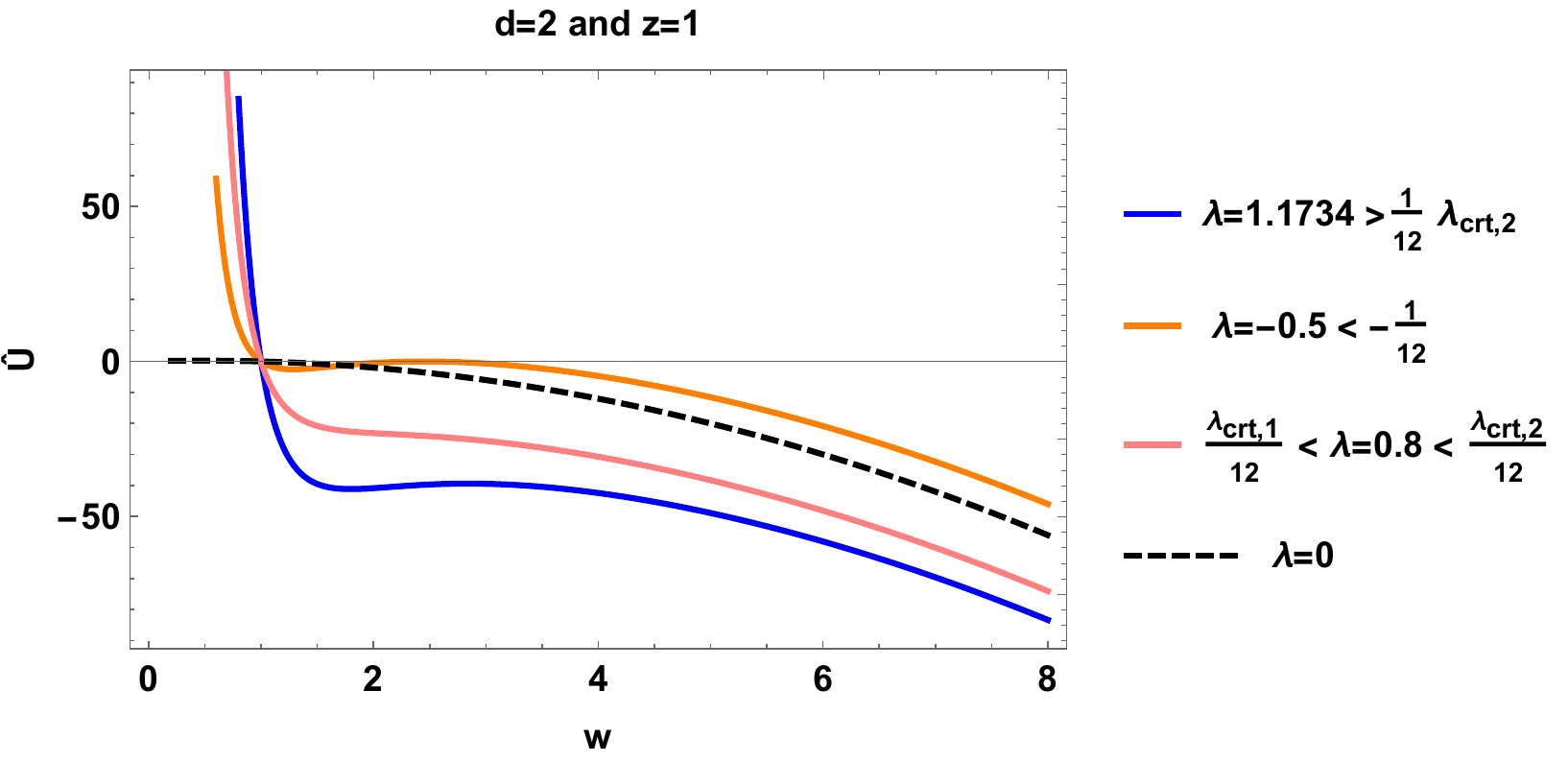}
\\
         \includegraphics[scale=0.26]{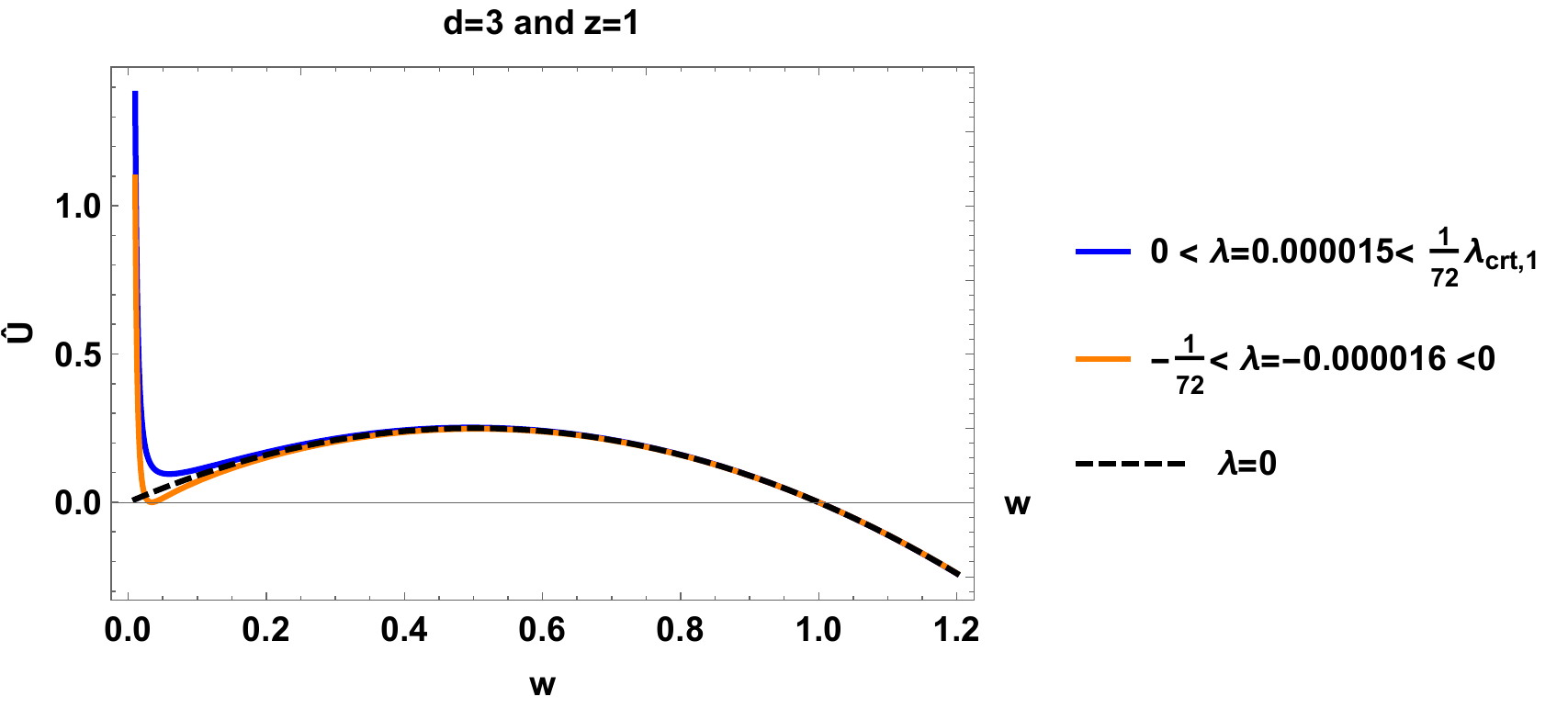}
         \hspace{0.1cm}
          \includegraphics[scale=0.26]{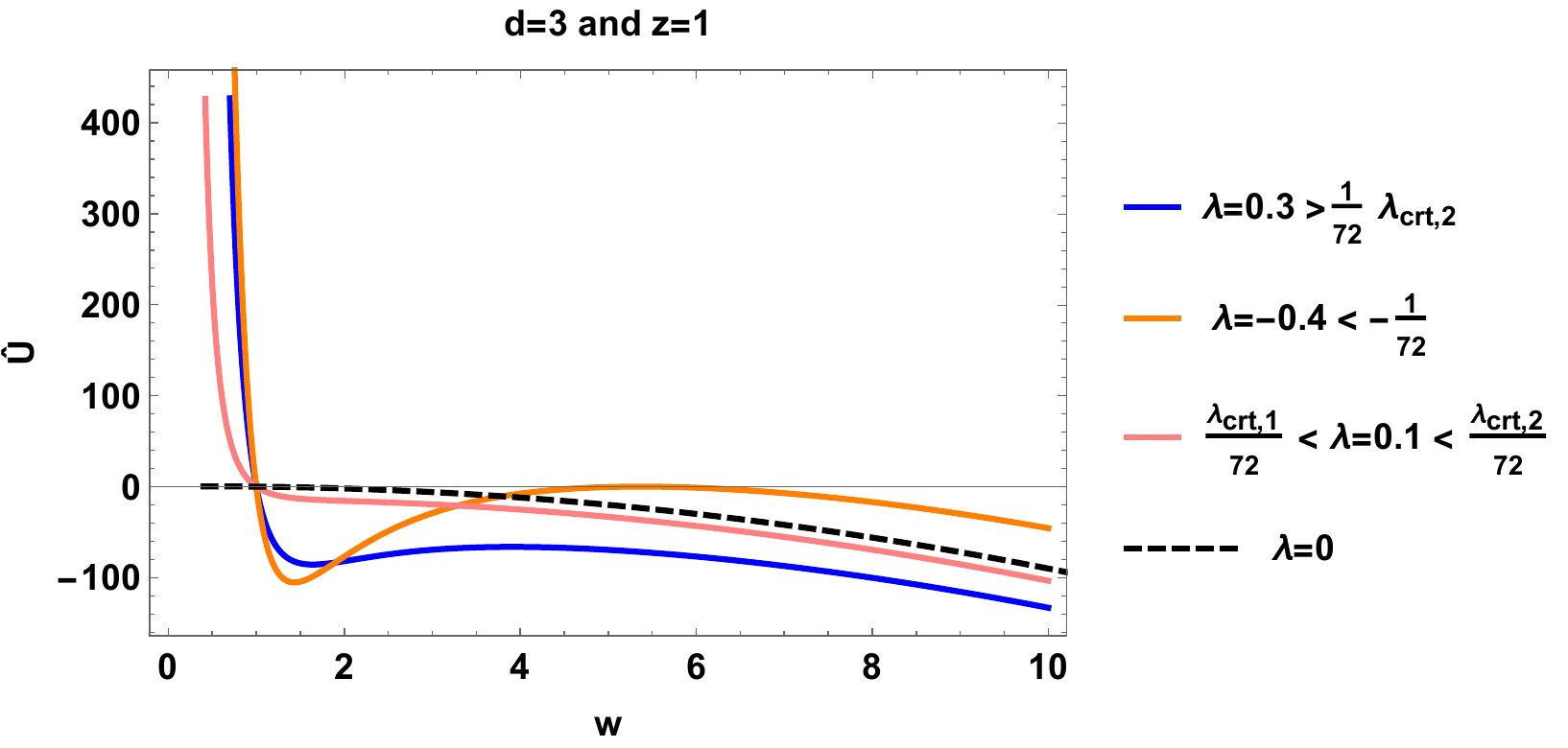}
	\end{center}
	\caption{ The effective potential $\hat{U}(w)$ as a function of $w= \left( \frac{r}{r_h} \right)^{d_e +z}$ for $\theta=0$, $z=1$ and different values of $d$ and $\lambda$. {\it Left}) when $\lambda=0$, $ 0 < \tilde{\lambda} < \lambda_{crt,1}$  and $ -1< \tilde{\lambda} < 0$, there is a local maximum $w_{fB}$ inside the horizon (See Table \ref{Table: wfB-inside}). {\it Right}) when  $ \tilde{\lambda} > \lambda_{crt,2}$ and $ \tilde{\lambda} < - 1$ there are two maxima $w_{fA1}$ and $w_{fB}$ which are located outside the horizon (See Tables \ref{Table: wfA1-outside} and \ref{Table: wfB-outside}). For $ \lambda_{crt,1} <  \tilde{\lambda} < \lambda_{crt,2}$ there is no maximum.
	 In the first row $d=2$ and in the second row $d=3$. Notice that for $d=2,3$ one has $\tilde{\lambda} = 12 \lambda$ and $\tilde{\lambda} = 72 \lambda$
% and $\tilde{\lambda} = 240 \lambda$
 , respectively. We also plotted the case $\lambda =0$, i.e. the volume-complexity, which is indicated by the dashed black curve. We set $ L= r_h =r_F=1$. Since for these values of $\theta$ and $z$, the HV black brane becomes a planar AdS-Schwarzschild black hole, these figures are the same as Figure 3 in ref. \cite{Belin:2021bga}. 
	%Notice that $ \tilde{\lambda}_{there}$.
	}
	\label{fig: theta=0-z=1}
\end{figure}

For $\theta=0$ and $z=1$, the background becomes a planar AdS-Schwarzschild black hole. In this case, the exact expression for $w_{fB}$ is easily obtained as follows
\bea
w_{fB} &=& \frac{1}{6} \Bigg( 1 + \frac{1 + 12 \tilde{\lambda}}{ \left( 1 -144 \tilde{\lambda} + 6 \sqrt{3} \sqrt{ - \tilde{\lambda} ( 4 \tilde{\lambda} (4 \tilde{\lambda} - 47) + 3 )} \right)^{\frac{1}{3}}} 
\cr && \cr
&& \;\;\;\;\;\;\;\;\;\;\;\;\;\;\;\;\;\;\;\;\;\;\;\;\;\;
+  \left( 1 -144 \tilde{\lambda} + 6 \sqrt{3} \sqrt{ - \tilde{\lambda} ( 4 \tilde{\lambda} (4 \tilde{\lambda} - 47) + 3 )} \right)^{\frac{1}{3}} \Bigg) ,
\label{wfB-lambda-neq-0-z=1}
\eea 
where the effective coupling constant $\tilde{\lambda}$ is defined as follows
\bea 
\tilde{\lambda} = d^2 (d^2 -1) \lambda.
%(d-1) d^2 (d+1) \lambda
\label{lambda-t-z=1}
\eea
It should be pointed out that $w_{fB}$ in eq. \eqref{wfB-lambda-neq-0-z=1} is the same as the maximum reported in ref. \cite{Belin:2021bga} for a planar AdS-Schwarzschild black hole. 
%Moreover, for $ \lambda < -1$, $w{fA}$ is a maximum outside the horizon (See the orange curve in the right panel of figure \ref{fig: z=1}).
Moreover, for $z=1$,  when $ -1 < \tilde{\lambda} < \lambda_{crt,1}$ and $ \tilde{\lambda} > \lambda_{crt,2}$, $w_{fB}$ is a maximum located inside and outside the horizon, respectively (See Figure \ref{fig: theta=0-z=1}) where the critical values for $\lambda$ are defined as follows (See also \cite{Belin:2021bga})
\bea
\lambda_{crt,1} = \frac{1}{8} ( 47 - 13 \sqrt{13}), \;\;\;\;\;\;\;\;\;\;\;\;\;\;\;\;\; \lambda_{crt,2} = \frac{1}{8} ( 47 + 13 \sqrt{13}).
\label{lambda-crt-z=1}
\eea 
Furthermore, for $ \lambda_{crt,1} < \tilde{\lambda} < \lambda_{crt,2}$, there is no real $w_{fB}$. 
%\\ In Figure \ref{fig: theta=0-z=1}, we plotted the effective potential for $\theta=0$ and $z=1$. In the left panels of the figure, the maxima are located inside the horizon. In the right panels, either there is no maximum or it is located outside the horizon. 
%%%%%%%%%%%%%%%%%%%%%%%%%%%%%%%%%
%%%%%%%%%%%%%%%%%%%%%%%%%
%%%%%%%%%%%%%%%%%%%%%%%%%%%%%%%%%%
%\begin{figure}
%	\begin{center}
%		\includegraphics[scale=0.28]{d=2-theta=0-z=2-inside}
%		%		\hspace{-0.1cm}
%		%\includegraphics[scale=0.28]{d=2-theta=0-z=2-outside}
%		%		\hspace{1cm}
%		%		\includegraphics[scale=0.34]{S-rc-de-1-l-5.pdf}
%		%		\vspace{-5mm}
%	\end{center}
%	\caption{ The effective potential $\hat{U}(w)$ as a function of $w$ for $\theta=0$ and $z=2$. 
%	%{\it Left}) when $\lambda=0$, $  \lambda_{crt,1} <\lambda < \lambda_{crt,2}$ .
%	%	 and $ -1<\lambda < 0$,
%%		  there is a local maximum inside the horizon. 
%%		  {\it Right}) when  $ \lambda= 1.2 \lambda_{crt,2} > \lambda_{crt,2}$ and $\lambda = - \frac{1}{2} < - \frac{1}{12}$ there is a maximum outside the horizon. For $ \lambda_{crt,1} < \lambda = 0.8 < \lambda_{crt,2}$ there is no maximum. 
%		  We also plotted the case $\lambda =0$. We set $L=r_h =r_f=1$ and $d=2$. 
%	}
%	\label{fig: z=2}
%\end{figure}
%%%%%%%%%%%%%%%%%%%%%%%%%%
%\\In Figures \ref{fig: theta=0-z=1} to \ref{fig: theta=0-z-wfB-inside}, 
%% \ref{fig: theta=0-z=1}, \ref{fig: theta=0-z-wfA-outside}, \ref{fig: theta=0-z-wfB-outside} and  \ref{fig: theta=0-z-wfB-inside},
%we plotted the effective potential $\hat{U}(w)$ as a function of $w$ for the case $\theta=0$ and different values of $d$ and $z$ to represent its maxima.
In Figure \ref{fig: theta=0-z=1}, we restricted ourselves to the case $z=1$. In the left panels of Figure \ref{fig: theta=0-z=1}, the maxima are located inside the horizon. However, in the right panels of Figure \ref{fig: theta=0-z=1}, either there is no maximum or it is located outside the horizon. Next, by plugging eq. \eqref{wfB-lambda-neq-0-z=1}, into eq. \eqref{dC-dtau-2}, one has
\bea
\lim_{ \tau \rightarrow \infty} \frac{d \mathcal{C}_{\rm gen} }{d \tau} = \frac{16 \pi M}{d} \sqrt{w_{fB} - w_{fB}^2} \left( 1 + \frac{ \tilde{\lambda}}{w_{fB}^2} \right).
\label{dC-dtau-lambda-neq-0-theta=0-z=1}
\eea 
Notice that for $\lambda=0$, one has $w_{fB}= \frac{1}{2}$, and hence eq. \eqref{dC-dtau-lambda-neq-0-theta=0-z=1} reduces to eq. \eqref{dC-dtau-lambda-0-theta-0-z-1}.
%\\In Figure \ref{fig: theta=0-z-wfA-outside}, $w_{fA}$ is a maximum of $\hat{U}(w)$ located outside the horizon.
%%\\In Figure \ref{fig: theta=0-z-wfB-outside}, we plotted the effective potential as a function of $w$ for $\theta=0$. 
%In Figure \ref{fig: theta=0-z-wfB-outside}, $w_{fB}$ is a maximum of $\hat{U}(w)$ located outside the horizon.
It should also be pointed out that for some values of $d$, $\theta$, $z$ and $\lambda$, the potential $\hat{U} (w)$ may have two maxima (See the left panel of Figures \ref{fig: dCgen-dtau-lambda-neq-0-maxima-inside-outside} and \ref{fig: dCgen-dtau-lambda-neq-0-maxima-two-outside}). In the left panel of Figure \ref{fig: dCgen-dtau-lambda-neq-0-maxima-inside-outside}, there is a maximum inside the horizon and there is another one outside the horizon. On the other hand, in the left panel of Figure \ref{fig: dCgen-dtau-lambda-neq-0-maxima-two-outside}, both maxima are located outside the horizon. 
\subsubsection{$d=2$, $\theta=1$ and $z=1$}
\label{Sec: d=2, theta=1 and z=1}
%%%%%%%%%%%%%%%%%%%%%%%%%%

In this case, equation $A(\lambda,w)=0$ has a solution which is a maximum outside the horizon. On the other hand, equation $B( \lambda, w)= 0$ has the following solution
\bea
w_{fB} = \frac{3 r_F^2 + 4 r_h^2 \lambda - \sqrt{9 r_F^2 - 408 r_F^2 r_h^2 \lambda + 16 r_h^4 \lambda^2}}{ 18 r_F^2}.
\label{wfB-d=1-theta=1-z=1}
\eea 
It is easy to verify that for 
\bea
0 < \lambda < \frac{3(17 - 12 \sqrt{2}) r_F^2}{4 r_h^2},
\label{lambda-d=2-theta=1-z=1}
\eea 
$w_{fB}$ is a maximum inside the horizon. Next, it is straightforward to calculate the late time value of $\mathcal{C}_{\rm gen}$ from eqs. \eqref{dC-dtau-2} and \eqref{wfB-d=1-theta=1-z=1}.
%%%%%%%%%%%%%%%%%%%%%%%%%%%
%\subsubsection{Maxima of the Effective Potential For $d=4$, $\theta=3$ and $z=1$}
%\label{Sec: Maxima of the Effective Potential For d=4, theta=3 and z=1}
%%%%%%%%%%%%%%%%%%%%%%%%%%%

%%%%%%%%%%%%%%%%%%%%%%%%%
\subsection{General Time Dependence for $\lambda \neq 0$ and Arbitrary Values of $\theta$ and $z$}
\label{Sec: General Time Dependence-lambda-neq-0}
%%%%%%%%%%%%%%%%%%%%%%%%%

\begin{figure}
	\begin{center}
		\includegraphics[scale=0.31]{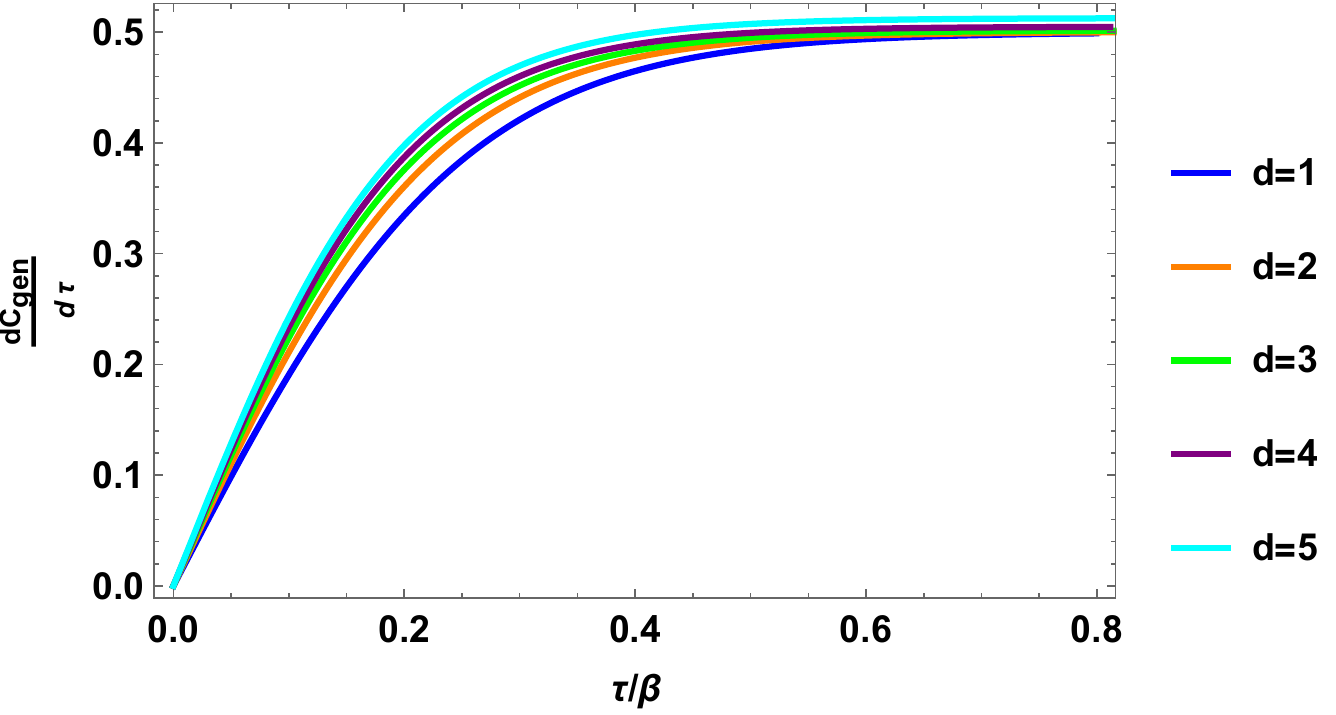}
		\hspace{0.5cm}
		\includegraphics[scale=0.31]{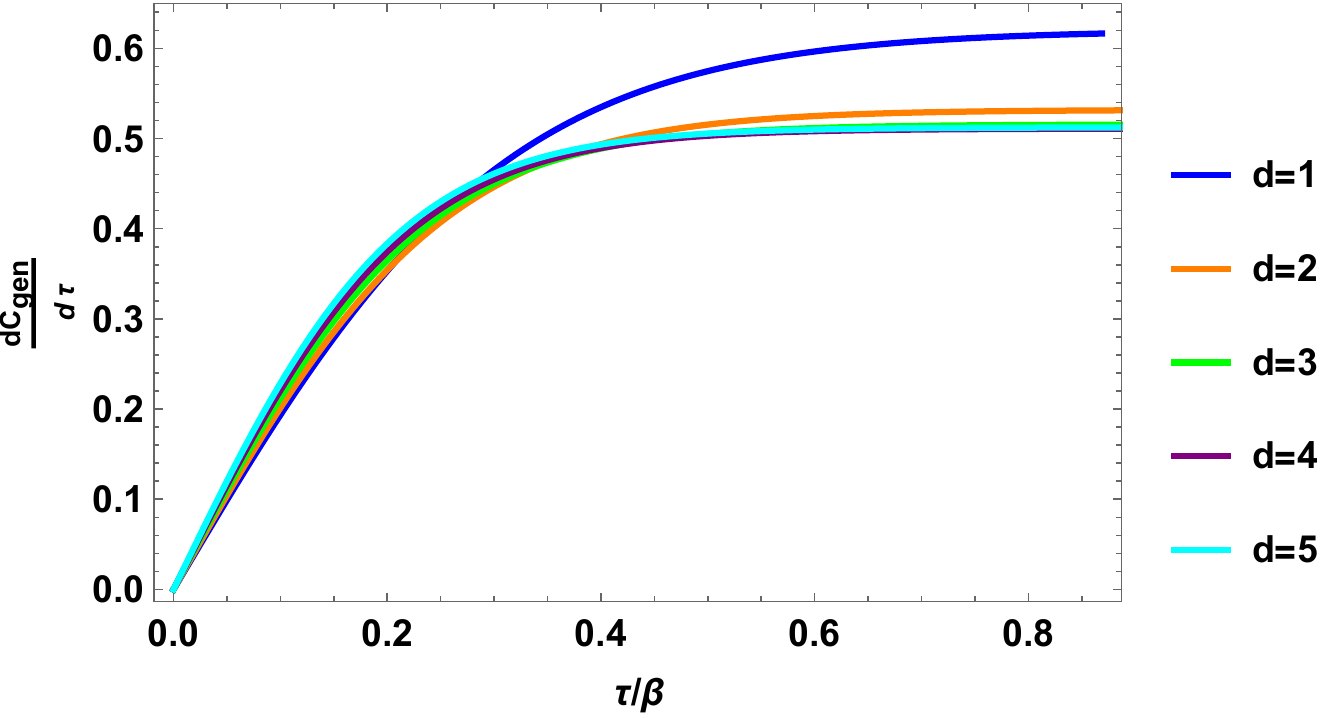}
		%		\\
		%		\includegraphics[scale=0.29]{dCgen-dtau-lambda-neq-0-d-4-theta-0-z.pdf}
		%		\hspace{0.5cm} 
		%		\includegraphics[scale=0.29]{dCgen-dtau-lambda-neq-0-d-5-theta-m1-z.pdf}
	\end{center}
	\caption{ The growth rate $ \frac{d \mathcal{C}_{\rm gen}}{d \tau}$ in eq. \eqref{dC-dtau-1}, as a function of $\tau / \beta$ for $\lambda \neq 0$ and different values of $d$, $\theta$ and $z$:
		{\it Left}) $\theta=0$, $z=1$,
		{\it Right}) $\theta= \frac{1}{2}$, $z= \frac{3}{2}$.
		%		{\it Down Left}) $d=4$ and $\theta=0$ and $\lambda= 10^{-5}$.
		%		{\it Down Right}) $d=5$ and $\theta=-1$ and $\lambda= 10^{-5}$.
		%	 We also plotted the case $\lambda =0$, i.e. the volume complexity, which is indicated by the dashed black curve. 
		We set $ L= r_h =r_F= G_N= V_d= 1$ and $\lambda= 10^{-5}$.
	}
	\label{fig: dCgen-dtau-lambda-neq-0-d}
\end{figure}

\begin{figure}
	\begin{center}
		\includegraphics[scale=0.31]{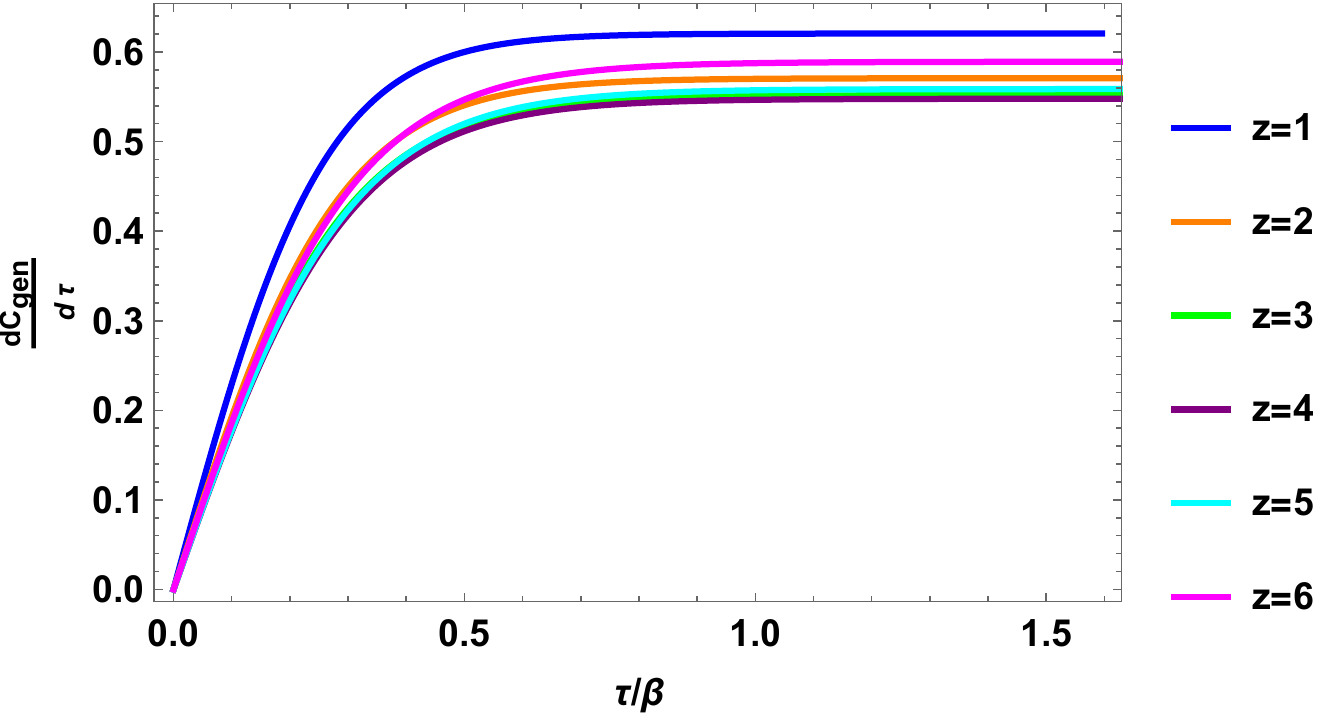}
		\hspace{0.5cm}
		\includegraphics[scale=0.31]{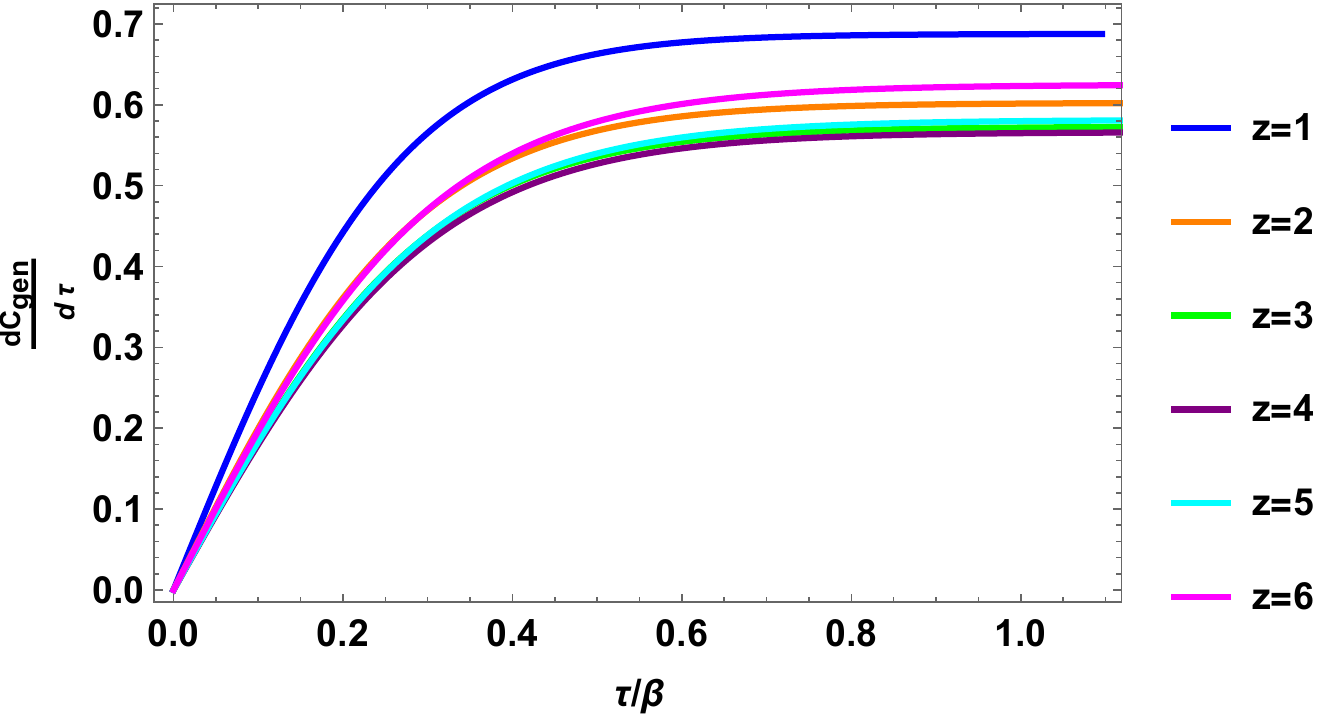}
		\\
		\includegraphics[scale=0.31]{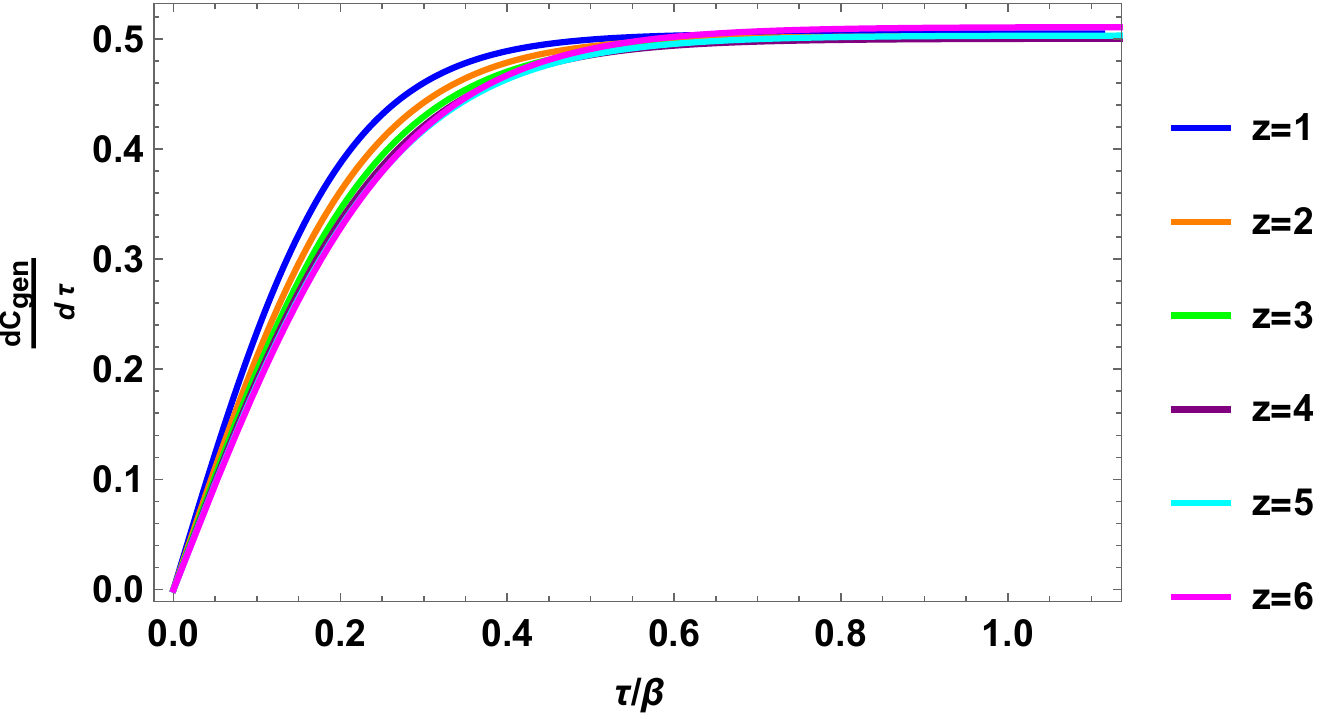}
		\hspace{0.5cm} 
		\includegraphics[scale=0.31]{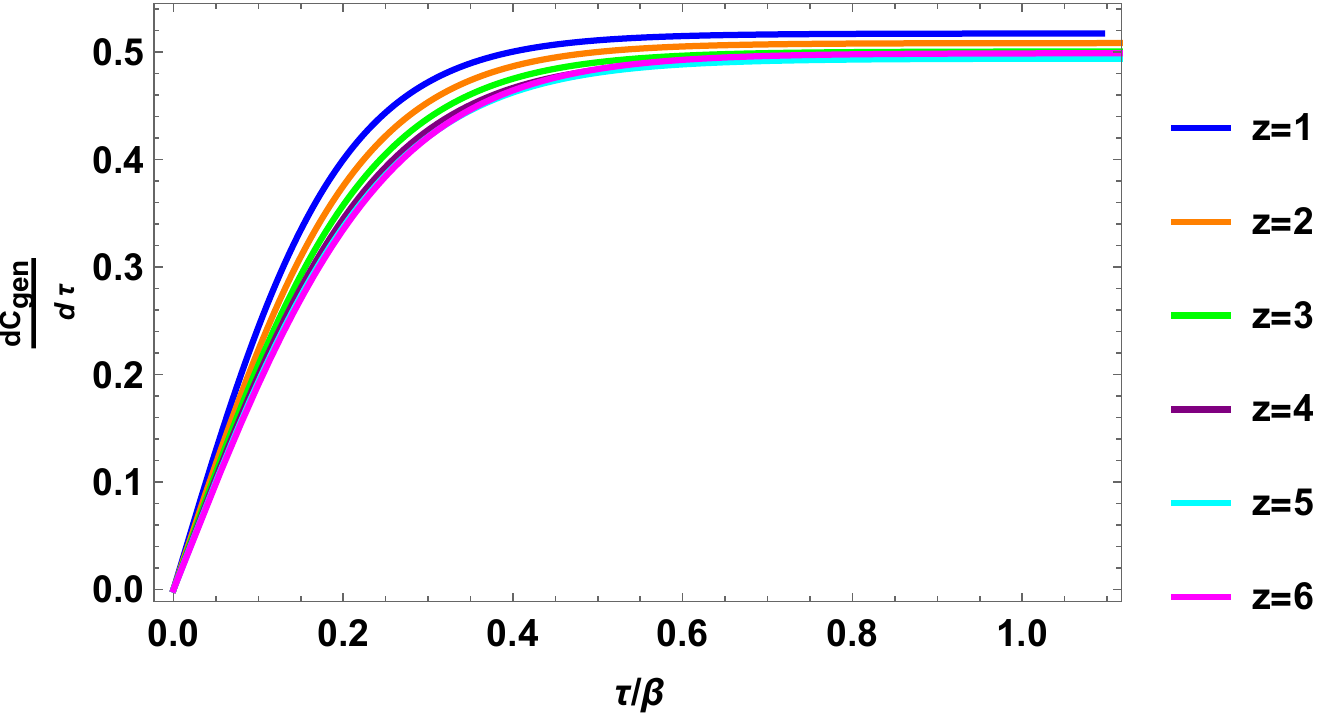}
	\end{center}
	\caption{ The growth rate $ \frac{d \mathcal{C}_{\rm gen}}{d \tau}$ as a function of $\tau / \beta$ for $\lambda \neq 0$ and different values of $d$, $\theta$ and $z$:
		{\it Top Left}) $d=2$, $\theta=1$ and $\lambda= 10^{-4}$.
		{\it Top Right}) $d=3$, $\theta=2$ and $\lambda= 10^{-4}$.
		{\it Down Left}) $d=4$ and $\theta=0$ and $\lambda= 10^{-5}$.
		{\it Down Right}) $d=5$ and $\theta=-1$ and $\lambda= 10^{-5}$.
		%	 We also plotted the case $\lambda =0$, i.e. the volume complexity, which is indicated by the dashed black curve. 
		We set $ L= r_h =r_F= G_N= V_d= 1$.
	}
	\label{fig: dCgen-dtau-lambda-neq-0-z}
\end{figure}

\begin{figure}
	\begin{center}
		\includegraphics[scale=0.31]{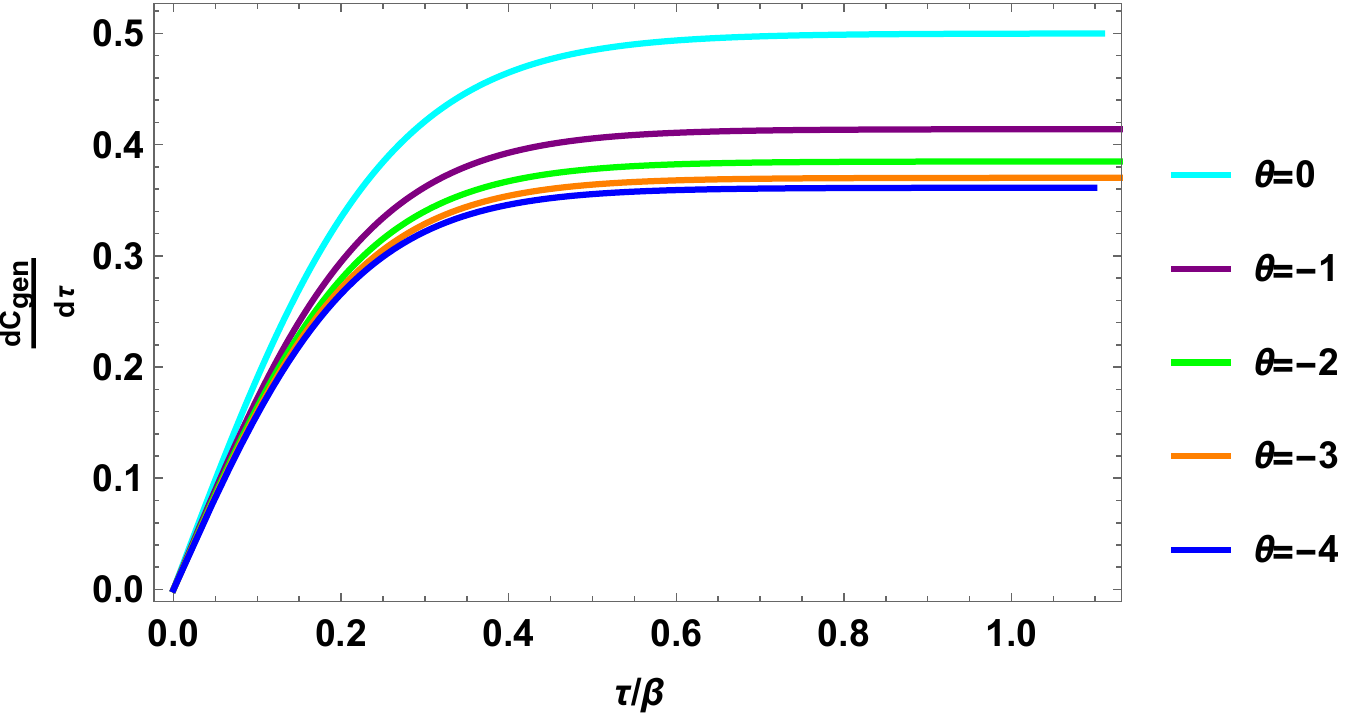}
		\hspace{0.5cm} 
		\includegraphics[scale=0.31]{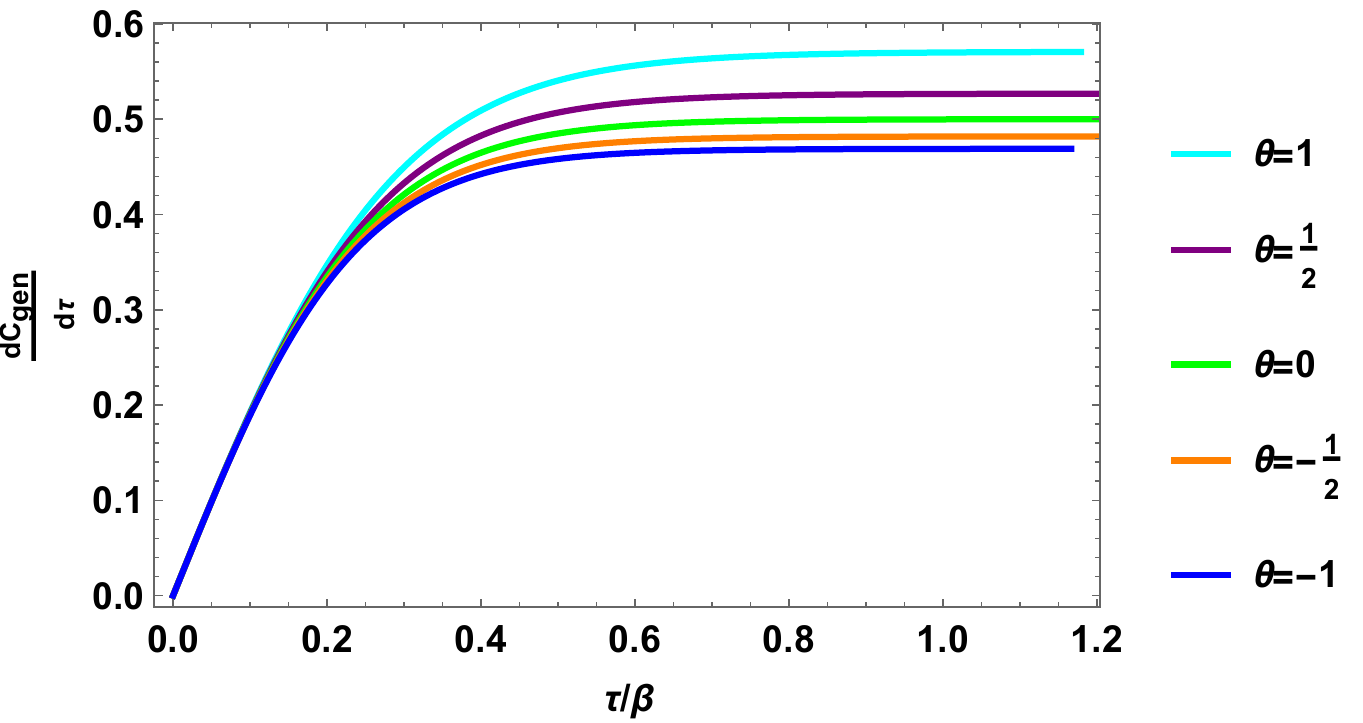}
		\\
		\includegraphics[scale=0.31]{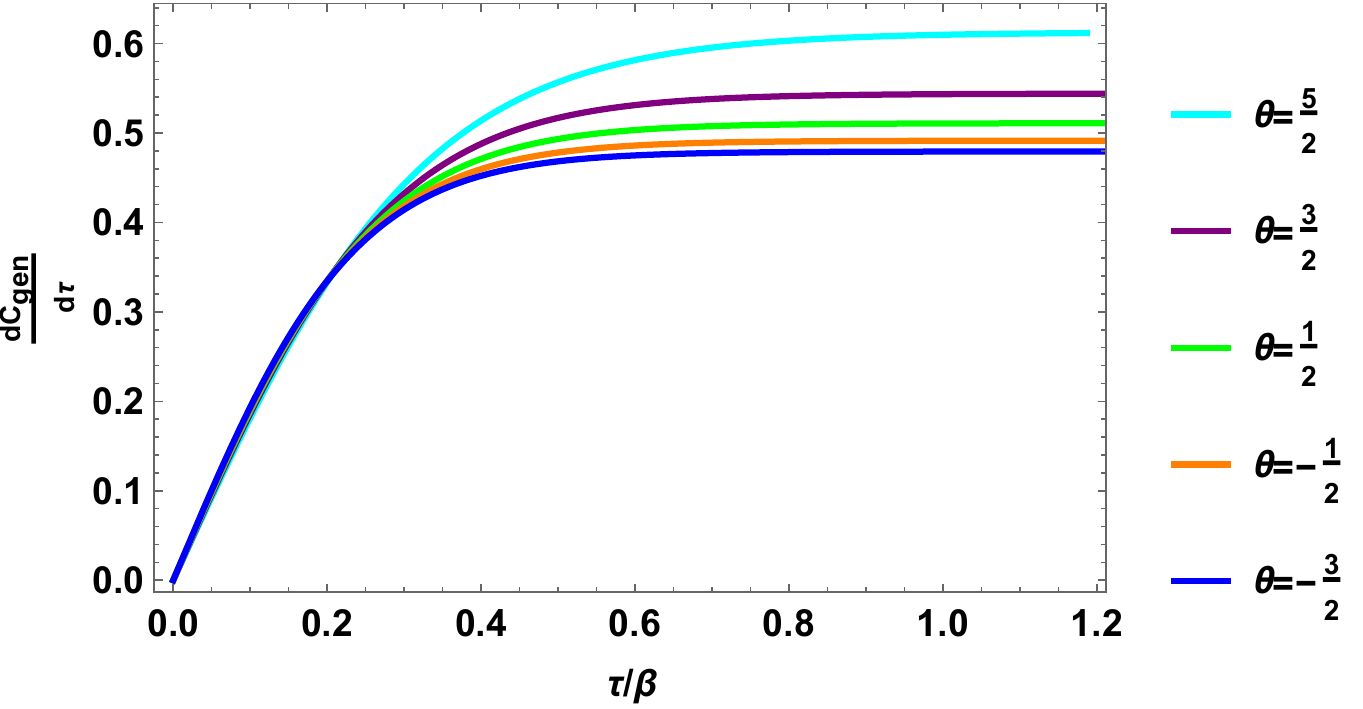}
		\hspace{0.5cm} 
		\includegraphics[scale=0.31]{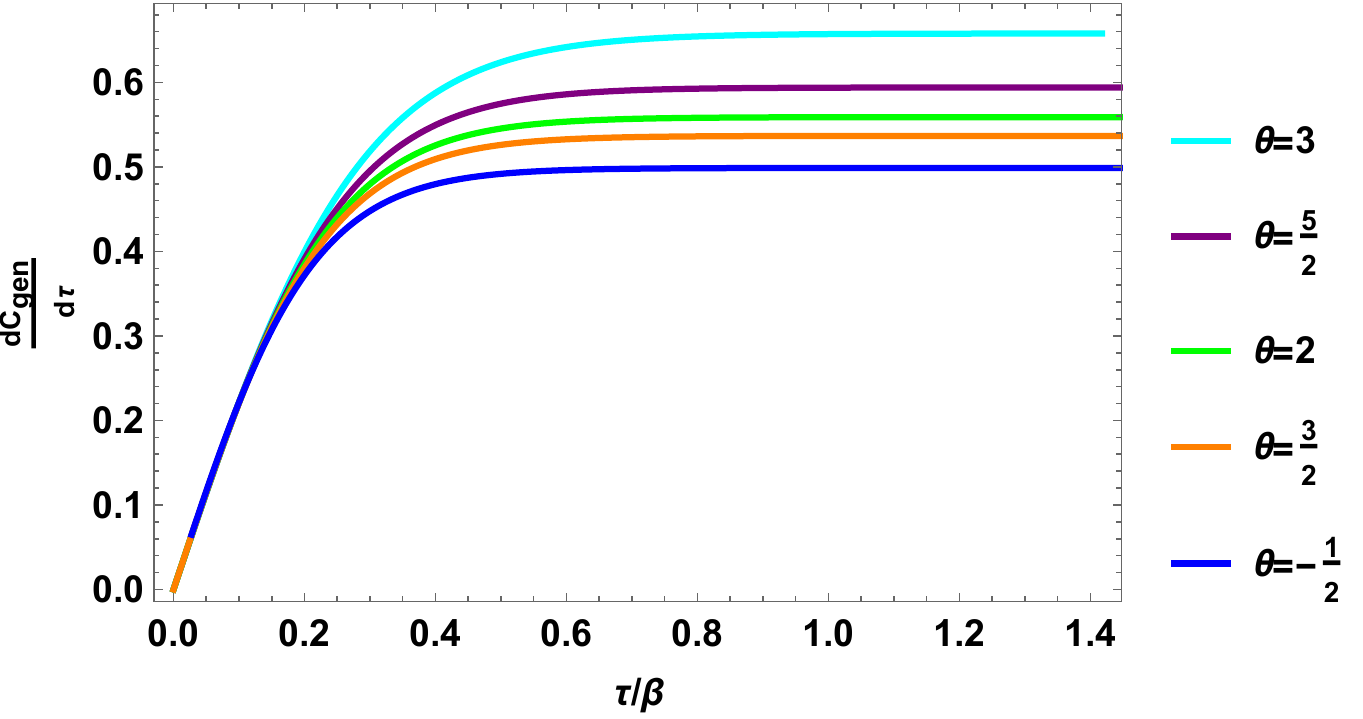}
	\end{center}
	\caption{ The growth rate $ \frac{d \mathcal{C}_{\rm gen}}{d \tau}$ as a function of $\tau / \beta$ for $\lambda \neq 0$ and different values of $d$, $\theta$ and $z$:
		{\it Top Left}) $d=1$, $z=1$ ,
		{\it Top Right}) $d=2$, $z=2$,
		{\it Down Left}) $d=3$ and $z=3$,
		{\it Down Right}) $d=4$ and $z=\frac{3}{2}$.
		%	 We also plotted the case $\lambda =0$, i.e. the volume complexity, which is indicated by the dashed black curve. 
		We set $ L= r_h =r_F= G_N= V_d= 1$ and $\lambda= 10^{-5}$.
		%Notice that $ \tilde{\lambda}_{there}$.
	}
	\label{fig: dCgen-dtau-lambda-neq-0-theta}
\end{figure}

\begin{figure}
	\begin{center}
		\includegraphics[scale=0.31]{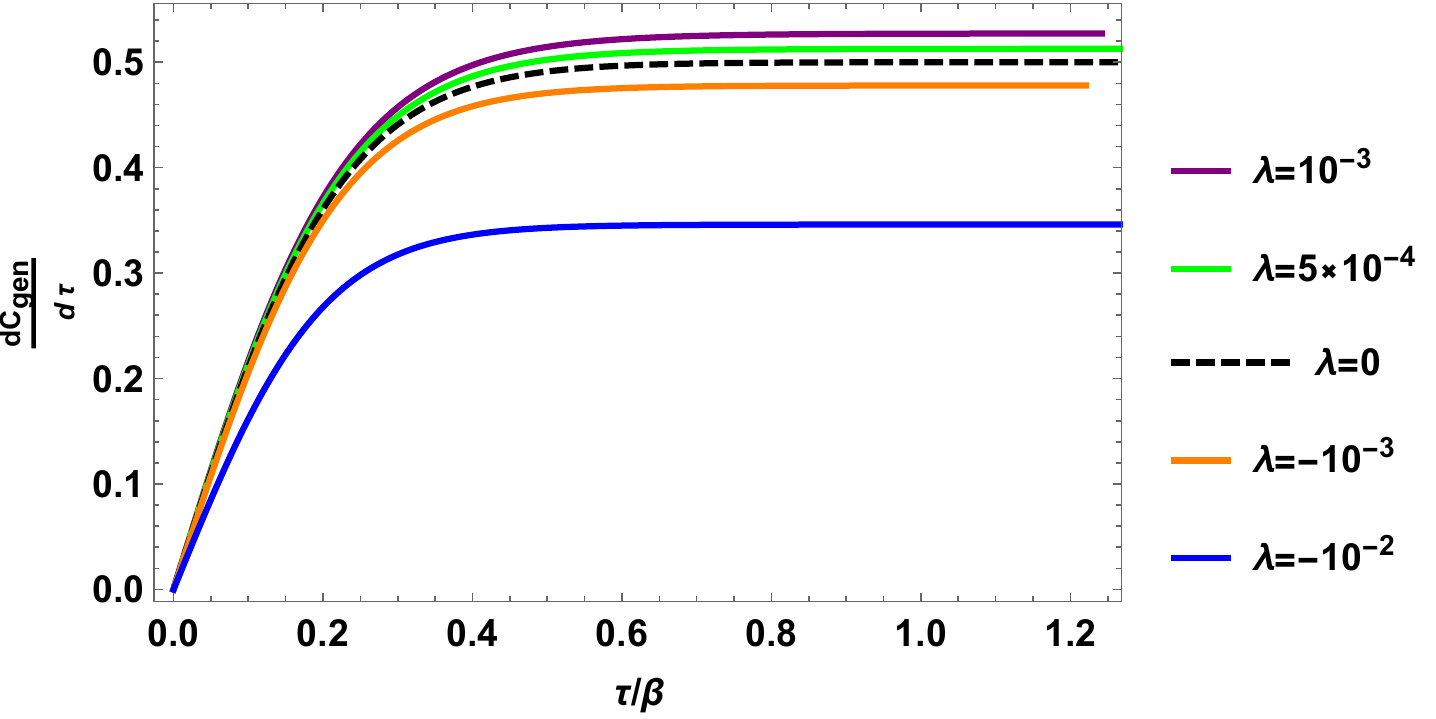}
		\hspace{0.1cm}
		\includegraphics[scale=0.31]{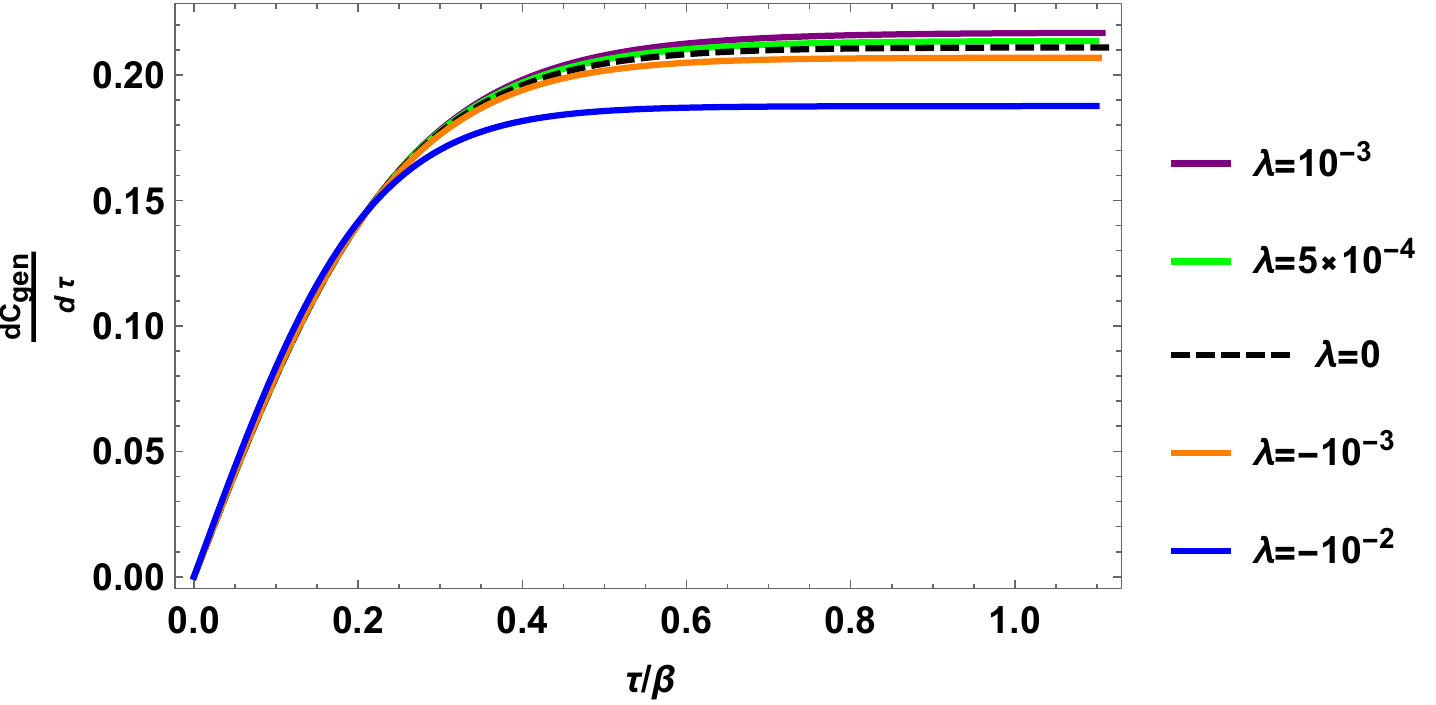}
		\\
		\includegraphics[scale=0.31]{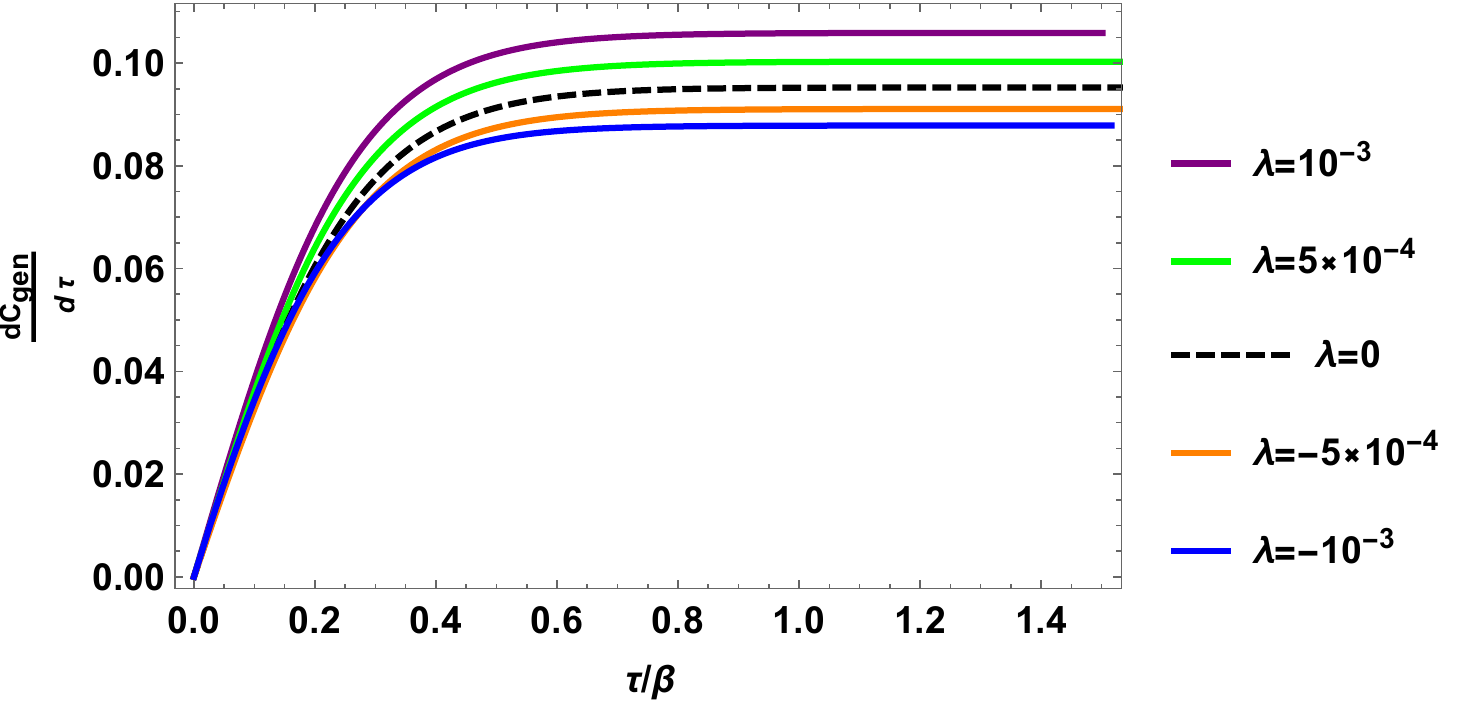}
		\hspace{0.1cm}
		\includegraphics[scale=0.31]{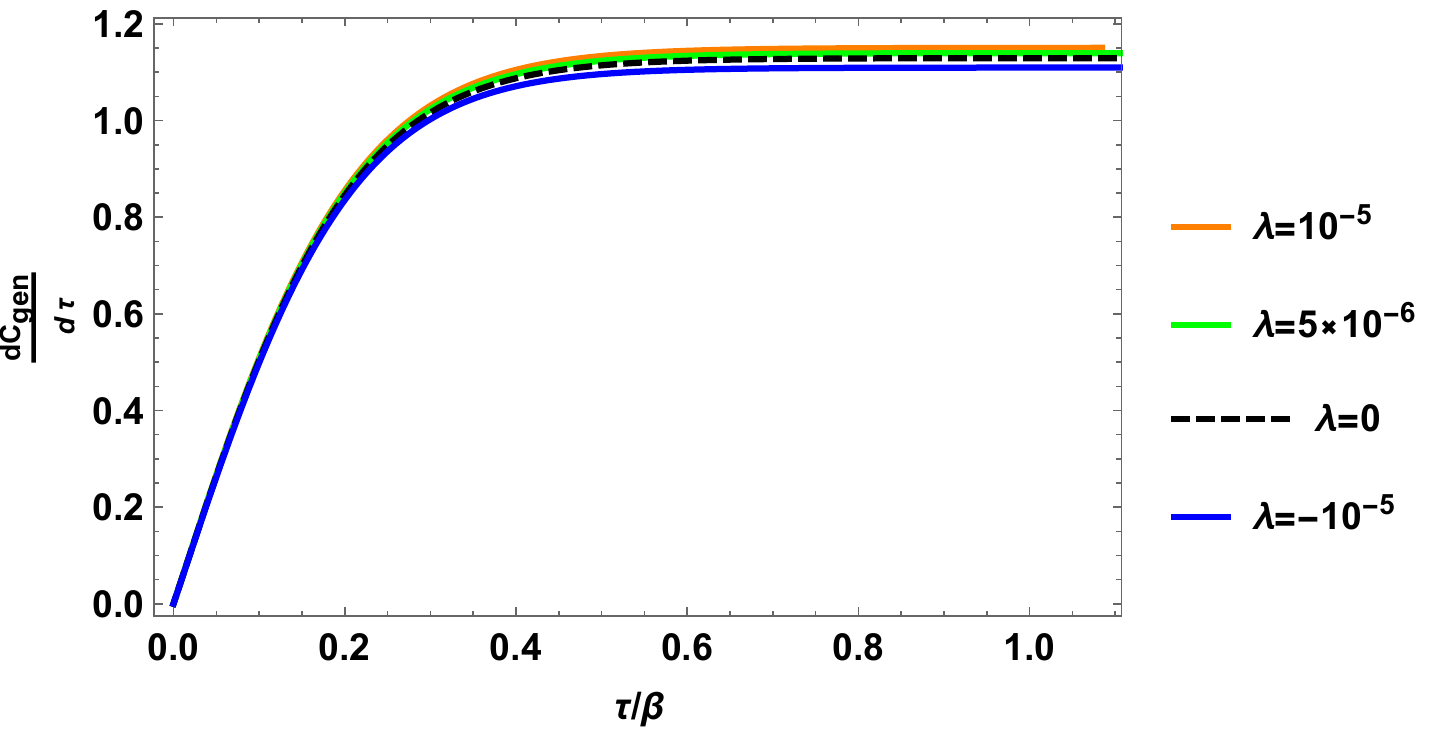}
	\end{center}
	\caption{The growth rate $ \frac{d \mathcal{C}_{\rm gen}}{d \tau}$ as a function of $\tau / \beta$ for different values of $\lambda \neq 0$: 
		{\it Top Left}) $d=2$, $\theta=0$ and $z=1$.
		{\it Top Right}) $d=3$, $\theta=1$ and $z=2$.
		{\it Down Left}) $d=4$, $\theta=2$ and $z=3$.
		{\it Down Right}) $d=5$, $\theta=-1$ and $z=2$.
		We also plotted it for the case $\lambda =0$, i.e. volume-complexity, which is indicated by the dashed black curve. 
		We set $ L= r_h = G_N= V_d= 1$ and $r_F= 0.5$.
		%%%%%%%%%%%%%%%%%%%%%%%%%%%%%%%%
		%	\footnote{We chose this value for $r_F$ to separate the curves better.}
		%%%%%%%%%%%%%%%%%%%%%%%%%%%%%%%%
		%Notice that $ \tilde{\lambda}_{there}$.
	}
	\label{fig: dCgen-dtau-lambda-neq-0-lambda}
\end{figure}

\begin{figure}
	%	\vspace{-2.5cm}
	\begin{center}
		\includegraphics[scale=0.25]{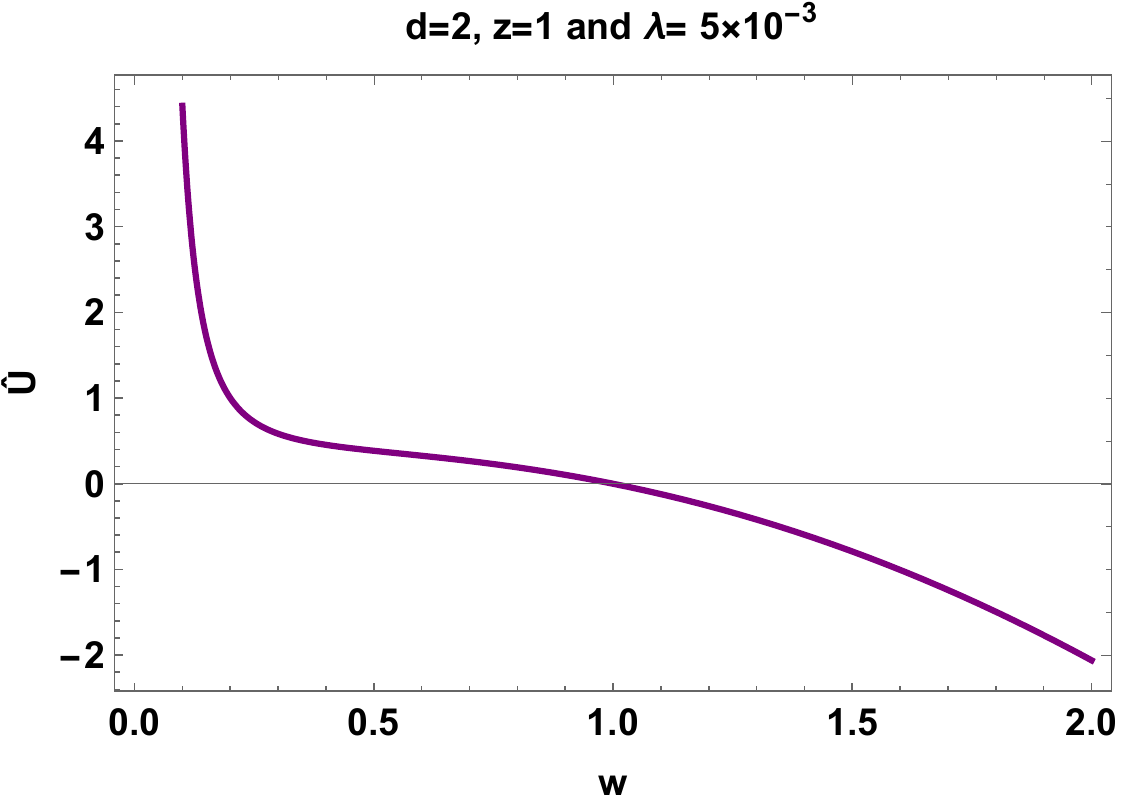}
		\hspace{0.1cm}
		\includegraphics[scale=0.25]{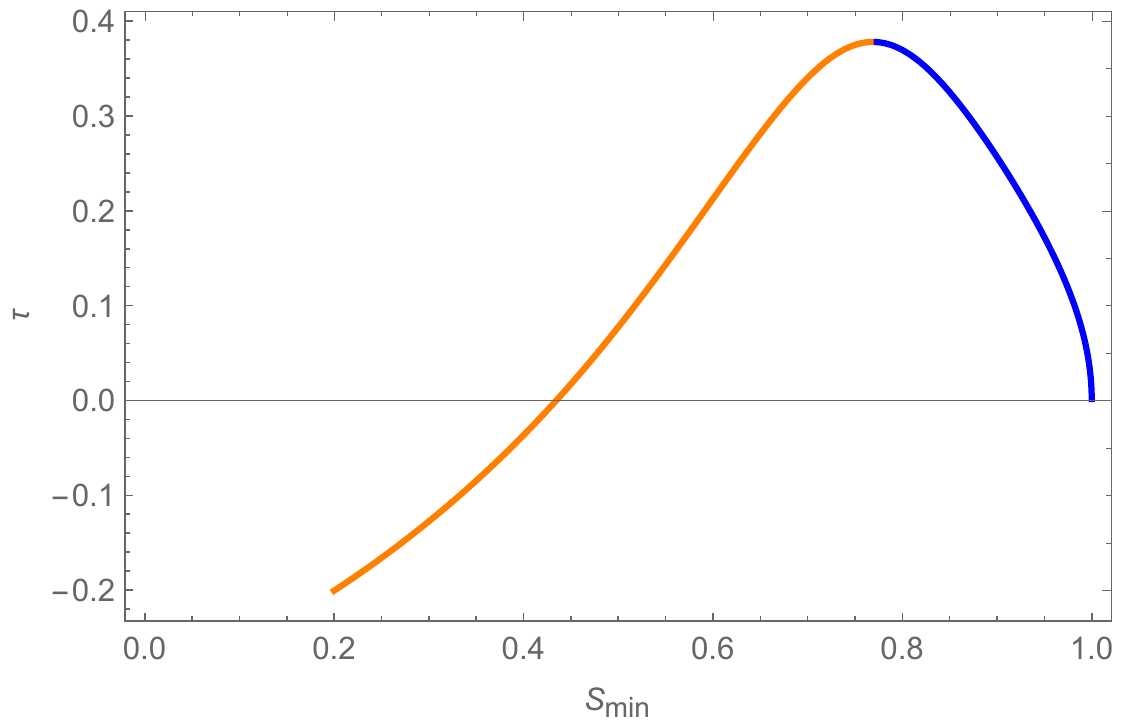}
		\hspace{0.1cm}
		\includegraphics[scale=0.25]{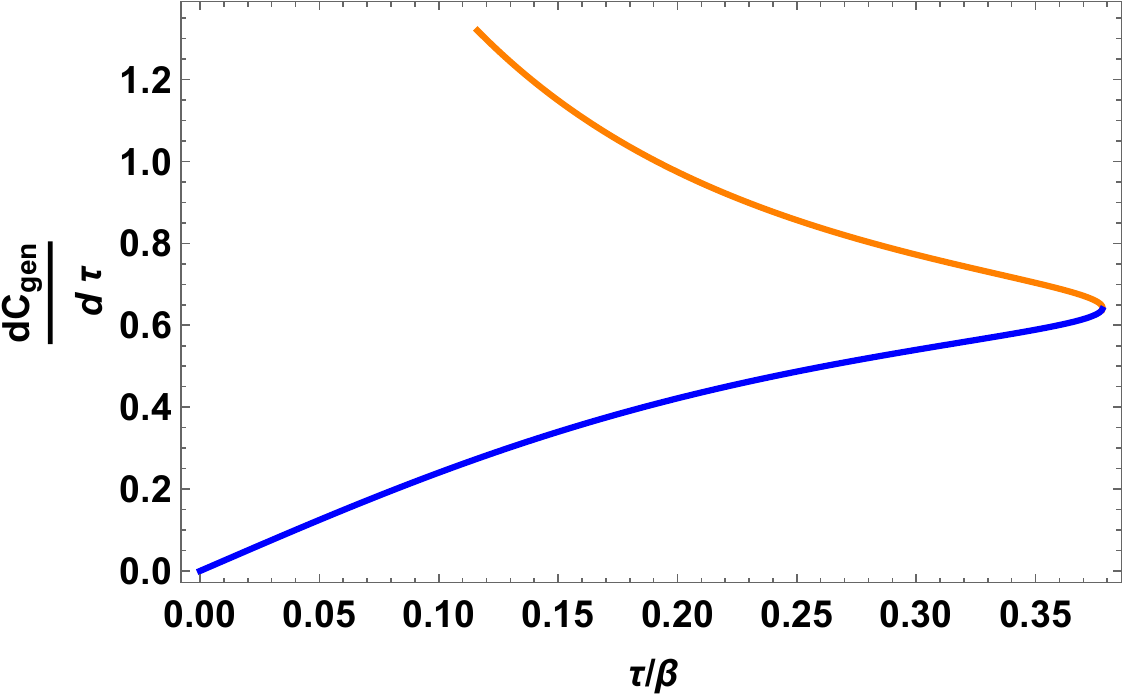}
	\end{center}
	\caption{The case $d=2$, $\theta=0$, $z=1$ and $\lambda = 5 \times 10^{-3}$: {\it Left}) The effective potential as a function of $w$. For this value of $\lambda$ there are no maxima neither inside or outside the horizon. 
		{\it Middle}) $\tau / \beta$ in terms of $s_{\rm min} = \frac{r_{\rm min}}{r_h}$. For each value of $\tau$ there are two values of $r_{\rm min}$. Therefore, it is expected that there are two branches for $\frac{ d \mathcal{C}_{\rm gen} }{d \tau}$ which one of them should be unphysical. The blue curve has larger values of $r_{\rm min}$ in comparison to the orange one. Moreover, according to Figure \ref{fig: Penrose-Extremal-Surface}, as time passes $r_{\rm min}$ decreases. Therefore, the correct values of $r_{\rm min}$ are given by the blue curve. 
		{\it Right}) The growth rate $ \frac{d \mathcal{C}_{\rm gen}}{d \tau}$ in terms of $\tau / \beta$. There are two branches indicated in blue and orange. The physical branch is the blue curve. It is evident that $\mathcal{C}_{\rm gen}$ does not grow linearly at late times. This is due to the fact that $\lambda > \frac{12}{\lambda_{\rm crt,1}}$, and it does not satisfy the given bound in Table \ref{Table: wfB-inside}. 
		%We set $ L= r_h =r_F= V_d= G_N= 1$.
	}
	\label{fig: dCgen-dtau-lambda-neq-0-too-large-lambda}
\end{figure}

\begin{figure}
	%	\vspace{-2.5cm}
	\begin{center}
		\includegraphics[scale=0.31]{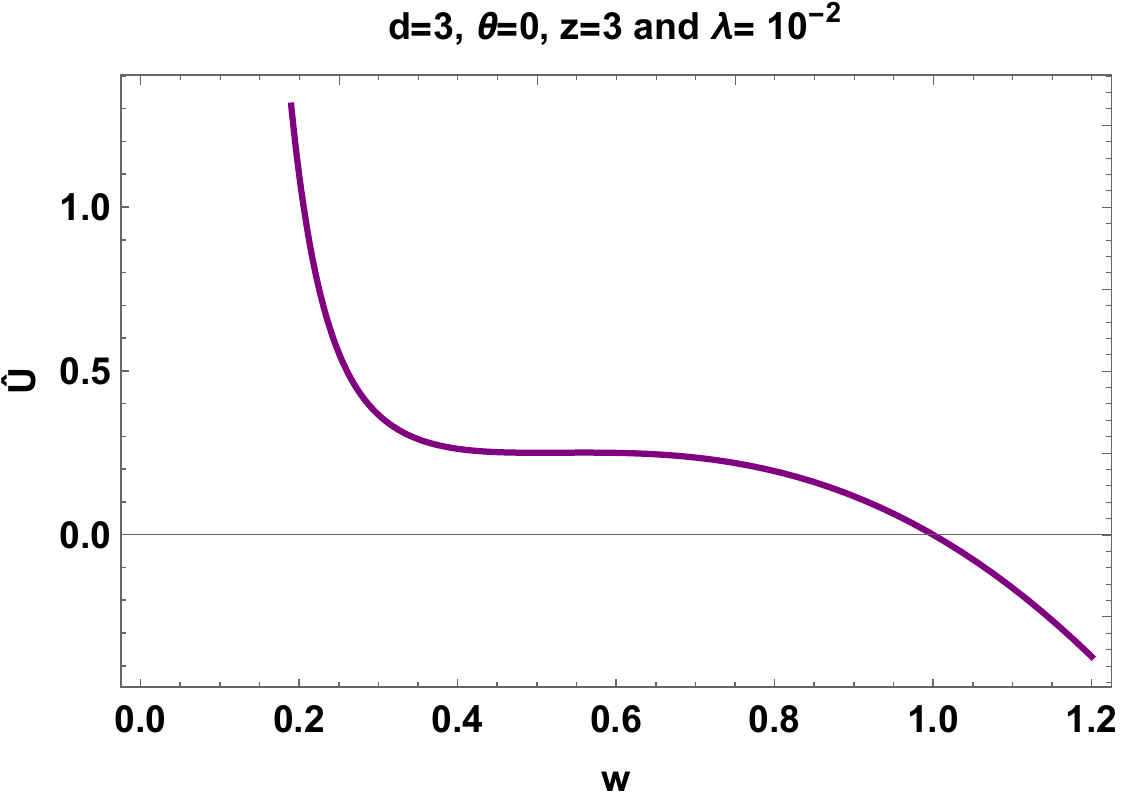}
		%\hspace{0.1cm}
		%\includegraphics[scale=0.25]{Cgen-d=3-theta=0-z=3-lambda-10m2.pdf}
		\hspace{1cm}
		\includegraphics[scale=0.31]{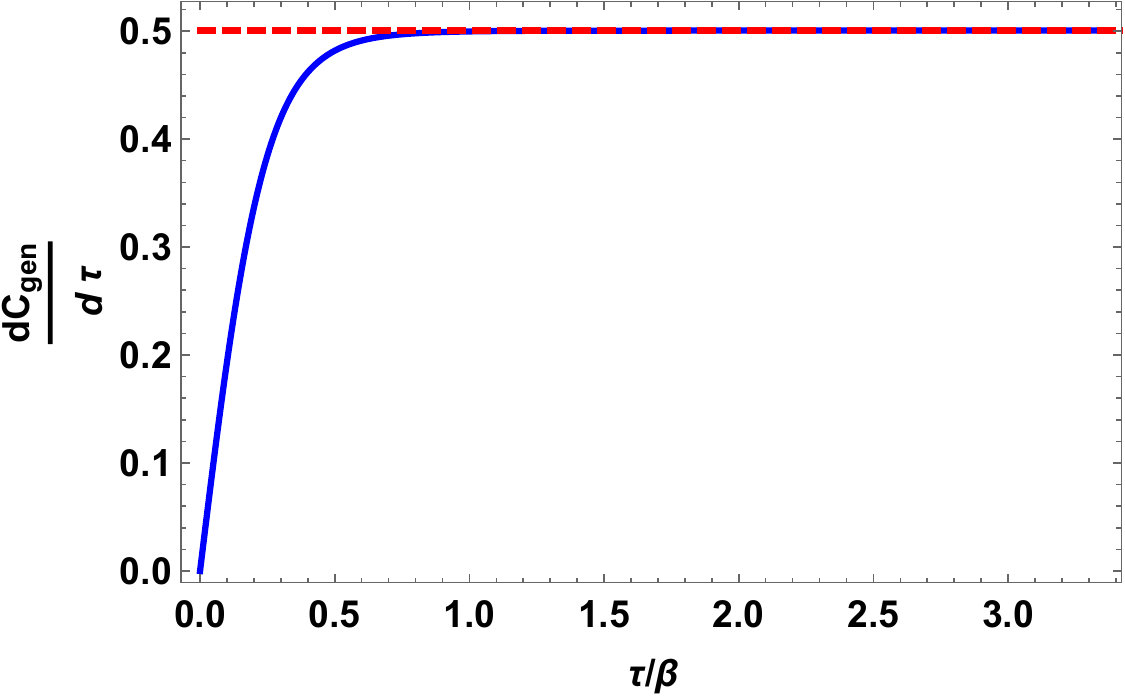}
		%		\hspace{0.1cm}
		%		\includegraphics[scale=0.25]{dCgen-dtau-lambda-neq-0-d-2-theta-0-z-1-large-lambda.pdf}
	\end{center}
	\caption{The case $d=3$, $\theta=0$, $z=3$ and $\lambda = 0.01$: {\it Left}) The effective potential as a function of $w$. For this value of $\lambda$ there is just one maximum $w_{fB1}$. It is located inside the horizon and is the solution to equation $B( \lambda, w)=0 $.
		{\it Right}) The growth rate $ \frac{d \mathcal{C}_{\rm gen}}{d \tau}$ in terms of $\tau / \beta$. The late time growth rate is linear at late times and its value indicated by the red dashed horizontal line is given by eq. \eqref{dC-dtau-2}.
		%		{\it Right}) $\Delta \mathcal{C}_{\rm gen} = \mathcal{C}_{\rm gen} - \mathcal{C}_{\rm gen}^{(0)}$. It is evident that at late times it grows linearly in time.  
	}
	\label{fig: dCgen-dtau-lambda-neq-0-maxima-one-inside}
\end{figure}

\begin{figure}
	%	\vspace{-2.5cm}
	\begin{center}
		\includegraphics[scale=0.31]{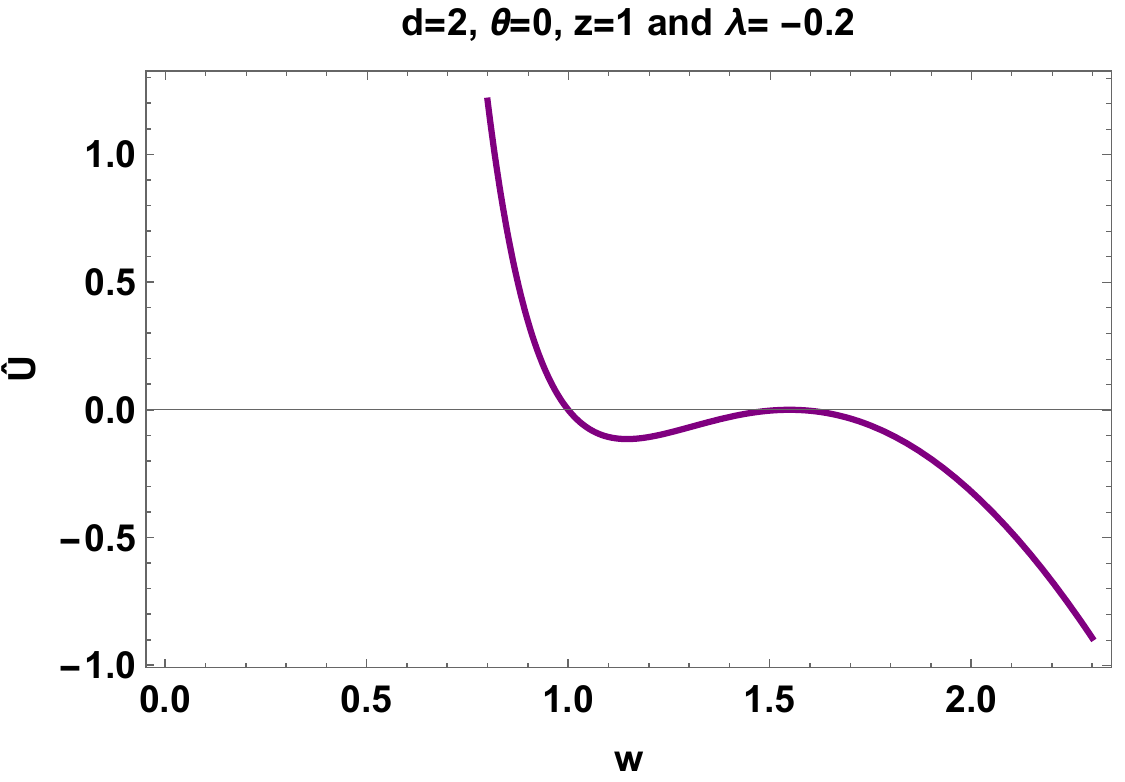}
		\hspace{1cm}
		\includegraphics[scale=0.31]{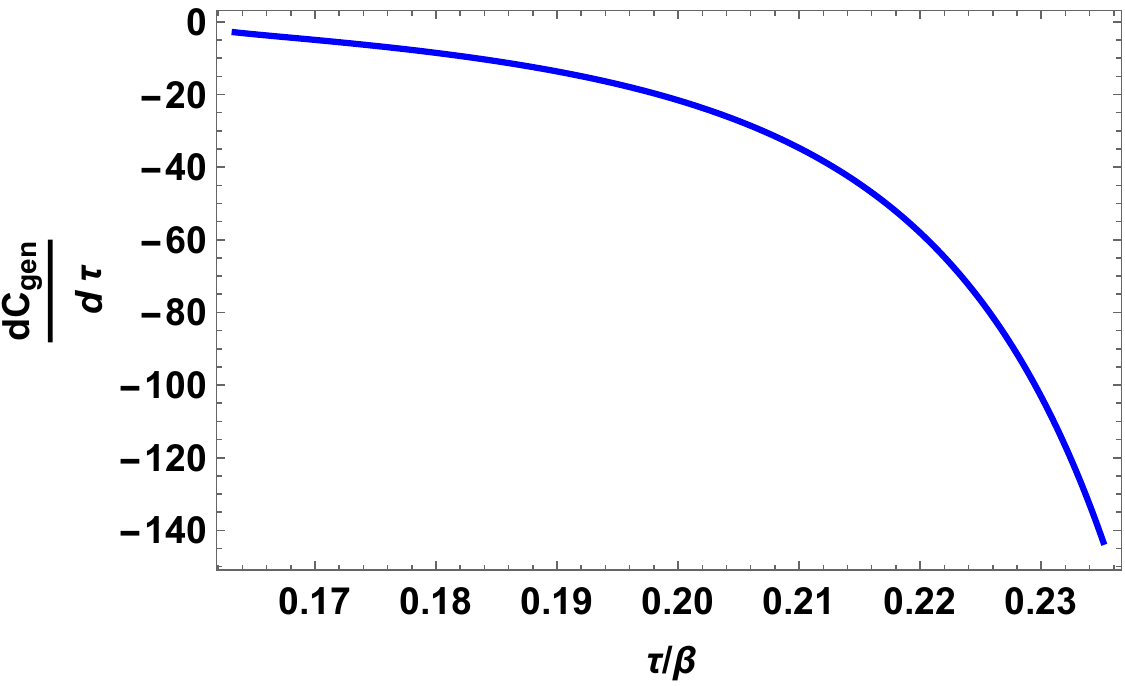}
		%		\hspace{0.1cm}
		%		\includegraphics[scale=0.25]{tau-s-min-d-2-theta-0-z-1.pdf}
		%		\hspace{0.1cm}
		%		\includegraphics[scale=0.25]{dCgen-dtau-lambda-neq-0-d-2-theta-0-z-1-large-lambda.pdf}
	\end{center}
	\caption{The case $d=2$, $\theta=0$, $z=1$ and $\lambda = -0.2$: {\it Left}) The effective potential as a function of $w$. For this value of $\lambda$ there is just one maximum $w_{fA1}$ which is located inside the horizon.
		{\it Right}) The growth rate $ \frac{d \mathcal{C}_{\rm gen}}{d \tau}$ in terms of $\tau / \beta$. The late time growth rate is negative and a decreasing function of $\tau$. Notice that since the growth rate is negative this case is not physical.
	}
	\label{fig: dCgen-dtau-lambda-neq-0-maxima-one-outside}
\end{figure}

\begin{figure}
	%	\vspace{-2.5cm}
	\begin{center}
		\includegraphics[scale=0.31]{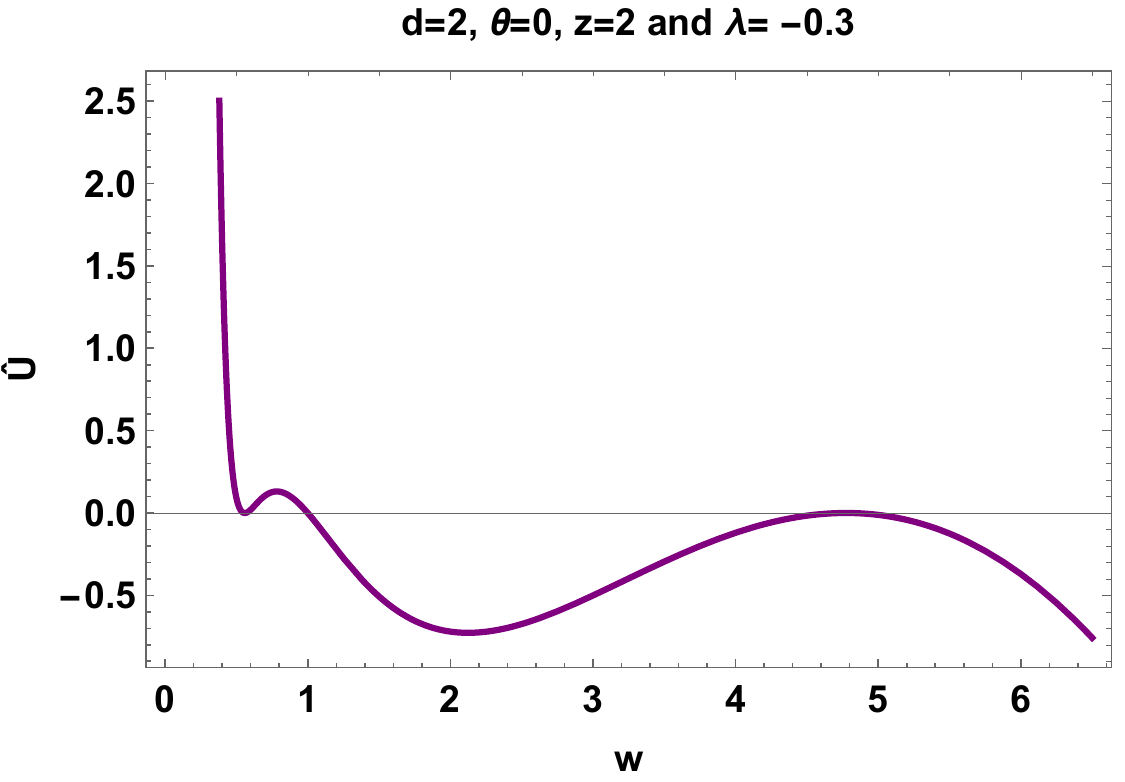}
		\hspace{1cm}
		\includegraphics[scale=0.31]{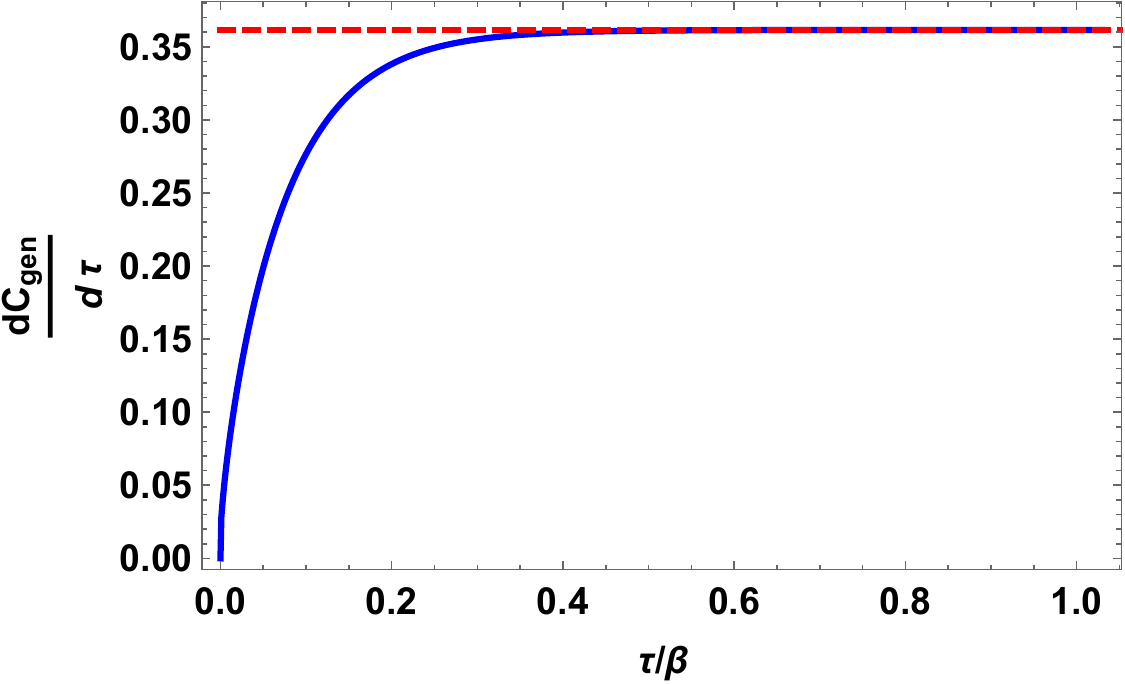}
%		\hspace{0.1cm}
%		\includegraphics[scale=0.25]{tau-s-min-d-2-theta-0-z-1.pdf}
%		\hspace{0.1cm}
%		\includegraphics[scale=0.25]{dCgen-dtau-lambda-neq-0-d-2-theta-0-z-1-large-lambda.pdf}
	\end{center}
	\caption{The case $d=2$, $\theta=0$, $z=2$ and $\lambda = -0.3$: {\it Left}) The effective potential as a function of $w$. For this value of $\lambda$, there are two maxima. There is a maximum $w_{fA2}$ inside the horizon which is the solution to equation $A( \lambda, w)=0 $ and  there is another one $w_{fB3}$ outside the horizon which is the solution to equation $B( \lambda, w) =0$.
		{\it Right}) The growth rate $ \frac{d \mathcal{C}_{\rm gen}}{d \tau}$ in terms of $\tau / \beta$. The late time growth rate is given by the value of $\hat{U}(w)$ at the maximum $w_{fA2}$ inside the horizon according to eq. \eqref{dC-dtau-2}. This value is shown by the horizontal dashed red line. Notice that $\hat{U}(wf_{B3}) =0$, and hence it gives a zero contribution to eq. \eqref{dC-dtau-2}.
	}
	\label{fig: dCgen-dtau-lambda-neq-0-maxima-inside-outside}
\end{figure}

\begin{figure}
	%	\vspace{-2.5cm}
	\begin{center}
		\includegraphics[scale=0.31]{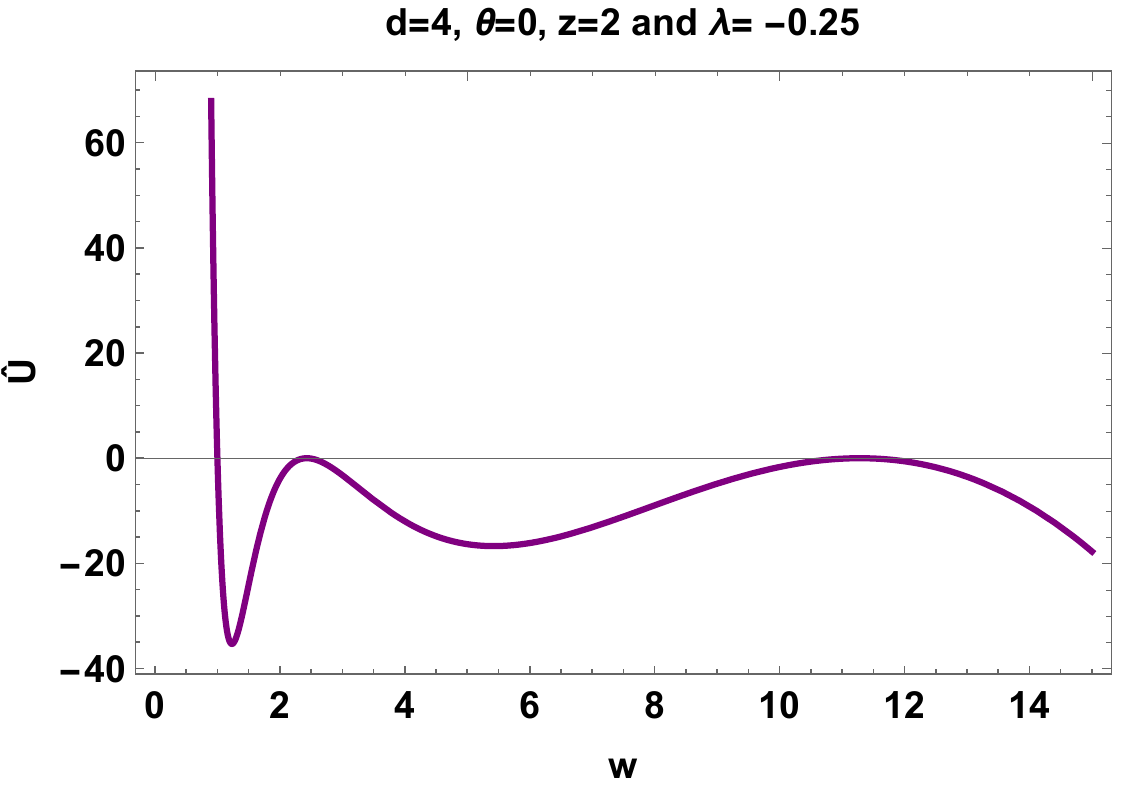}
		\hspace{1cm}
		\includegraphics[scale=0.31]{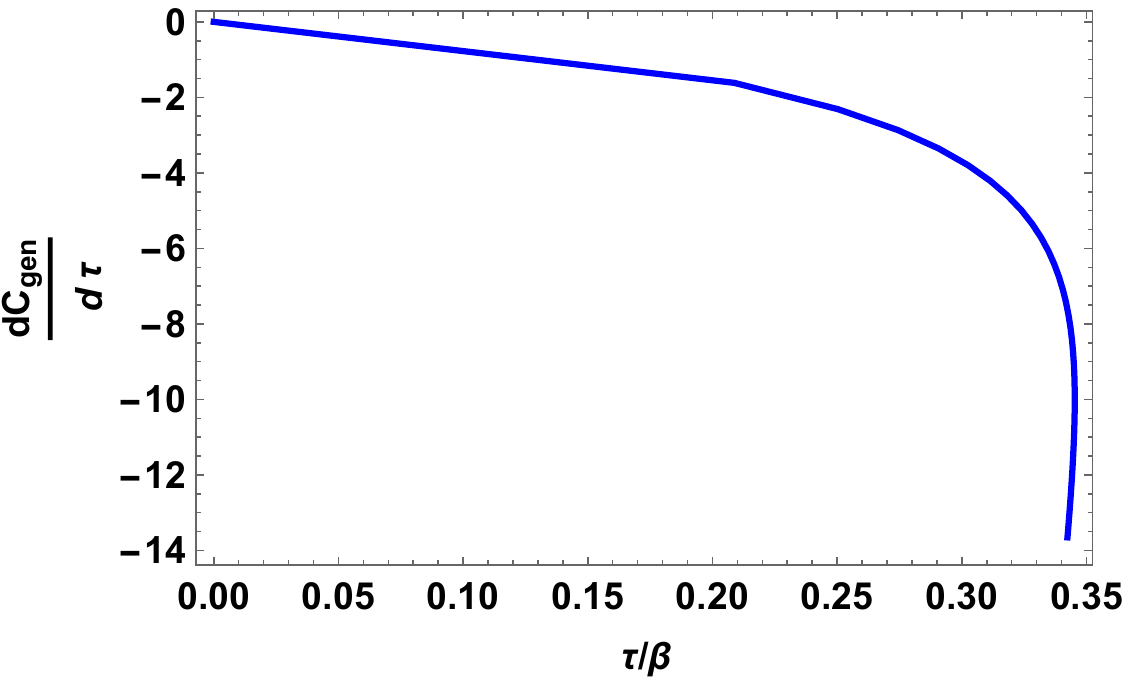}
		%		\hspace{0.1cm}
		%		\includegraphics[scale=0.25]{tau-s-min-d-2-theta-0-z-1.pdf}
		%		\hspace{0.1cm}
		%		\includegraphics[scale=0.25]{dCgen-dtau-lambda-neq-0-d-2-theta-0-z-1-large-lambda.pdf}
	\end{center}
	\caption{The case $d=4, \theta=0, z=2$ and $\lambda= - 0.25$ : {\it Left}) The effective potential as a function of $w$. For this value of $\lambda$, there are two maxima $w_{fA1,2}$ which are located outside the horizon. These are the solutions to equation $A( \lambda, w) =0$.
		{\it Right}) The growth rate $ \frac{d \mathcal{C}_{\rm gen}}{d \tau}$ in terms of $\tau / \beta$. It is negative and a decreasing function of $\tau / \beta$, and hence it is not physical.
%		The late time growth rate is given by the value of $\hat{U}(w)$ at the maximum $w_{fA2}$ inside the horizon according to eq. \eqref{dC-dtau-2}. This value is shown by the horizontal dashed red line. Notice that $\hat{U}(wf_{B3}) =0$, and hence it gives a zero contribution to eq. \eqref{dC-dtau-2}.
	}
	\label{fig: dCgen-dtau-lambda-neq-0-maxima-two-outside}
\end{figure}

In this section, we calculate the time dependence of the growth rate of $\mathcal{C}_{\rm gen}$. For $\lambda \neq 0$, it is straightforward to rewrite eq. \eqref{dC-dtau-1} as follows
\bea
\frac{d \mathcal{C}_{\rm gen}}{d \tau} = \frac{V_d r_h^{d_e+z - \theta_e} r_F^{\theta_e (d+1) }}{G_N L^{d+z+1}} s_{\rm min}^{ \frac{d_e +z}{2} -\theta_e} \sqrt{1 - s_{\rm min}^{d_e+ z}} a(s_{\rm min}).
\label{dCgen-dtau-lambda neq 0}
\eea 
Moreover,
\bea
\tau &=& \frac{(d_e +z) \beta a(s_{\rm min}) s_{\rm min}^{ \frac{de+z}{2} - \theta_e } \sqrt{ 1 -s_{\rm min}^{d_e+z}} }{2 \pi}
\cr && \cr
&& 
\;\;\;\;\; \times \!
\int_{s_{\rm min}}^{\infty} \! ds \frac{s^{d_e -1} }{ \big(1- s^{d_e+z} \big) \sqrt{ s_{\rm min}^{d_e+z- 2 \theta_e} \big(1- s_{\rm min}^{d_e+z}  \big)  a(s_{\rm min} )^2 - s^{d_e +z - 2 \theta_e} \big( 1- s^{d_e +z} \big) a(s)^2 } },
\nonumber
\\
\label{tau-lambda neq 0}
\eea 
where
\bea
a(s) = 1 + \lambda \frac{(d-1)}{(d+1)} \left( \frac{r_h}{r_F} \right)^{4 \theta_e} s^{4 \theta_e -2 (d_e +z)} 
\left( d_e (d_e -z +2) -2 z (z-1) s^{d_e +z} \right)^2.
\label{a(s)-lambda-neq-0}
\eea 
In Figures \ref{fig: dCgen-dtau-lambda-neq-0-d}, \ref{fig: dCgen-dtau-lambda-neq-0-z}, \ref{fig: dCgen-dtau-lambda-neq-0-theta} and \ref{fig: dCgen-dtau-lambda-neq-0-lambda}, we numerically plotted the growth rate $\frac{d \mathcal{C}_{\rm gen}}{d \tau} $ as a function of $\tau / \beta$ for different values of $d$, $\theta$, $z$ and $\lambda$. From these figures, it is clear that for all values of $d$, $\theta$ and $z$, the growth rate of generalized volume-complexity saturates at late times as it was argued below eq. \eqref{dC-dtau-expansion} (See also ref. \cite{Belin:2021bga}). Moreover, it reaches its late time value from below as it was expected. 
%%%%%%%%%%%%%%%%%%%%%%%%%%%%%%%%%
%%%%%%%%%%%%%%%%%%%%%%%%%
%%%%%%%%%%%%%%%%%%%%%%%%%%%%%%%%%
%%%%%%%%%%%%%%%%%%%%%%%%%
%%%%%%%%%%%%%%%%%%%%%%%%%%%%%%%%%
%%%%%%%%%%%%%%%%%%%%%%%%%
%%%%%%%%%%%%%%%%%%%%%%%%%%%%%%%%%
%%%%%%%%%%%%%%%%%%%%%%%%%
\\Furthermore, from Figures \ref{fig: dCgen-dtau-lambda-neq-0-d} and \ref{fig: dCgen-dtau-lambda-neq-0-z}, it is observed that the growth rate is neither an increasing nor a decreasing function of $d$ and $z$. Furthermore, from Figure \ref{fig: dCgen-dtau-lambda-neq-0-theta}, it is evident that $\frac{d \mathcal{C}_{\rm gen}}{d \tau} $ is an increasing function of $\theta$ which is in contrast to the case for $\lambda =0$ (See Figure \ref{fig: dCV-dtau-lambda=0-theta}).
\\On the other hand, from eq. \eqref{a(s)-lambda-neq-0}, it is clear that for a given value of $d$, $\theta$ and $z$, $a(s)$ is an increasing function of $\lambda$. Therefore, $\frac{d \mathcal{C}_{\rm gen}}{d \tau} $ is also an increasing function of $\lambda$. 
%%%%%%%%%%%%%%%%%%%%%%%%%%%%
\footnote{This behavior can also be seen from eqs. \eqref{U-w} and \eqref{dC-dtau-1}. It is clear that $\hat{U}(w)$ is an increasing function of $\lambda$.}
%%%%%%%%%%%%%%%%%%%%%%%%%%%%
To verify this behavior, we also numerically plotted the growth rate for different values of $\lambda$ in Figure \ref{fig: dCgen-dtau-lambda-neq-0-lambda}.
% It is clear that it is an increasing function of $\lambda$. 
\\It should be pointed out that to have the so-called linear growth rate at late times, one has to choose $\lambda$ properly such that the effective potential $\hat{U}(w)$ has {\it a local maximum inside the horizon}.
%by increasing the coupling $\lambda$ above a critical value $\lambda_0$, $\mathcal{C}_{\rm gen}$ do not show linear growth at late times. The value of $\lambda_0$, depends on $d$, $\theta$ and $z$.
In ref. \cite{Belin:2021bga}, it was argued that
% to have a late time linear growth, $\hat{U}(r)$ must have a local maximum inside the horizon and this maximum only exists for small values of $\lambda$. 
for small values of $\lambda$, such maximum always exists. 
%%%%%%%%%%%%%%%%%%%%%%%%%%%%
\footnote{For an AdS-Schwarzschild black brane, depending on the value of the coupling constant, $\hat{U}(w)$ might have 1) no maxima neither inside nor outside the horizon 2) one maximum inside the horizon 3) one maximum outside the horizon. Refer to Figure 3 in ref. \cite{Belin:2021bga}.}
%%%%%%%%%%%%%%%%%%%%%%%%%%%%
In our case, depending on the value of $d, \theta, z$ and $\lambda$, the effective potential might have: 
\begin{enumerate}
	\item no maxima neither inside nor outside the horizon (See Figure \ref{fig: dCgen-dtau-lambda-neq-0-too-large-lambda})
	\item one maximum inside the horizon and no maximum outside the horizon (See Figure \ref{fig: dCgen-dtau-lambda-neq-0-maxima-one-inside})
	\item no maximum inside the horizon and one maximum outside the horizon (See Figure \ref{fig: dCgen-dtau-lambda-neq-0-maxima-one-outside})
	\item one maximum inside the horizon and another one outside the horizon (See Figure \ref{fig: dCgen-dtau-lambda-neq-0-maxima-inside-outside})
	\item two maxima outside the horizon (See Figure \ref{fig: dCgen-dtau-lambda-neq-0-maxima-two-outside}).
\end{enumerate}
It is interesting to know what happens to the growth rate when there is no maximum inside the horizon.
%%%%%%%%%%%%%%%%%%%%%%%%%%%
\footnote{We would like to thank the referee for raising this intriguing question. We are also very grateful to Mohsen Alishahiha and Robert Myers for their very helpful discussions in this regard.}
%%%%%%%%%%%%%%%%%%%%%%%%%%%
To address this question, in Figures
%%%%%%%%%%%%%%%%%%%%%%%%%%%
\footnote{In all of the figures, we set $ L= r_h =r_F= V_d= G_N= 1$.}
%%%%%%%%%%%%%%%%%%%%%%%%%%%
\ref{fig: dCgen-dtau-lambda-neq-0-too-large-lambda}, \ref{fig: dCgen-dtau-lambda-neq-0-maxima-one-inside}, \ref{fig: dCgen-dtau-lambda-neq-0-maxima-one-outside}, \ref{fig: dCgen-dtau-lambda-neq-0-maxima-inside-outside} and \ref{fig: dCgen-dtau-lambda-neq-0-maxima-two-outside} we considered an example for each of the above cases, respectively. 
%\\As an example, we considered the case $d=2$, $\theta=0$, $z=1$ and $\lambda= 5 \times 10^{-3} > \frac{1}{12} \lambda_{\rm crt,1}$ 
In Figure \ref{fig: dCgen-dtau-lambda-neq-0-too-large-lambda}, we considered the case $d=2$, $\theta=0$, $z=1$ and $\lambda= 5 \times 10^{-3} > \frac{1}{12} \lambda_{\rm crt,1}$. In this case, according to Table \ref{Table: wfB-inside}, there is no maximum inside the horizon, and hence the growth rate does not show the so-called linear growth at late times. In the left panel of Figure \ref{fig: dCgen-dtau-lambda-neq-0-too-large-lambda}, we plotted the effective potential which shows that there are no maxima neither inside nor outside the horizon. In the middle panel of the figure, we plotted $\tau$ as a function of $s_{\rm min} = \frac{r_{\rm min}}{r_h}$. It is observed that when $s_{\rm min}$ is large enough, for each value of $\tau$ there are two values of $r_{\rm min}$. In the right panel of the figure, we plotted $\frac{d \mathcal{C}_{\rm gen} }{d \tau}$ as a function of $\tau / \beta$. There are two branches which are indicated in orange and blue. Therefore, for each value of $\tau$, there are two values of $\frac{d \mathcal{C}_{\rm gen} }{d \tau}$, and hence two branches for $\frac{d \mathcal{C}_{\rm gen} }{d \tau}$. 
%%%%%%%%%%%%%%%%%%%%%%%%%%%%%%
\footnote{We would like to thank the referee very much for her/his very helpful comments about the two branches in Figure \ref{fig: dCgen-dtau-lambda-neq-0-too-large-lambda}.}
%%%%%%%%%%%%%%%%%%%%%%%%%%%%%%
%%%%%%%%%%%%%%%%%%%%%%%%%%%%%%%%%
%%%%%%%%%%%%%%%%%%%%%%%%%
%%%%%%%%%%%%%%%%%%%%%%%%%%%%%%%%%
%%%%%%%%%%%%%%%%%%%%%%%%%
In the middle panel of the figure, the blue curve has a larger value of $r_{\rm min}$ in comparison to the orange curve. Moreover, by looking at Figure \ref{fig: Penrose-Extremal-Surface}, as the time increases the endpoints of the extremal surface, e.g. the orange curve in Figure \ref{fig: Penrose-Extremal-Surface}, moves upward. In this case, the value of $r_{\rm min}$ decreases and reaches $r_f$ at late times. Next, by looking at the middle panel in Figure \ref{fig: dCgen-dtau-lambda-neq-0-too-large-lambda}, it is evident that on the orange curve as $\tau$ increases $r_{\rm min}$ increases. On the other hand, on the blue curve, as $\tau$ increases, $r_{\rm min}$ decreases. Therefore, the correct values of $r_{\rm min}$ are given by the blue curve in the middle panel. Consequently, it is expected that the correct values of $\frac{d \mathcal{C}_{\rm gen} }{d \tau}$ are given by the blue curve in the right panel. It should also be pointed out that the blue curve for $\frac{d \mathcal{C}_{\rm gen} }{d \tau}$ in the right panel does not show the linear growth rate at late times, and even is faster than linear. It is widely believed that the quantum complexity and hence its gravity dual show a linear growth rate at late times (See for example \cite{Susskind:2018pmk}). Therefore, this case in which the late time behavior is not linear seems not to be physical.
%\\Furthermore, from Figure \ref{fig: dCgen-dtau-lambda-neq-0-d}, it is evident that the growth rate is an increasing function of $d$.
\\In Figure \ref{fig: dCgen-dtau-lambda-neq-0-maxima-one-inside}, we considered the case, $d=3, \theta=0, z=3$ and $\lambda=10^{-2}$. In this case, there is just one maximum $w_{fB1}$ and it is located inside the horizon. The growth rate is linear at late times and its value is given by the value of $\hat{U}(w)$ at $w_{fB1}$ according to eq. \eqref{dC-dtau-2}.
\\In Figure \ref{fig: dCgen-dtau-lambda-neq-0-maxima-one-outside}, we considered the case $d=2, \theta=0, z=1$ and $\lambda= - 0.2$. In this case, there is just one maximum $w_{fA1}$ which is located outside the horizon. The growth rate is negative and is a decreasing function of $\tau / \beta$. Therefore, this case is not physical.
%%%%%%%%%%%%%%%%%%%%%%%%%%%%%%%%%
%%%%%%%%%%%%%%%%%%%%%%%%%
\\ In Figure \ref{fig: dCgen-dtau-lambda-neq-0-maxima-inside-outside}, we considered the case $d=2, \theta=0 , z=2$ and $\lambda =- 0.3$. In this case, there is a maximum $w_{fA2}$ inside the horizon and another one $w_{fB3}$ outside the horizon. The growth rate at late times is linear and given by the value of $\hat{U}(w)$ at $w_{fA2}$ according to eq. \eqref{dC-dtau-2}. It should be pointed out that when there are two maxima, one should choose the maximum which gives a larger value for $\mathcal{C}_{\rm gen}$.
%%%%%%%%%%%%%%%%%%%%%%%%%%%%%%%%%
%%%%%%%%%%%%%%%%%%%%%%%%%
\\ In Figure \ref{fig: dCgen-dtau-lambda-neq-0-maxima-two-outside}, we considered the case $d=4, \theta=0 , z=2$ and $\lambda =- 0.25$. In this case, there are two maxima $w_{fA1,2}$ which both of them are located outside the horizon. The growth rate at late times is not linear and it is negative. Therefore, this case is not physical.
%%%%%%%%%%%%%%%%%%%%%%%%%%%%%%%%%
%%%%%%%%%%%%%%%%%%%%%%%%%
\\Therefore, from Figures \ref{fig: dCgen-dtau-lambda-neq-0-too-large-lambda}, \ref{fig: dCgen-dtau-lambda-neq-0-maxima-one-inside}, \ref{fig: dCgen-dtau-lambda-neq-0-maxima-one-outside}, \ref{fig: dCgen-dtau-lambda-neq-0-maxima-inside-outside} and \ref{fig: dCgen-dtau-lambda-neq-0-maxima-two-outside}, one might conclude that only when the coupling constant $\lambda$ is chosen such that the effective potential has a {\it maximum inside the horizon}, the growth rate behaves {\it linearly} at late times. If there is not a maximum inside the horizon, either the growth rate is negative or increases faster than linear at late times. For this reason, throughout the paper we focused on the cases where there is a maximum inside the horizon.

%%%%%%%%%%%%%%%%%%%%%%%%
\section{Complexity of Formation}
\label{Sec: Complexity of Formation}
%%%%%%%%%%%%%%%%%%%%%%%%

In this section, we calculate the complexity of formation obtained form volume-complexity. In ref. \cite{Chapman:2016hwi}, the complexity of formation for a two-sided AdS-black hole was defined as follows
\bea
\Delta \mathcal{C} = \mathcal{C}^{\rm BH} - 2 \mathcal{C}^{\rm Vac},
\label{Delta-C}
\eea 
where $\mathcal{C}^{\rm BH}$ and $\mathcal{C}^{\rm Vac} $ are the complexity of the two-sided black hole and the corresponding vacuum, respectively. In the above formula, $\mathcal{C}$ can be either volume-complexity \cite{Chapman:2016hwi} or action-complexity \cite{Chapman:2016hwi,Carmi:2017jqz,Akhavan:2019zax}. Here, we apply it for $\mathcal{C}_{\rm V}$. In this case, the vacuum, i.e. the zero-temperature HV geometry, is given by eq. \eqref{metric-vacuum}. To calculate $\Delta \mathcal{C}_{\rm V}$, we consider an extremal hypersurface which is anchored at $t_L=t_R= 0$. By symmetry, the extremal hypersurface is a bulk surface for which $t=0$ (See the magenta straight line in Figure \ref{fig: Penrose-Extremal-Surface}). In this case, from eq. \eqref{C-gen}, one simply obtains $\mathcal{C}_{\rm gen}$ of the black brane as follows
\bea
\mathcal{C}_{\rm gen} = \frac{2 V_d r_F^{ \theta_e (d+1)}}{G_N L^d} \int_{r_h}^{r_{\rm max}} 
%\left( \frac{r}{r_F}\right)^{ -\theta_e (d+1)} \left( \frac{r}{L}\right)^{d-1} \frac{a(r)}{\sqrt{f(r)}}
 \frac{r^{d_e - \theta_e -1} }{\sqrt{f(r)}} a(r) dr.
\label{C-gen-t=0-lambda-neq-0-BB}
\eea 
Next, by setting $\lambda=0$, one obtains the corresponding  volume-complexity
\bea
\mathcal{C}_{\rm V}= \frac{2 V_d r_F^{ \theta_e (d+1)}}{G_N L^d} \int_{r_h}^{r_{\rm max}^{\rm BB}} \;\;  \frac{r^{d_e - \theta_e -1} }{\sqrt{f(r)}} dr.
%\cr && \cr
% \!\!\!\! &=& \!\!\!\! \frac{2 V_d}{G_N d} \left( \frac{r}{L} \right)^{d} \; {}_2 F_{1} \left( \frac{1}{2}; - \frac{d}{d+z}, \frac{z}{d+z}; \left( \frac{r_h}{r} \right)^{d+z} \right) \arrowvert_{r_h}^{r_{\rm max}^{\rm BB}},
\label{C-gen-t=0-lambda-0-BB}
\eea 
In the following, we consider the two cases $\theta=0$ and $\theta \neq 0$, separately. It should be pointed out that the radial cutoff $r_{\rm max}$ in the black brane (BB) geometry and the vacuum are different. Therefore, to calculate the complexity of formation in eq. \eqref{Delta-C}, one has to find the relation between the two cutoffs $r_{\rm max}^{\rm BB}$ and $r_{\rm max}^{\rm Vac}$. However, both geometries have the same radial cutoff at $z= \delta$ in the Fefferman-Graham (FG) coordinate. We investigate this connection in Appendix \ref{Appendix A: Relating the Radial cutoffs}.
%%%%%%%%%%%%%%%%%%%%%%%%%%%%%%%%%
\begin{figure}
	\begin{center}
		\includegraphics[scale=1.1]{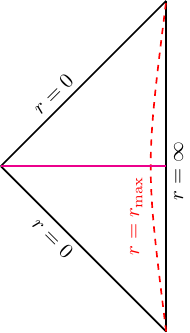}
	\end{center}
	\caption{Penrose diagram of the vacuum whose metric is given in eq. \eqref{metric-vacuum}.
%		The extremal codimension-one bulk hypersurface $\Sigma(\tau)$ denoted in orange, starts from the left boundary at time $t_L$, then goes inside the horizon and reaches the minimal radius $r_{\rm min}$.
		The red dashed curve is the radial cutoff at $r = r_{\rm max}$. Moreover, the magenta straight line denotes the extremal hypersurface anchored at $t= 0$ which is applied in the calculation of the complexity of formation.
	 }
	\label{fig: Penrose-Vacuum-Extremal-Surface}
\end{figure}
%%%%%%%%%%%%%%%%%%%%%%%%%
%For the case $\lambda=0$, $a(r)=1$, and hence one has
%\bea
%\mathcal{C}_{\rm V}= \frac{2 V_d r_F^{ \theta_e (d+1)}}{G_N L^d} \int_{r_h}^{r_{\rm max}^{\rm BB}} \;\;  \frac{r^{d_e - \theta_e -1} }{\sqrt{f(r)}} dr.
%%\cr && \cr
%% \!\!\!\! &=& \!\!\!\! \frac{2 V_d}{G_N d} \left( \frac{r}{L} \right)^{d} \; {}_2 F_{1} \left( \frac{1}{2}; - \frac{d}{d+z}, \frac{z}{d+z}; \left( \frac{r_h}{r} \right)^{d+z} \right) \arrowvert_{r_h}^{r_{\rm max}^{\rm BB}},
%\label{C-gen-t=0-lambda-0-BB}
%\eea 

%%%%%%%%%%%%%%%%%%%%%%%%
\subsection{$\theta=0$}
\label{Sec: Delta-CV-theta=0}
%%%%%%%%%%%%%%%%%%%%%%%%

For the case $\theta=0$, from eq. \eqref{C-gen-t=0-lambda-0-BB}, one obtains
\bea
\mathcal{C}_{\rm V} =
%\!\!\!\! &=& \!\!\!\! \frac{2 V_d}{G_N L} \int_{r_h}^{r_{\rm max}^{\rm BB}} \;\; \frac{r_F^{ \theta_e (d+1)}}{L^{d-1}} \frac{r^{d_e - \theta_e -1} }{\sqrt{f(r)}} dr
%\cr && \cr
%\!\!\!\! &=& \!\!\!\! 
\frac{2 V_d}{G_N d} \left( \frac{r}{L} \right)^{d} \; {}_2 F_{1} \left( \frac{1}{2}; - \frac{d}{d+z}, \frac{z}{d+z}; \left( \frac{r_h}{r} \right)^{d+z} \right)  \biggr|_{r_h}^{r_{\rm max}^{\rm BB}}.
\label{C-gen-t=0-lambda-0-BB-theta-0}
\eea 
Next, by plugging eq. \eqref{r-max-BB-r-max-Vac-theta-0-2} into the above expression, and expanding it in powers of $\delta \ll 1$, one has
\bea
\mathcal{C}_{\rm V} &=& 
\frac{2 V_d}{G_N} \Bigg[
\frac{1}{d} \left( \frac{L}{\delta} \right)^d + \frac{\kappa_0}{8} \left( \frac{r_h }{L} \right)^d + \mathcal{O} \left( \delta^z \right)
\Bigg]
\cr && \cr
&=& \frac{2 V_d}{G_N d} \left( \frac{L}{\delta} \right)^d + \kappa_0 S +  \mathcal{O} \left( \delta^z \right),
\label{C-gen-t=0-lambda-0-BB-theta-0-1}
\eea 
where $S$ is the thermal entropy given in eq. \eqref{T-S}. Moreover, $\kappa_0$ is a dimensionless constant defined as follows
\bea
\kappa_0 = 
\begin{cases}
	0 &~~ d=z, \\
	- \frac{8 \sqrt{\pi} \Gamma \left( \frac{z}{d+z}\right)}{d \;\Gamma \left( \frac{z-d}{2(d+z)}\right)} &~~ d \neq z.
\end{cases}
\label{kappa-0}
\eea 
For large values of $d$, it approaches a constant
\bea
\kappa_0 = \frac{4}{z} + \mathcal{O} \left( \frac{1}{d} \right).
\eea 
Furthermore, for the case $z=1$, one has
\bea
\kappa_0 = 
\begin{cases}
	0 &~~ d=1, \\
	- \frac{8 \sqrt{\pi} \Gamma \left( \frac{1}{d+1}\right)}{d \;\Gamma \left( \frac{1-d}{2(d+1)}\right)} = 
	\frac{8 \sqrt{\pi} \Gamma \left( \frac{-d}{d+1}\right)}{ (d+1) \Gamma \left( \frac{-d}{2(d+1)}\right)}&~~ d \neq 1,
\end{cases}
\label{kappa-0-z-1}
\eea 
and eq. \eqref{C-gen-t=0-lambda-0-BB-theta-0-1} reduces to the volume-complexity of a planar AdS-Schwarzschild black hole (See eq. (5.6) in ref. \cite{Chapman:2016hwi}.) On the other hand, for the vacuum, by setting $f(r)=1$ in eq. \eqref{C-gen-t=0-lambda-0-BB}, one has (See Figure \ref{fig: Penrose-Vacuum-Extremal-Surface})
\bea
\mathcal{C}_{\rm V}^{\rm Vac} &=&  \frac{ V_d r_F^{ \theta_e (d+1)}}{G_N L^d} \int_{0}^{r_{\rm max}^{\rm Vac}}  r^{d_e - \theta_e -1} dr
\label{C-gen-t=0-lambda-0-vacuum-1}
%\cr && \cr
\\
&=& \frac{V_d r_F^{\theta_e (d+1)}}{  (d_e - \theta_e) G_N L^d} \left( r_{\rm max}^{\rm Vac} \right)^{d_e - \theta_e}.
\label{C-gen-t=0-lambda-0-vacuum-2}
\eea 
For the case $\theta=0$, it is simplified as follows
\bea
\mathcal{C}_{\rm V}^{\rm Vac} 
%\!\!\!\! &=& \!\!\!\!  \frac{ V_d r_F^{ \theta_e (d+1)}}{G_N L^d} \int_{0}^{r_{\rm max}^{\rm Vac}}  r^{d_e - \theta_e -1} dr
%\cr && \cr
%\!\!\!\! &=& \!\!\!\! 
= \frac{V_d }{  G_N d} \left( \frac{r_{\rm max}^{\rm Vac}}{L} \right)^{d}.
\label{C-gen-t=0-lambda-0-vacuum-theta-0-1}
\eea 
Next, by plugging eq. \eqref{r-max-vac-delta} into the above equation, one has
\bea
\mathcal{C}_{\rm V}^{\rm Vac} = \frac{V_d }{  G_N d} \frac{L^d}{\delta^d}.
\label{C-gen-t=0-lambda-0-vacuum-theta-0-2}
\eea 
Then by substituting eqs. \eqref{C-gen-t=0-lambda-0-BB-theta-0-1} and \eqref{C-gen-t=0-lambda-0-vacuum-theta-0-2} into eq. \eqref{Delta-C}, and talking the limit $\delta \rightarrow 0$, one has
\bea
\Delta \mathcal{C}_V= \mathcal{C}_V^{\rm BB} - 2 \; \mathcal{C}_V^{\rm Vac}= \kappa_0 S.
\label{Delta-C-lambda-0-theta-0}
\eea 
%where $S$ is the thermal entropy of the black brane. 
Therefore, the UV divergent terms in $\mathcal{C}_V^{\rm BB}$ and $\mathcal{C}_V^{\rm Vac}$ cancel each others, and the result is proportional to the thermal entropy $S$.

%%%%%%%%%%%%%%%%%%%%%%%%
\subsection{$\theta \neq 0$}
\label{Sec: Delta-CV-theta-neq-0}
%%%%%%%%%%%%%%%%%%%%%%%%
 
For $\theta \neq 0$, from eq. \eqref{C-gen-t=0-lambda-0-BB} it is clear that one has to consider the two cases $d_e \neq \theta_e$ and $d_e = \theta_e$, separately. The reason is that for the former the UV divergent terms are power law and for the latter they are logarithmic.

%%%%%%%%%%%%%%%%%%%%%%%%
\subsubsection{$d_e \neq \theta_e$}
\label{Sec: Delta-CV-theta-neq-0-de-neq-theta-e}
%%%%%%%%%%%%%%%%%%%%%%%%

In this case, it is straightforward to check that the integral in eq. \eqref{C-gen-t=0-lambda-0-BB} has no analytical results for arbitrary values of $d$, $\theta$ and $z$. However, for some cases
%%%%%%%%%%%%%%%%%%%%%%%%%%%%%%
\footnote{For example, it can be calculated for the following cases
	\begin{itemize}
		\item $d=1$ and $\theta= \frac{1}{3}, \frac{2}{3}, \frac{3}{4}, \frac{1}{5}, \frac{2}{5}, \frac{3}{5}, \frac{4}{5}, \frac{1}{6}, \frac{5}{6}, \frac{1}{7}, \frac{2}{7}, \frac{3}{7}, \frac{4}{7}, \frac{5}{7}, \frac{6}{7}, \frac{1}{8}, \frac{3}{8}, \frac{5}{8}, \frac{7}{8} , \frac{1}{9}, \cdots$
		\item $d=2$ and $\theta= \frac{3}{2}, \frac{5}{3}, \frac{5}{4}, \frac{7}{4}, \frac{6}{5}, \frac{7}{5}, \frac{8}{5}, \frac{9}{5}, \frac{7}{6}, \frac{11}{6}, \frac{8}{7}, \frac{9}{7}, \frac{10}{7}, \frac{11}{7}, \frac{12}{7}, \frac{13}{7}, \frac{9}{8}, \frac{11}{8},  \cdots$
		\item $d=3$ and $\theta= \frac{5}{2}, \frac{7}{3}, \frac{8}{3}, \frac{11}{4}, \frac{11}{5}, \frac{12}{5}, \frac{14}{5}, \frac{13}{6}, \frac{17}{6}, \frac{15}{7}, \frac{16}{7}, \frac{18}{7}, \frac{19}{7}, \frac{20}{7}, \frac{17}{8}, \frac{19}{8}, \cdots$
	\end{itemize}
Moreover, for the following cases, $\{d=1, \theta=\frac{1}{4} \}, \{d=2, \theta = \frac{1}{3},-3 \}, \{d=3, \theta= \frac{9}{8}, \frac{15}{8} \}$, volume-complexities are again given by hypergeometric functions. However, they cannot be written in the form given in eq. \eqref{C-gen-t=0-lambda-0-BB-theta-neq-0}. For these cases, 
%eq. \eqref{Delta-C-lambda-0-theta-neq-0} is still valid, but with a different value for $\xi$.
one still has $\Delta \mathcal{C}_V \propto S T^{- \frac{ \theta_e}{z}} $. 
}
%%%%%%%%%%%%%%%%%%%%%%%%%%%%%%
it can be calculated analytically as follows
\bea
\mathcal{C}_{\rm V} =
%\!\!\!\! &=& \!\!\!\! \frac{2 V_d}{G_N L} \int_{r_h}^{r_{\rm max}^{\rm BB}} \;\; \frac{r_F^{ \theta_e (d+1)}}{L^{d-1}} \frac{r^{d_e - \theta_e -1} }{\sqrt{f(r)}} dr
%\cr && \cr
%\!\!\!\! &=& \!\!\!\! 
\frac{2 V_d r_F^{\theta_e (d+1)}}{G_N (d_e- \theta_e)} \left( \frac{r^{d_e - \theta_e} }{L^d} \right) \; {}_2 F_{1} \left( \frac{1}{2}; - \frac{(d_e - \theta_e)}{(d_e+z)}, \frac{z+ \theta_e}{d_e+z}; \left( \frac{r_h}{r} \right)^{d_e+z} \right)  \biggr|_{r_h}^{r_{\rm max}^{\rm BB}}.
\label{C-gen-t=0-lambda-0-BB-theta-neq-0}
\eea 
Next, by plugging eq. \eqref{r-max-BB-r-max-Vac-theta-neq-0-2} into the above expression, and expanding it in powers of $\delta \ll 1$, one has
\bea
\mathcal{C}_{\rm V} &=& 
\frac{2 V_d r_F^{\theta_e (d+1)}}{G_N } \Bigg[
\frac{L^{d_e - \theta_e (d+2)}}{(d_e - \theta_e)} \frac{1}{ \delta^{d_e -\theta_e}} + \frac{\kappa}{8} \;  \frac{r_h^{d_e - \theta_e}}{L^d} + \mathcal{O} \left( \delta^{z +\theta_e} \right) + \cdots
\Bigg]
\cr && \cr
&=& \frac{2 V_d }{G_N (d_e - \theta_e)} \left( \frac{r_F}{L} \right)^{ \theta_e (d+1)} \left( \frac{L}{\delta} \right)^{d_e - \theta_e} 
\cr && \cr
&& 
%\;\;\;\;\;\;\;\;\;\;\;
+ \kappa \;  \left( \frac{4 \pi L^{z+1}}{(d_e +z) r_F^z} \right)^{- \frac{\theta_e}{z}} S T^{- \frac{ \theta_e}{z}}+  \mathcal{O} \left( \delta^{z+ \theta_e} \right) + \cdots,
\label{C-gen-t=0-lambda-0-BB-theta-neq-0-1}
\eea 
where 
%%%%%%%%%%%%%%%%%%%%%%%%%%%%%
\footnote{Notice that when $\theta=0$, $\kappa$ reduces to $\kappa_0$ for the case $d \neq z$.}
%%%%%%%%%%%%%%%%%%%%%%%%%%%%%
\bea
\kappa = - \frac{8 \sqrt{\pi} \Gamma \left( \frac{z + \theta_e}{d_e+ z} \right)}{(d_e - \theta_e) \Gamma \left( \frac{z- d_e + 2 \theta_e }{2(d_e +z)}\right)}.
\label{kappa}
\eea  
%\bea
%\kappa = 
%\begin{cases}
%	0 &~~ d=z, \\
%	- \frac{8 \sqrt{\pi} \Gamma \left( \frac{z + \theta_e}{d_e+ z} \right)}{(d_e - \theta_e) \Gamma \left( \frac{z- d_e + 2 \theta_e }{2(d_e +z)}\right)} &~~ d \neq z,
%\end{cases}
%\label{kappa}
%\eea  
Moreover, for large values of $d$, it approaches a constant
\bea
\kappa = \frac{4}{z} + \mathcal{O} \left( \frac{1}{d} \right).
\eea 
On the other hand, for the vacuum, by plugging eq. \eqref{delta-r-max-theta-neq-0-2} into eq. \eqref{C-gen-t=0-lambda-0-vacuum-2}, one arrives at
\bea
\mathcal{C}_{\rm V}^{\rm Vac} &=&  
\frac{V_d r_F^{\theta_e (d+1)}}{  (d_e - \theta_e) G_N L^d} \left( r_{\rm max}^{\rm Vac} \right)^{d_e - \theta_e}
\cr && \cr
&=&
\frac{V_d r_F^{\theta_e (d+1)}}{  (d_e - \theta_e) G_N} \frac{L^{d_e- \theta - 2 \theta_e}}{\delta^{d_e - \theta_e}}.
\label{C-gen-t=0-lambda-0-thtea-neq-0-vacuum}
\eea 
Next, by plugging eqs. \eqref{C-gen-t=0-lambda-0-BB-theta-neq-0-1} and \eqref{C-gen-t=0-lambda-0-thtea-neq-0-vacuum} into eq. \eqref{Delta-C}, and taking the limit $\delta \rightarrow 0$, one obtains
\bea
\Delta \mathcal{C}_V =\kappa \;  \left( \frac{4 \pi L^{z+1}}{(d_e +z) r_F^z} \right)^{- \frac{\theta_e}{z}} S T^{- \frac{ \theta_e}{z}} = \xi \; S T^{- \frac{\theta_e}{z}},
%\mathcal{C}_V^{\rm BH} - 2 \; \mathcal{C}_V^{\rm Vac}= \kappa S.
\label{Delta-C-lambda-0-theta-neq-0}
\eea 
where
\bea
\xi = \kappa \left( \frac{4 \pi L^{z+1}}{(d_e +z) r_F^z } \right)^{ - \frac{\theta_e}{z}}.
\eea 
Therefore, similar to the case $\theta=0$, the UV divergent terms in $\mathcal{C}_V^{\rm BB}$ and $\mathcal{C}_V^{\rm Vac}$ cancel each others exactly. Moreover, $\Delta \mathcal{C}_V$ is proportional to $S T^{- \frac{\theta_e}{z}}$. Furthermore, for $\theta=0$, eq. \eqref{Delta-C-lambda-0-theta-neq-0} reduces to eq. \eqref{Delta-C-lambda-0-theta-0}.

%%%%%%%%%%%%%%%%%%%%%%%%
\subsubsection{$d_e = \theta_e$}
\label{Sec: Delta-CV-theta-neq-0-de=theta-e}
%%%%%%%%%%%%%%%%%%%%%%%%

For $d_e = \theta_e$ or equivalently
\bea
\theta= \frac{d^2}{d+1},
\eea
by applying eqs. \eqref{C-gen-t=0-lambda-0-BB} and \eqref{r-max-BB-r-max-Vac-theta-neq-0-2}, one easily obtains
\bea
\mathcal{C}_{\rm V} &=&  \frac{2 V_d r_F^d}{G_N L^d} \int_{r_h}^{r_{\rm max}^{\rm BB}} \;\;  \frac{1}{ r \sqrt{f(r)}} dr
\cr && \cr
%\!\!\!\! &=& \!\!\!\! \frac{4 (d+1)V_d}{G_N \big( d +z (d+1) \big)} \left( \frac{r_F}{L} \right)^{d} \tanh^{-1} \sqrt{1 - \left( \frac{r_h}{r} \right)^{ \frac{d}{d+1}+ z} } \arrowvert_{r_h}^{r_{\rm max}^{\rm BB}},
&=& \frac{4 V_d}{G_N \big( z +  \theta_e \big)} \left( \frac{r_F}{L} \right)^{d} \tanh^{-1} \sqrt{1 - \left( \frac{r_h}{r_{\rm max}^{\rm BB}} \right)^{ z + \theta_e} } 
\cr && \cr
&=& \frac{2 V_d}{G_N \big( d+ z(d+1) \big)} \left( \frac{r_F}{L} \right)^{d} \log \Bigg[ 2^{2(d+1)} \left( \frac{L^2}{r_h \delta} \right)^{d+ z(d+1)} \Bigg] + \mathcal{O} \left( \delta^{z + \theta_e} \right).
\label{C-gen-t=0-lambda-0-BB-de=theta-e}
\eea 
%By applying eq. \eqref{r-max-BB-r-max-Vac-theta-neq-0-2} and expanding eq. \eqref{C-gen-t=0-lambda-0-BB-de=theta-e}, one has
Therefore, the UV divergent term is logarithmic in this case. On the other hand, for the vacuum, from eq. \eqref{C-gen-t=0-lambda-0-vacuum-1}, one has
\bea
\mathcal{C}_{\rm V}^{\rm Vac} &=&  \frac{ V_d r_F^{ d}}{G_N L^d} \int_{\epsilon}^{r_{\rm max}^{\rm Vac}} \frac{dr}{r}
%\cr && \cr
%\!\!\!\! &=& \!\!\!\! 
= \frac{V_d r_F^d}{G_N L^d} \log \left( \frac{r_{\rm max}^{\rm Vac}}{\epsilon} \right) 
\cr && \cr
&=& \frac{V_d r_F^d}{G_N L^d} \log \left( \frac{L^2}{\epsilon \; \delta} \right), 
\label{C-gen-t=0-lambda-0-vacuum-d-e=theta-e}
\eea 
where $\epsilon \rightarrow 0$ is an IR cutoff. Next, from eqs. \eqref{C-gen-t=0-lambda-0-BB-de=theta-e} and \eqref{C-gen-t=0-lambda-0-vacuum-d-e=theta-e}, one can find $ \Delta \mathcal{C}_V$ as follows
\bea
\Delta \mathcal{C}_V &=&
%\frac{2 V_d}{ G_N \big( d + z(d+1) \big) } \left( \frac{r_F}{L} \right)^d \log \Bigg[ 2^{2(d+1) }\left( \frac{\epsilon}{r_h} \right)^{d+ z (d+1)}  \Bigg] 
\frac{2 V_d}{ G_N } \left( \frac{r_F}{L} \right)^d \log \left( \frac{2^{\frac{2}{z+ \theta_e}} \epsilon}{r_h} \right)  
\cr && \cr
&=& \frac{2 V_d}{ G_N } \left( \frac{r_F}{L} \right)^d \log \left( \frac{4^{ \frac{d_e}{d_e +z}} \epsilon^{d_e} V_d r_F^\theta}{4 G_N L^d S} \right).
\label{Delta-C-lambda-0-theta-neq-0-d-e=theta-e}
\eea 
It is observed that the logarithmic UV divergent terms in eqs. \eqref{C-gen-t=0-lambda-0-BB-de=theta-e} and \eqref{C-gen-t=0-lambda-0-vacuum-d-e=theta-e} are canceled. Moreover, $\Delta \mathcal{C}_V$ depends on $\log S$. 
%This behavior is surprising.

%%%%%%%%%%%%%%%%%%%%%%%%
\section{Discussion}
\label{Sec: Discussion}
%%%%%%%%%%%%%%%%%%%%%%%%

In this paper, we studied the generalized volume-complexity $\mathcal{C}_{\rm gen}$ which is an extension of volume-complexity for a two-sided uncharged HV black brane in $d+2$ dimensions. We considered the case where the two functionals $F_1$ and $F_2$ are equal to each other such that $F_1=F_2= 1 + \lambda C^2$. Here $C^2$ is the square of the Weyl tensor of the background and $\lambda$ is a dimensionless coupling constant. We investigated the two cases $\lambda =0$ and $\lambda \neq 0$, separately.
%, where $\lambda$ is the coupling constant of the higher curvature term in $\mathcal{C}_{\rm gen}$ (See eq. \eqref{F1=F2}). 
For both cases, it was proved that the growth rate of generalized volume-complexity at late times, i.e. $\lim_{\tau \to \infty} \frac{d \mathcal{C}_{\rm gen}}{d \tau}$, is determined by the maxima of the effective potential $\hat{U}(r)$ which are located inside the horizon (See eq. \eqref{dC-dtau-2}). Therefore, we also investigated these maxima. 
\\ For the case, $\lambda =0$, $\mathcal{C}_{\rm gen}$ reduces to the volume-complexity $\mathcal{C}_V$ in the CV proposal. In this case, the effective potential $\hat{U}(w)$ has a simple maximum inside the horizon. Therefore, one is able to analytically calculate the late time growth rate (See eq. \eqref{dC-dtau-lambda-0-2}). We also numerically calculated the growth rate $\frac{d \mathcal{C}_{\rm V}}{d \tau}$ of volume-complexity (See Figures \ref{fig: dCV-dtau-lambda=0-d}, \ref{fig: dCV-dtau-lambda=0-z} and \ref{fig: dCV-dtau-lambda=0-theta}) for different values of $d$, $\theta$ and $z$. It was observed that it always saturates to a constant value at late times and reaches this value from {\it below}. Furthermore, it is an increasing function of $d$, and is a decreasing function of the dynamical exponent $z$ (See Figures \ref{fig: dCV-dtau-lambda=0-d} and \ref{fig: dCV-dtau-lambda=0-z}). However, it is neither an increasing nor a decreasing function of $\theta$ (See Figure \ref{fig: dCV-dtau-lambda=0-theta}).
\\For the case $\lambda \neq 0$, we calculated numerically the growth rate $\frac{d \mathcal{C}_{\rm gen}}{d \tau}$ of generalized volume-complexity (See Figures \ref{fig: dCgen-dtau-lambda-neq-0-d}, \ref{fig: dCgen-dtau-lambda-neq-0-z}, \ref{fig: dCgen-dtau-lambda-neq-0-theta} and \ref{fig: dCgen-dtau-lambda-neq-0-lambda}) for different values of $d$, $\theta$, $z$ and $\lambda$. It was observed that for appropriate ranges of $\lambda$, it saturates to a constant value at late times and reaches this value from {\it below}. Furthermore, it is neither an increasing nor a decreasing function of $d$ and $z$ (See Figures \ref{fig: dCgen-dtau-lambda-neq-0-d} and \ref{fig: dCgen-dtau-lambda-neq-0-z}). However, it is an increasing function of the exponent $\theta$ (See Figure \ref{fig: dCgen-dtau-lambda-neq-0-theta}). Moreover, it is an increasing function of $\lambda$ (See Figure \ref{fig: dCgen-dtau-lambda-neq-0-lambda}). 
%\\Moreover, it was proved that the value of $\frac{d \mathcal{C}_{\rm gen}}{d \tau}$ at late times is determined by the maxima of the effective potential $\hat{U}(r)$ which are located inside the horizon (See eq. \eqref{dC-dtau-2}). Therefore, we also investigated these maxima of $\hat{U}(w)$. 
\\On the other hand, for the case $\lambda \neq 0$, we also examined the maxima of $\hat{U}(w)$. It was observed that for arbitrary values of $d$, $\theta$ and $z$, there are no analytical solutions to equation $\hat{U}''(w)=0$. However, for the following cases one can find analytical solutions: 1) $\theta=0$ and arbitrary values of $d$ and $z$, 2) $d=2$, $\theta=1$ and $z=1$. It was observed that for the case $\theta=0$, i.e. Lifshitz black branes, there are three possible local maxima $w_{fA1}$, $w_{fA2}$ and $w_{fB}$
%for the effective potential $\hat{U}(w)$, 
depending on the value of the coupling constant $\lambda$. Theses values were reported in Tables \ref{Table: wfA1-outside}, \ref{Table: wfA2-outside}, \ref{Table: wfB-outside} and \ref{Table: wfB-inside}. The maxima at $w= w_{fA1,2}$ are always located outside the horizon (See Figures \ref{fig: theta=0-z-wfA1-outside} and \ref{fig: theta=0-z-wfA2-outside}), and hence have no significance. On the other hand, the maximum at $w= w_{fB}$ is located either inside or outside the horizon depending on the value of $\lambda$ (See Figures \ref{fig: theta=0-z-wfB-outside} and \ref{fig: theta=0-z-wfB-inside}). 
%As mentioned before, the maximum which is located inside the horizon, determines $\lim_{\tau \to \infty} \frac{d \mathcal{C}_{\rm gen}}{d \tau}$. 
Moreover, for the case $d=2$, $\theta=1$ and $z=1$, the appropriate range of $\lambda$ is given by eq. \eqref{lambda-d=2-theta=1-z=1}. 
%For both cases, there is an analytic expression for 
It should be pointed out that for $\theta=0$ and $z=1$, the HV black brane becomes a planar AdS-Schwarzschild black hole and all of our results reduce to those reported in ref. \cite{Belin:2021bga}. 
\\On the other hand, we calculated the growth rate of generalized complexity in Appendix \ref{Appendix B: Other Higher Derivative Functionals} for the case where $F_1= F_2$ and they are a linear combination of the Ricci scalar, square of the Ricci tensor and square of the Riemann tensor (See eq. \ref{F1-F2-R2-Ric2-Riemann2}). It was observed that by choosing the coupling constants $\lambda_{1,2,3}$ properly such that the effective potential has a maximum inside the horizon, the growth rate again has a linear behavior at late times.
\\Furthermore, we calculated the complexity of formation $\Delta \mathcal{C}_V$ obtained from volume-complexity. We observed that the UV divergent terms in $\mathcal{C}_V$ both for the black brane and the vacuum geometries cancel each other exactly. Therefore, in the limit $\delta \rightarrow 0$, where $\delta$ is the radial cutoff in the Fefferman-Graham coordinate, $\Delta \mathcal{C}_V$ is finite. For the case $\theta=0$, it is proportional to the thermal entropy $S$ of the black brane (See eq. \eqref{Delta-C-lambda-0-theta-0}). On the other hand, for the case $\theta \neq 0$ and $d_e \neq \theta_e$, it can be analytically calculated for some values of $d$ and $\theta$. For these values, it was observed that $\Delta \mathcal{C}_V$ is proportional to $S T^{- \frac{\theta_e}{z}}$, where $T$ is the temperature of the black brane (See eq. \eqref{Delta-C-lambda-0-theta-neq-0}). Moreover, for the case $\theta \neq 0$ and $d_e = \theta_e$, it depends on $\log S$ (See eq. \eqref{Delta-C-lambda-0-theta-neq-0-d-e=theta-e}). It should be pointed out that for AdS black holes, one has $\Delta \mathcal{C}_V \propto S$ \cite{Chapman:2016hwi}. Therefore, the behavior of $\Delta \mathcal{C}_V$ for the case $\theta \neq 0$ is very intriguing. In particular, the logarithmic behavior for the case $d_e \neq \theta_e$, and it would be very interesting to further investigate it.
\\ Eventually, in ref. \cite{Carmi:2017jqz}, the effect of nonzero charge on volume-complexity $\mathcal{C}_V$ was explored for charged planar AdS black holes. It was observed that for a given temperature, by increasing the charge, the growth rate of $\mathcal{C}_V$ decreases. It would be interesting to investigate the effect of charge on generalized volume-complexity $\mathcal{C}_{\rm gen}$ by considering charged HV or AdS black holes.
%\\It would be interesting to investigate the effect of charge on generalized volume-complexity $\mathcal{C}_V$ by considering charged HV black branes. In ref. \cite{Carmi:2017jqz}, this effect on $\mathcal{C}_V$ was explored for charged planar AdS black holes. It was observed for a given temperature, by increasing the charge, the growth rate of $\mathcal{C}_V$ decreases.
% that having a non-zero charge, reduces the late time value of the growth rate of volume-complexity in comparison to that for uncharged AdS black branes.
Moreover, it would be also interesting to study different scalar functional $F_1$ of the background curvature as well as cases in which the functionals $F_1$ and $F_2$ are not equal to each other. 
%One-sided case.
Another interesting direction to pursue is to extend these calculations to the HV black hole solutions found in ref. \cite{Pedraza:2018eey} or Kerr-AdS black holes whose volume-complexity were recently studied in ref. \cite{Bernamonti:2021jyu}.

%%%%%%%%%%%%%%%%%%%%%%%%%%%%%%%%%%%%%%%%%%%
\section*{Acknowledgment}
%%%%%%%%%%%%%%%%%%%%%%%%%%%%%%%%%%%%%%%%%%%

We would like to thank Mohsen Alishahiha very much for his support and illuminating discussions during this work. 
We would also like to thank Mohsen Alishahiha, Robert Myers and Ali Naseh very much for their very valuable comments on the draft. We are also very grateful to Ghadir Jafari for helpful discussions. 
%This work is supported by Iran Science Elites Federation (ISEF).
%%We would like to thank Mohsen Alishahiha very much for his very helpful comments on the draft. We also would like to thank the referee for her/his helpful comments which enriched this work. 
The work of FO is supported by the School of Physics at IPM and Iran Science Elites Federation (ISEF). 
%%The work of the author is supported by a grant from ISEF.
%%We are also very grateful to M. R. Tanhayi for his helpful comments on the manuscript.
%
%%%%%%%%%%%%%%%%%%%%%%%%%%%%%%%%%%
\appendix
\addcontentsline{toc}{section}{Appendix}
\section*{Appendix}
\section{Connection between the Radial Cutoffs}
\label{Appendix A: Relating the Radial cutoffs}
%%%%%%%%%%%%%%%%%%%%%%%%%%%%%%%%%

In this section, we investigate the connection between the radial cutoffs $r_{\rm max}^{\rm BB}$ and $r_{\rm max}^{\rm Vac}$. To this end, we proceed in the same manner as Appendix A in ref. \cite{Chapman:2016hwi}, and work in the Fefferman-Graham (FG) coordinate. We assume that both the black brane geometry and the vacuum have the same radial cutoff at $z= \delta$ in the FG coordinate. We write the metric \eqref{metric-BB} in the FG coordinate
\bea
ds^2 = \frac{L^2}{z^2} \left( dz^2 + g_{ij} (z, x^i) dx^i dx^j \right),
\label{metric-FG}
\eea 
where $g_{ij} (z, x^i)$ is the boundary metric. For an asymptotically AdS spacetime the coordinates $z$ and $r$ are connected as follows
%%%%%%%%%%%%%%%%%%%%%%%%%%%%
\footnote{Notice that we work in $d+2$ dimensions. Moreover, for $z \rightarrow 0$, one has $r \rightarrow \infty$. Therefore, one can write an expansion for $z$ in inverse powers of $r$. }
%%%%%%%%%%%%%%%%%%%%%%%%%%%%
\bea
z = \frac{L^2}{r} + \frac{a_1}{r^2}+ \cdots + \frac{a_d}{r^{d+2}} + \frac{a_{d+2}}{r^{d+3}} + \cdots,
\label{z-r-FG}
\eea 
Therefore, at leading order one has $z= \frac{L^2}{r}$. Moreover, the first non-zero coefficient that is different between the black hole geometry and the vacuum is $a_{d}$ \cite{Chapman:2016hwi}. In the following, we first investigate how the above formula is changed for a HV black brane. Next, we find the connection between $r_{\rm max}^{\rm BB}$ and $r_{\rm max}^{\rm Vac}$. Asking for the radial part of the black brane metric in eqs. \eqref{metric-BB} and \eqref{metric-FG} to be equal to each other, one has
\bea
\frac{dz}{z} = - \frac{r_F^{\theta_e}}{r^{\theta_e+1} \sqrt{f(r)}} dr.
\label{dz-dr-FG}
\eea 
In the following, we consider the two cases $\theta=0$ and $\theta \neq 0$, separately.

%%%%%%%%%%%%%%%%%%%%%%%%
\subsection{$\theta=0$}
\label{Sec: theta-0}
%%%%%%%%%%%%%%%%%%%%%%%%

For very large $r$, one has
\bea
f(r) \approx 1,
\label{f(r)-leading}
\eea 
and hence by taking the integral of eq. \eqref{dz-dr-FG}, one obtains
\bea
\log z = - \log r + \tilde{c},
\eea 
or equivalently
\bea
z= \frac{L^2}{r}.
\label{z-r-leading-theta-0-1}
\eea 
It should be pointed out that we choose the integration constant such that at leading order, one has $z = \frac{L^2}{r}$ (See also \cite{Chapman:2016hwi}). Moreover, to find the subleading corrections to eq. \eqref{z-r-leading-theta-0-1}, one can expand $f(r)$ in powers of $r$ in eq. \eqref{dz-dr-FG} and taking the integrals on both sides. It is straightforward to see that
%\bea
%z= \frac{c}{r} + \frac{1}{r^{d+z+1}} + \cdots,
%\label{z-r-leading-theta-0-2}
%\eea 
%For $\theta=0$, one has
\bea
z = \frac{L^2}{r} +
% \frac{a_1}{r^2}+ \cdots + 
\frac{a_1}{r^{d+z+1}} + \frac{a_2}{r^{2(d+z)+ 1}} + \frac{a_3}{r^{3(d+z)+ 1}} + \cdots.
\label{z-r-FG-theta-0}
\eea 
Therefore, at $z= \delta$, one can write $\delta$ in terms of $r_{\rm max}$ as follows
\bea
\delta = \frac{L^2}{r_{\rm max}} + \frac{a_1}{r_{\rm max}^{d+z+1}} + \frac{a_2}{r_{\rm max}^{2(d+z)+ 1}} + \frac{a_3}{r_{\rm max}^{3(d+z)+ 1}}+ \cdots.
\eea 
Now one can invert the above formula and write $r_{\rm max}$ in terms of $\delta$
\bea
r_{\rm max} = \frac{L^2}{\delta} + \tilde{a_1} \delta^{d+z-1}+  \tilde{a_2} \delta^{2(d+z) -1} + \cdots.
\label{r-max-delta-theta-0}
\eea 
Next, we find the connection between $r_{\rm max}^{\rm BB}$ and $r_{\rm max}^{\rm Vac}$. By taking the integral on the left hand side of eq. \eqref{dz-dr-FG}, one has
\bea
\log \frac{\delta}{L^2} = \log \int^{\delta} \frac{dz}{z} = - \int^{r_{\rm max}} \frac{dr}{r \sqrt{f(r)}}.
\label{dz-dr-FG-theta-0-1}
\eea 
It should be pointed out that one can apply eq. \eqref{dz-dr-FG-theta-0-1}, both for the black brane and the vacuum geometries. Having this in mind that the radial cutoff $\delta$ in the FG coordinate is the same for both geometries, and subtracting eq. \eqref{dz-dr-FG-theta-0-1} for both of them, one obtains
\bea
0 =  \int^{r_{\rm max}^{\rm BB}} \frac{dr}{r \sqrt{f(r)}} -  \int^{r_{\rm max}^{\rm Vac}} \frac{dr}{r}.
\label{int-rmax-BB-rmax-Vac-1}
\eea 
Then, we assume that (See also \cite{Chapman:2016hwi})
\bea
r_{\rm max}^{BB} = r_{\rm max}^{\rm Vac} + \delta \; r_{\rm max}.
\label{r-max-r-Vac-delta}
\eea 
By plugging the above equation into the first integral in eq. \eqref{int-rmax-BB-rmax-Vac-1}, one can make the following approximation at leading order in $\delta$ and $r_{\rm max}^{\rm Vac}$
\bea
\int^{r_{\rm max}^{\rm BB}} \frac{dr}{r \sqrt{f(r)}} &=& \int^{r_{\rm max}^{\rm Vac} + \delta \; r_{\rm max}} \frac{dr}{r \sqrt{f(r)}} 
\cr && \cr 
&=&
\int^{r_{\rm max}^{\rm Vac}} \frac{dr}{r \sqrt{f(r)}} + \frac{1}{r \sqrt{ f(r)}} \biggr|_{r_{\rm max}^{\rm Vac} + \delta \; r_{\rm max}} \delta \; r_{\rm max}
\cr && \cr
&=&
\int^{r_{\rm max}^{\rm Vac}} \frac{dr}{r \sqrt{f(r)}} + \frac{1}{r_{\rm max}^{\rm Vac}} \left( r_{\rm max}^{\rm BB} - r_{\rm max}^{\rm Vac} \right),
\eea 
where we applied eq. \eqref{r-max-r-Vac-delta} in the last line. Next, by substituting the above equation into eq. \eqref{int-rmax-BB-rmax-Vac-1}, one arrives at
\bea
\frac{1}{r_{\rm max}^{\rm Vac}} \left( r_{\rm max}^{\rm BB} - r_{\rm max}^{\rm Vac} \right)
&=&
\int^{r_{\rm max}^{\rm Vac}} \left( 1- \frac{1}{\sqrt{f(r)}} \right) \frac{dr}{r} 
\cr && \cr 
&=&
- \frac{r_h^{d+z}}{2} \int^{r_{\rm max}} \frac{dr}{r^{d+z+1}} + \cdots 
\cr && \cr 
& \simeq & \frac{r_h^{d+z}}{2(d+z)} \left( \frac{1}{r_{\rm max}^{\rm Vac}} \right)^{d+z} + \cdots.
\eea 
From the above equation, one easily obtains
\bea
r_{\rm max}^{\rm BB} 
%\!\!\!\!  & = &  \!\!\!\! 
= r_{\rm max}^{\rm Vac} + \frac{r_h^{d+z}}{2(d+z)}  \left( \frac{1}{r_{\rm max}^{\rm Vac}} \right)^{d+z-1} + \cdots.
%\cr && \cr
%\!\!\!\!  & = &  \!\!\!\! 
%r_{\rm max}^{\rm Vac} + \frac{r_h^{d+z}}{2(d+z) L^{2(d+z-1)}} \delta^{d+z-1} + \cdots,
\label{r-max-BB-r-max-Vac-theta-0-1}
\eea 
%where in the last line we applied the fact that at leading order $r_{\rm max}^{\rm Vac} = \frac{L^2}{\delta}$. 
Next, we find the connection between $r_{\rm max}^{\rm Vac}$ and $\delta$. From eq. \eqref{dz-dr-FG-theta-0-1}, one can find
\bea
\log \frac{\delta}{L^2} = - \int^{r_{\rm max}^{\rm Vac}} \frac{dr}{r} = - \log r + \tilde{c},
\label{dz-dr-FG-theta-0-Vac}
\eea 
or equivalently,
\bea
r_{\rm max}^{\rm Vac} = \frac{L^2}{ \delta}.
\label{r-max-vac-delta}
\eea 
Therefore, by plugging eq. \eqref{r-max-vac-delta} into eq. \eqref{r-max-BB-r-max-Vac-theta-0-1}, one can write
\bea
r_{\rm max}^{\rm BB} = \frac{L^2}{ \delta} + \frac{r_h^{d+z}}{2(d+z) L^{2(d+z-1)}} \delta^{d+z-1} + \cdots.
\label{r-max-BB-r-max-Vac-theta-0-2}
\eea 
Notice that for $z=1$, it reduces to eq. (A.13) in ref. \cite{Chapman:2016hwi} for a planar AdS-Schwarzschild black hole in $d+2$ dimensions as it was expected.

%%%%%%%%%%%%%%%%%%%%%%%%
\subsection{$\theta \neq 0$}
\label{Sec: theta-neq-0}
%%%%%%%%%%%%%%%%%%%%%%%%

For very large $r$, one again has $f(r) \approx 1$, and hence
%\bea
%\log \frac{z}{L^2} = \frac{1}{\theta_e} \left( \frac{r_F}{r} \right)^{\theta_e}
%\label{z-r-leading-theta-neq-0}
%\eea 
\bea
\log \frac{\delta}{L^2} = \frac{1}{\theta_e} \left( \frac{r_F}{r_{\rm max}} \right)^{\theta_e},
\label{z-r-leading-theta-neq-0}
\eea 
or equivalently,
\bea
z= L^2 e^{\frac{1}{\theta_e} \left( \frac{r_F}{r_{\rm max}}\right)}.
\eea 
Notice that form eq. \eqref{z-r-leading-theta-neq-0} for $\theta>0$ when $r_{\rm max} \rightarrow \infty$, one has $\delta \rightarrow L^2$. Moreover, for $\theta<0$ when $r_{\rm max} \rightarrow \infty$, one has $\delta \rightarrow \infty$. These results are not consistent with the fact that for $r_{\rm max} \rightarrow \infty$, one must have $\delta \rightarrow 0$. To remedy the issue, we apply the following ansatz for the metric in the FG coordinate
\bea
ds^2 = \frac{L^2}{z^2} \left( \frac{z_F}{z} \right)^{-2 \theta_e} \left( dz^2 + g_{ij} (z, x^i) dx^i dx^j \right).
\label{metric-FG-theta-neq-0}
\eea 
Notice that for $\theta=0$, eq. \eqref{metric-FG-theta-neq-0} reduces to eq. \eqref{metric-FG}. Asking for the radial elements in eqs. \eqref{metric-BB} and \eqref{metric-FG-theta-neq-0} to be equal to each other, changes eq. \eqref{dz-dr-FG} as follows
\bea
 \frac{dz}{z_F^{\theta_e}\; z^{1- \theta_e}} = - \frac{r_F^{\theta_e}}{r^{ 1 + \theta_e} \sqrt{f(r)}} dr.
\label{dz-dr-FG-theta-neq-0-1}
\eea 
Next, by expanding the blackening factor $f(r)$ in powers of $r$ and taking the integrals on both sides of eq. \eqref{dz-dr-FG-theta-neq-0-1}, one arrives at
\bea
z^{\theta_e} &=& (r_F z_F)^{\theta_e} \Bigg[
\frac{1}{ r^{\theta_e}} + \frac{ \theta_e r_h^{d_e+z}}{2(d_e+z+ \theta_e)} \frac{1}{r^{d_e+z+ \theta_e}} + 
\frac{ 3 \theta_e r_h^{ 2 (d_e+z) }}{8 \big( 2(d_e+z) + \theta_e) \big)} \frac{1}{r^{2 (d_e+z) + \theta_e}} 
\cr && \cr
&& 
\;\;\;\;\;\;\;\;\;\;\;\;\;\;\;\;\;\;\;\;\;\;\;\;\;\;\;  + \frac{ 5 \theta_e r_h^{ 3 (d_e+z) }}{16 \big( 3(d_e+z) + \theta_e) \big)} \frac{1}{r^{3 (d_e+z) + \theta_e}}
+ \cdots 
\Bigg],
\label{dz-dr-FG-theta-neq-0-2}
\eea 
which can be recast in the following form 
\bea
z = \frac{L^2}{r} +
% \frac{a_1}{r^2}+ \cdots + 
\frac{a_1}{r^{d_e+z+1}} + \frac{a_2}{r^{2(d_e+z)+ 1}} + \frac{a_3}{r^{3(d_e+z)+ 1}} + \cdots.
\label{z-r-FG-theta-neq-0}
\eea 
It should be pointed out that in eq. \eqref{dz-dr-FG-theta-neq-0-2} we set the integration constant to zero. Moreover, we assume that $z_F = \frac{L^2}{r_F}$. Therefore, at $z= \delta$, one can write $\delta$ in terms of $r_{\rm max}$ as follows
\bea
\delta = \frac{L^2}{r_{\rm max}} + \frac{a_1}{r_{\rm max}^{d_e+z+1}} + \frac{a_2}{r_{\rm max}^{2(d_e+z)+ 1}} + \frac{a_3}{r_{\rm max}^{3(d_e+z)+ 1}}+ \cdots.
\eea 
Now, one can invert the above formula and write $r_{\rm max}$ in terms of $\delta$ as follows
\bea
r_{\rm max} = \frac{L^2}{\delta} + \tilde{a_1} \delta^{d_e+z-1}+  \tilde{a_2} \delta^{2(d_e+z) -1} + \cdots.
\label{r-max-delta-theta-neq-0}
\eea 
Then, we find the connection between $r_{\rm max}^{\rm BB}$ and $r_{\rm max}^{\rm Vac}$. By taking the integral on the left hand side of eq. \eqref{dz-dr-FG-theta-neq-0-1}, one has
\bea
\frac{1}{\theta_e} \left( \frac{\delta}{z_F} \right)^{\theta_e} = \frac{1}{z_F^{\theta_e}} \int^{\delta} \frac{dz}{z^{1- \theta_e}} = - r_F^{\theta_e} \int^{r_{\rm max}} \frac{dr}{r^{1+ \theta_e} \sqrt{f(r)}}.
\label{int-dz-dr-FG-theta-neq-0}
\eea 
As mentioned before, the cutoff $\delta$ in the FG coordinate is the same both for the black brane and the vacuum geometries. Therefore, by subtracting eq. \eqref{int-dz-dr-FG-theta-neq-0} for both geometries, one obtains
\bea
0 =  \int^{r_{\rm max}^{\rm BB}} \frac{dr}{r^{1+ \theta_e} \sqrt{f(r)}} -  \int^{r_{\rm max}^{\rm Vac}} \frac{dr}{r^{1 + \theta_e}}.
\label{int-rmax-BB-rmax-Vac-theta-neq-0-1}
\eea
Next, by plugging eq. \eqref{r-max-r-Vac-delta} into the first integral in eq. \eqref{int-rmax-BB-rmax-Vac-theta-neq-0-1}, one can make the following approximation to the integral at leading order in $\delta$ and $r_{\rm max}^{\rm Vac}$
\bea
\int^{r_{\rm max}^{\rm BB}} \frac{dr}{r^{1 + \theta_e} \sqrt{f(r)}} &=&
%\int^{r_{\rm max}^{\rm Vac} + \delta \; r_{\rm max}} \frac{dr}{r \sqrt{f(r)}} 
%\cr && \cr 
%\!\!\!\! & \simeq & \!\!\!\! 
\int^{r_{\rm max}^{\rm Vac}} \frac{dr}{r^{1 + \theta_e} \sqrt{f(r)}} + r_{\rm max} \; \delta \; \left( \frac{1}{r^{1 + \theta_e} \sqrt{ f(r)}} \right) \biggr|_{r_{\rm max}^{\rm Vac} + \delta \; r_{\rm max}} 
\cr && \cr
&=&
\int^{r_{\rm max}^{\rm Vac}} \frac{dr}{r^{1 + \theta_e} \sqrt{f(r)}} + \frac{1}{ \left( r_{\rm max}^{\rm Vac} \right)^{1 + \theta_e}} \left( r_{\rm max}^{\rm BB} - r_{\rm max}^{\rm Vac} \right).
\eea 
%In the last line, we applied eq. \eqref{r-max-r-Vac-delta}. 
Then, by substituting the above equation into eq. \eqref{int-rmax-BB-rmax-Vac-theta-neq-0-1}, and expanding the blackening factor in powers of $r$, one has
\bea
\frac{1}{ \left( r_{\rm max}^{\rm Vac} \right)^{1 + \theta_e} } \left( r_{\rm max}^{\rm BB} - r_{\rm max}^{\rm Vac} \right)
&\simeq& 
\int^{r_{\rm max}^{\rm Vac}} \left( 1- \frac{1}{\sqrt{f(r)}} \right) \frac{dr}{r^{1 + \theta_e}} 
\cr && \cr 
&=&
- \frac{r_h^{d_e+z}}{2} \int^{r_{\rm max}} \frac{dr}{r^{d_e+z+ \theta_e+1}} + \cdots 
\cr && \cr 
&=& \frac{r_h^{d_e+z}}{2(d_e+z+ \theta_e)} \left( \frac{1}{r_{\rm max}^{\rm Vac}} \right)^{d_e+z+ \theta_e} + \cdots.
\eea 
From the above equation, one easily obtains
\bea
r_{\rm max}^{\rm BB} 
%\!\!\!\!  & = &  \!\!\!\! 
= r_{\rm max}^{\rm Vac} + \frac{r_h^{d_e+z}}{2(d_e+z+ \theta_e)}  \left( \frac{1}{r_{\rm max}^{\rm Vac}} \right)^{d_e+z-1} + \cdots.
%\cr && \cr
%\!\!\!\!  & = &  \!\!\!\! 
%r_{\rm max}^{\rm Vac} + \frac{r_h^{d+z}}{2(d+z) L^{2(d+z-1)}} \delta^{d+z-1} + \cdots,
\label{r-max-BB-r-max-Vac-theta-neq-0-1}
\eea 
Now, we find the connection between $\delta$ and $r_{\rm max}^{\rm Vac}$. For the vacuum one has $f(r)=1$, and from eq. \eqref{int-dz-dr-FG-theta-neq-0}, one arrive at
\bea
\frac{1}{\theta_e} \left( \frac{\delta}{z_F} \right)^{\theta_e} = \frac{1}{z_F^{\theta_e}} \int^{\delta} \frac{dz}{z^{1- \theta_e}} = - r_F^{\theta_e} \int^{r_{\rm max}^{\rm Vac}} \frac{dr}{r^{1+ \theta_e}}.
\label{dz-dr-FG-theta-neq-0-vac}
\eea 
By taking the integral on the right hand side of the above equation, one finds
\bea
\delta = \frac{r_F z_F}{r_{\rm max}^{\rm Vac}},
\label{delta-r-max-theta-neq-0-1}
\eea 
where we set the integration constant to zero again. Moreover, by setting $z_F = \frac{L^2}{r_F}$, one can rewrite eq. \eqref{delta-r-max-theta-neq-0-1} as follows
\bea
r_{\rm max}^{\rm Vac} = \frac{L^2}{\delta}.
\label{delta-r-max-theta-neq-0-2}
\eea 
At the end, by plugging eq. \eqref{delta-r-max-theta-neq-0-2} into eq. \eqref{r-max-BB-r-max-Vac-theta-neq-0-1}, one has
\bea
r_{\rm max}^{\rm BB} 
%\!\!\!\!  & = &  \!\!\!\! 
= \frac{L^2}{\delta} + \frac{r_h^{d_e+z}}{2(d_e+z+ \theta_e) L^{2 (de+z-1)}} \delta^{d_e+z -1} + \cdots.
%\cr && \cr
%\!\!\!\!  & = &  \!\!\!\! 
%r_{\rm max}^{\rm Vac} + \frac{r_h^{d+z}}{2(d+z) L^{2(d+z-1)}} \delta^{d+z-1} + \cdots,
\label{r-max-BB-r-max-Vac-theta-neq-0-2}
\eea 

%%%%%%%%%%%%%%%%%%%%%%%%%%%%%%%%%
\section{Other Higher Derivative Functionals $F_{1,2} (g_{\mu \nu} ; X^{\mu} ( \sigma))$}
\label{Appendix B: Other Higher Derivative Functionals}
%%%%%%%%%%%%%%%%%%%%%%%%%%%%%%%%%

\begin{figure}
	\begin{center}
		\includegraphics[scale=0.31]{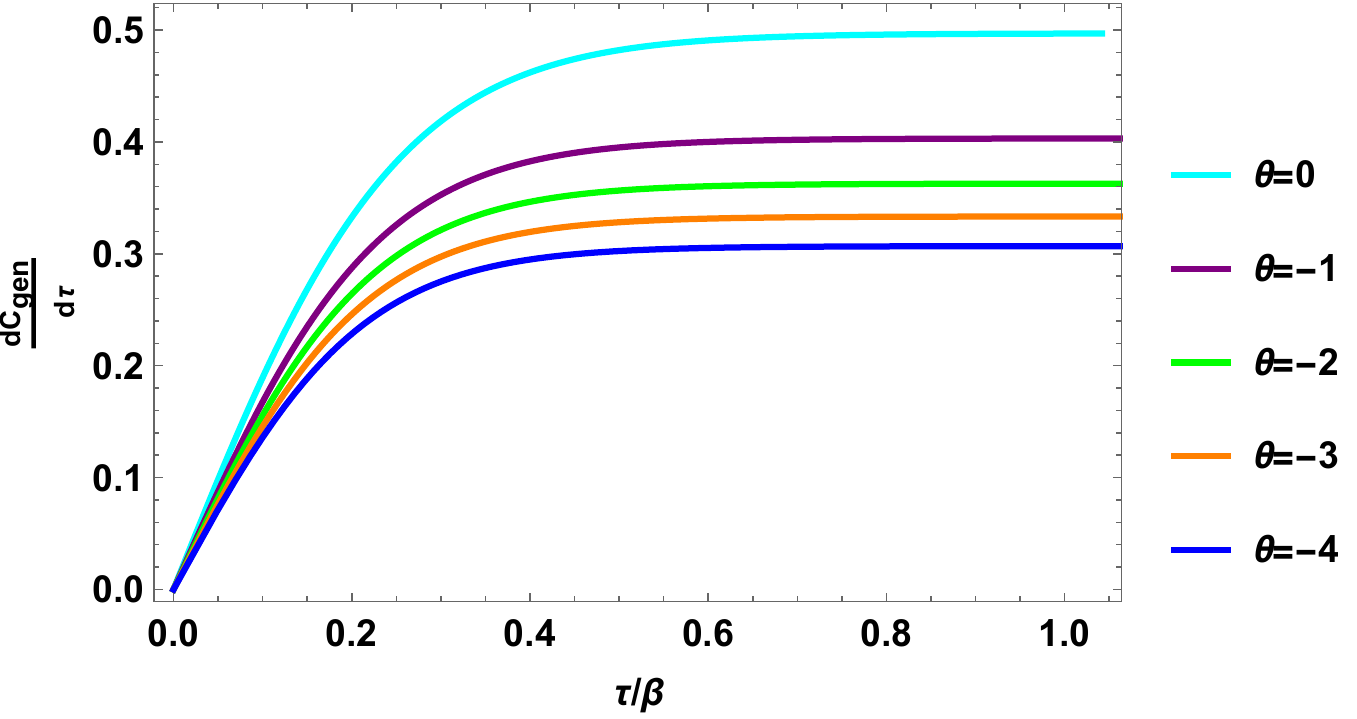}
		\hspace{0.5cm}
		\includegraphics[scale=0.31]{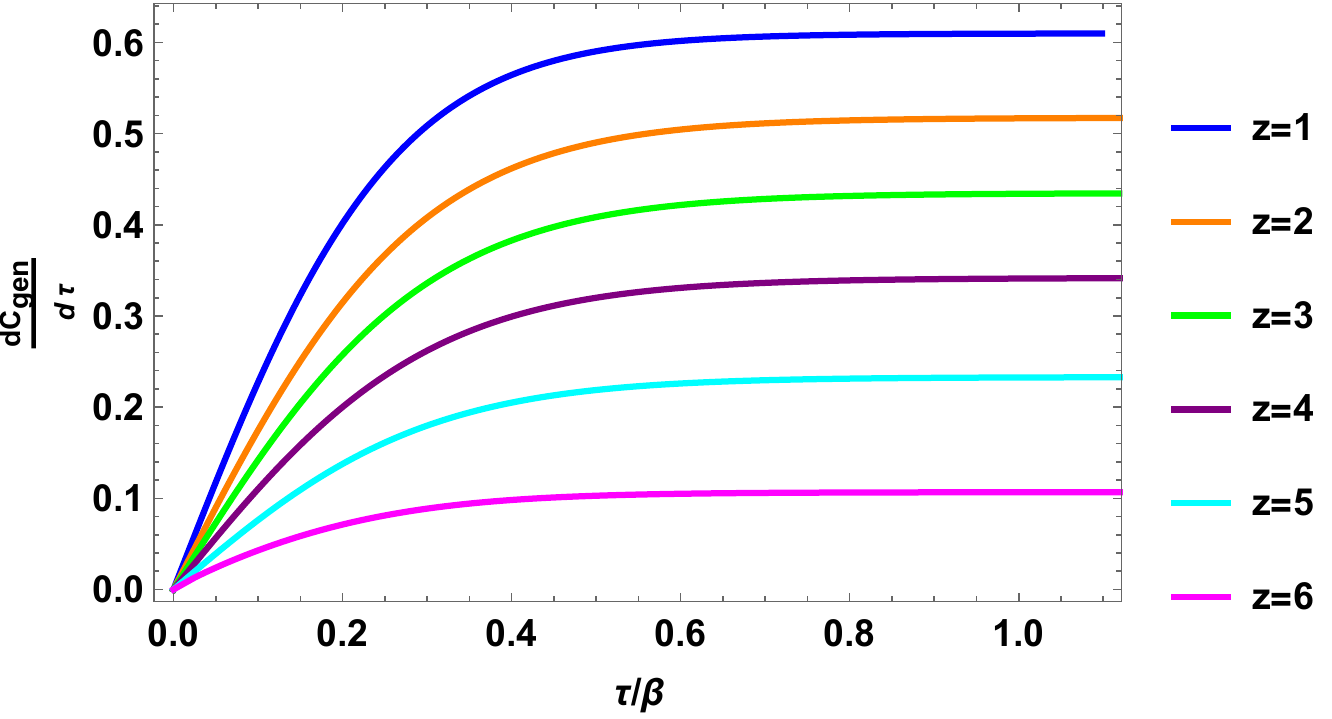}
	\end{center}
	\caption{The growth rate $ \frac{d \mathcal{C}_{\rm gen}}{d \tau}$ as a function of $\tau / \beta$ for $a(r)= 1 + \lambda_1 L^4 R$  and different values of $d$, $\theta$ and $z$:
		{\it Left}) $d=1$, $z=1$ and $\lambda_1= 10^{-3}$.
		{\it Right}) $d=2$, $\theta=1$ and $\lambda_1= 10^{-2}$.
%		{\it Down Left}) $d=4$ and $\theta=0$ and $\lambda= 10^{-5}$.
%		{\it Down Right}) $d=5$ and $\theta=-1$ and $\lambda= 10^{-5}$.
		%	 We also plotted the case $\lambda =0$, i.e. the volume complexity, which is indicated by the dashed black curve. 
%		We set $ L= r_h =r_F= G_N= V_d= 1$.
	}
	\label{fig: dCgen-dtau-lambda-theta-R}
\end{figure}

\begin{figure}
	\begin{center}
		\includegraphics[scale=0.31]{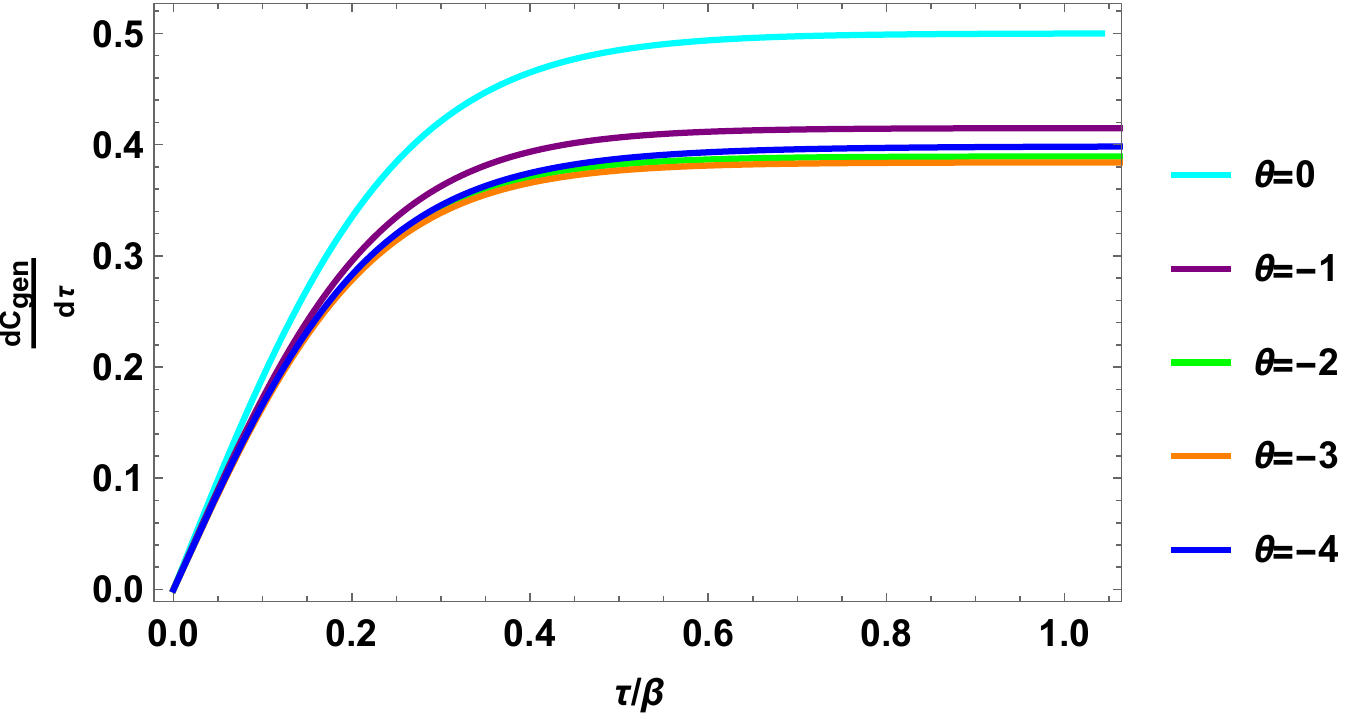}
		\hspace{0.5cm}
		\includegraphics[scale=0.31]{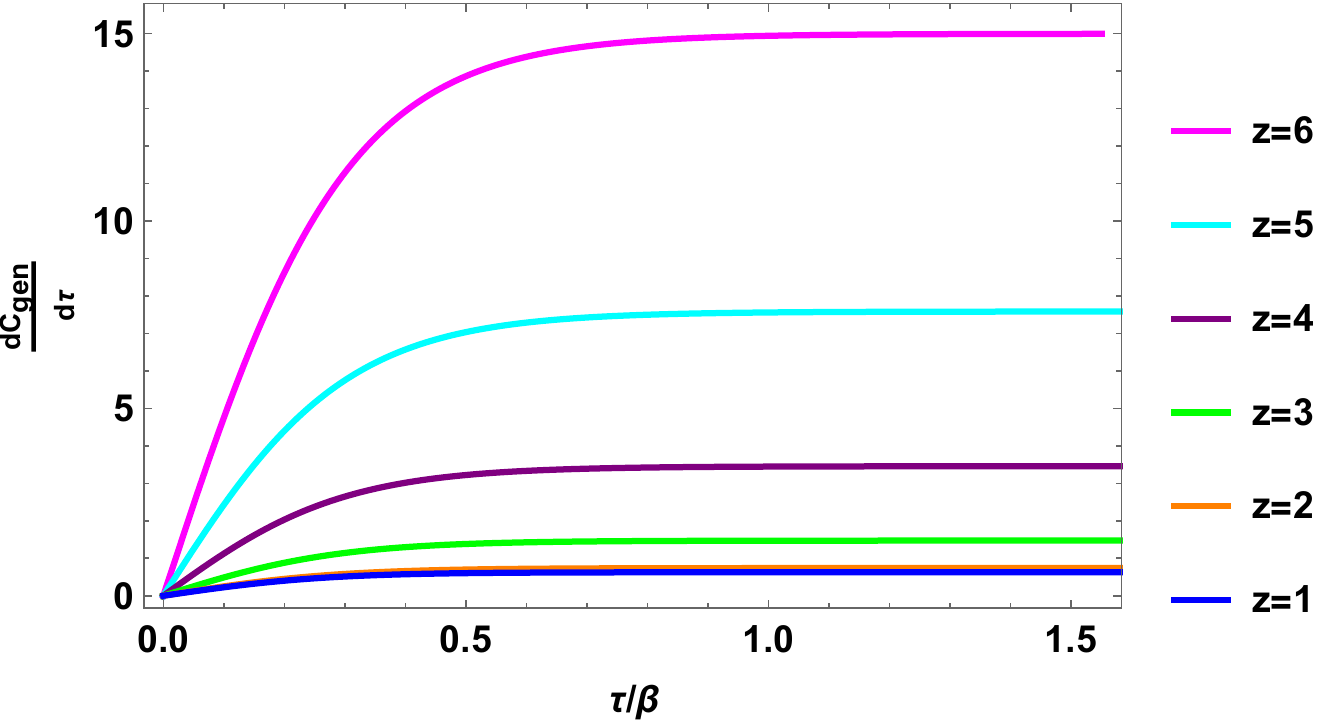}
		%		\\
		%		\includegraphics[scale=0.31]{dCgen-dtau-lambda-neq-0-d-4-theta-0-z.pdf}
		%		\hspace{0.5cm} 
		%		\includegraphics[scale=0.31]{dCgen-dtau-lambda-neq-0-d-5-theta-m1-z.pdf}
	\end{center}
	\caption{ The growth rate $ \frac{d \mathcal{C}_{\rm gen}}{d \tau}$ as a function of $\tau / \beta$ for $a(r)= 1 + \lambda_2 L^4 R_{\mu \nu} R^{\mu \nu }$ and different values of $d$, $\theta$ and $z$:
		{\it Left}) $d=1$, $z=1$ and $\lambda_2= 10^{-5}$.
		{\it Right}) $d=2$, $\theta=1$ and $\lambda_2= 10^{-2}$.
%		{\it Down Left}) $d=4$ and $\theta=0$ and $\lambda= 10^{-5}$.
%		{\it Down Right}) $d=5$ and $\theta=-1$ and $\lambda= 10^{-5}$.
		%	 We also plotted the case $\lambda =0$, i.e. the volume complexity, which is indicated by the dashed black curve. 
%		We set $ L= r_h =r_F= G_N= V_d= 1$.
	}
	\label{fig: dCgen-dtau-lambda-theta-Ric}
\end{figure}

\begin{figure}
	\begin{center}
		\includegraphics[scale=0.31]{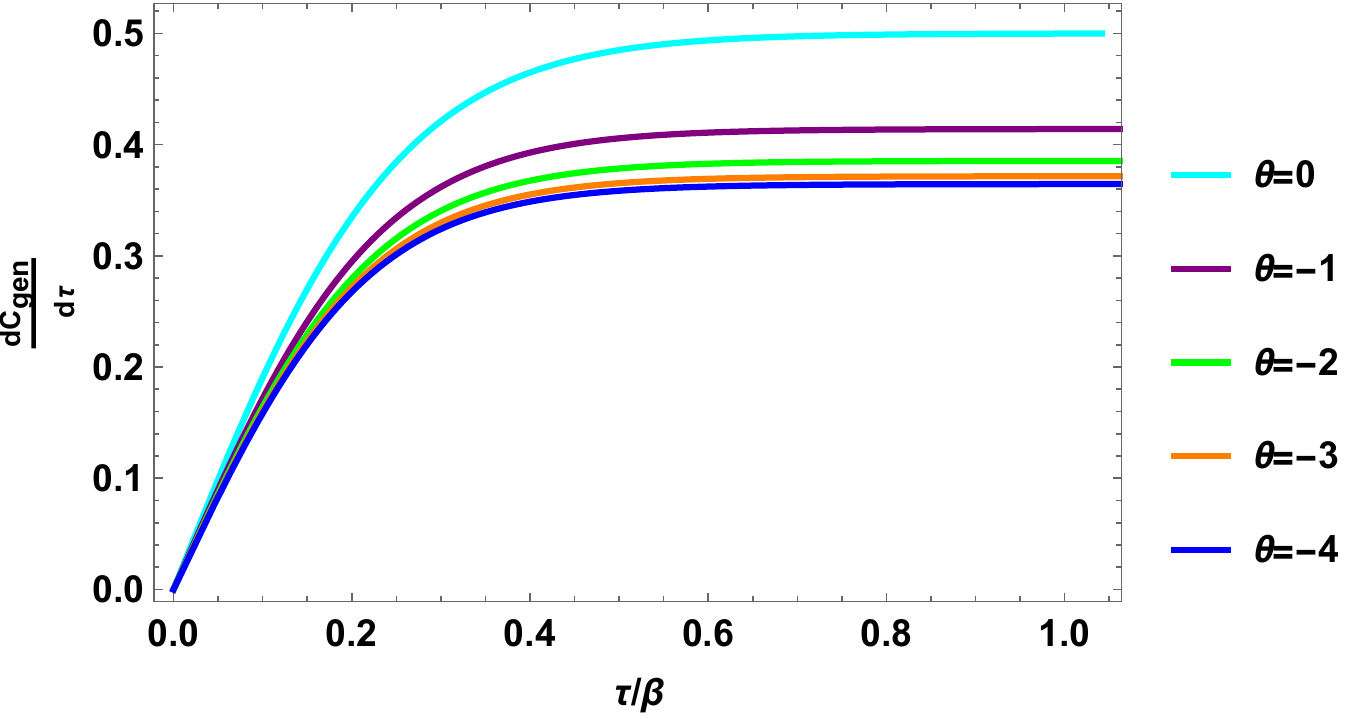}
		\hspace{0.5cm}
		\includegraphics[scale=0.31]{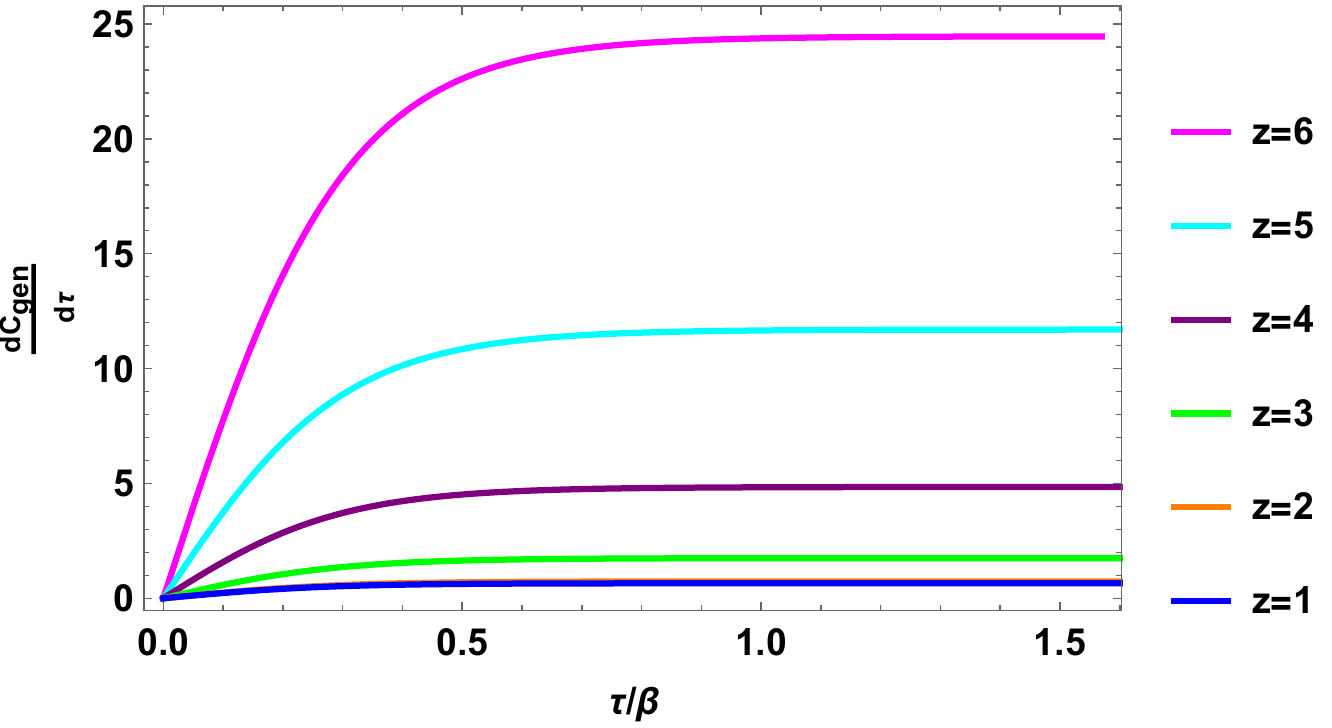}
		%		\\
		%		\includegraphics[scale=0.31]{dCgen-dtau-lambda-neq-0-d-4-theta-0-z.pdf}
		%		\hspace{0.5cm} 
		%		\includegraphics[scale=0.31]{dCgen-dtau-lambda-neq-0-d-5-theta-m1-z.pdf}
	\end{center}
	\caption{ The growth rate $ \frac{d \mathcal{C}_{\rm gen}}{d \tau}$ as a function of $\tau / \beta$ for $a(r)= 1 + \lambda_3 L^4 R_{\mu \nu \rho \sigma} R^{\mu \nu \rho \sigma}$ and different values of $d$, $\theta$ and $z$:
		{\it Left}) $d=1$, $z=1$ and $\lambda_3= 10^{-6}$.
		{\it Right}) $d=2$, $\theta=1$ and $\lambda_3= 10^{-2}$.
		%		{\it Down Left}) $d=4$ and $\theta=0$ and $\lambda= 10^{-5}$.
		%		{\it Down Right}) $d=5$ and $\theta=-1$ and $\lambda= 10^{-5}$.
		%	 We also plotted the case $\lambda =0$, i.e. the volume complexity, which is indicated by the dashed black curve. 
%		We set $ L= r_h =r_F= G_N= V_d= 1$.
	}
	\label{fig: dCgen-dtau-lambda-theta-Riemann}
\end{figure}

\begin{figure}
	\begin{center}
		\includegraphics[scale=0.31]{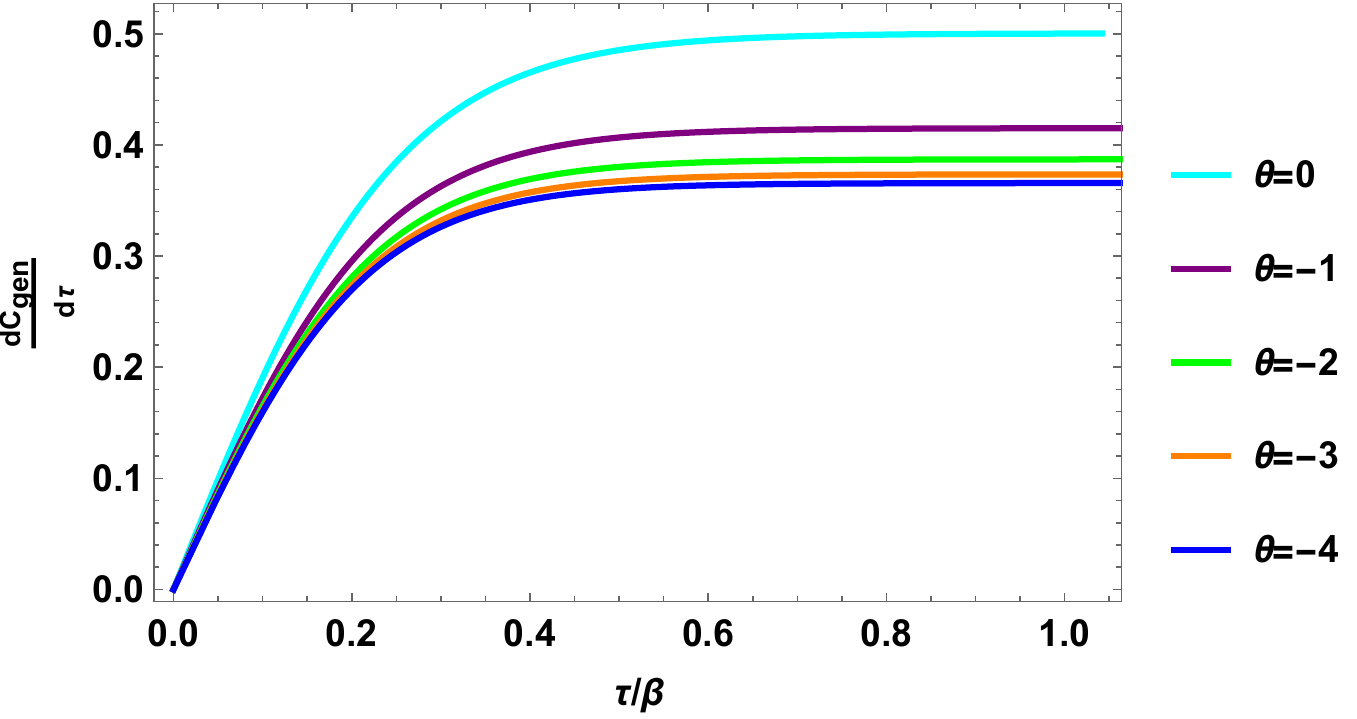}
		\hspace{0.5cm}
		\includegraphics[scale=0.31]{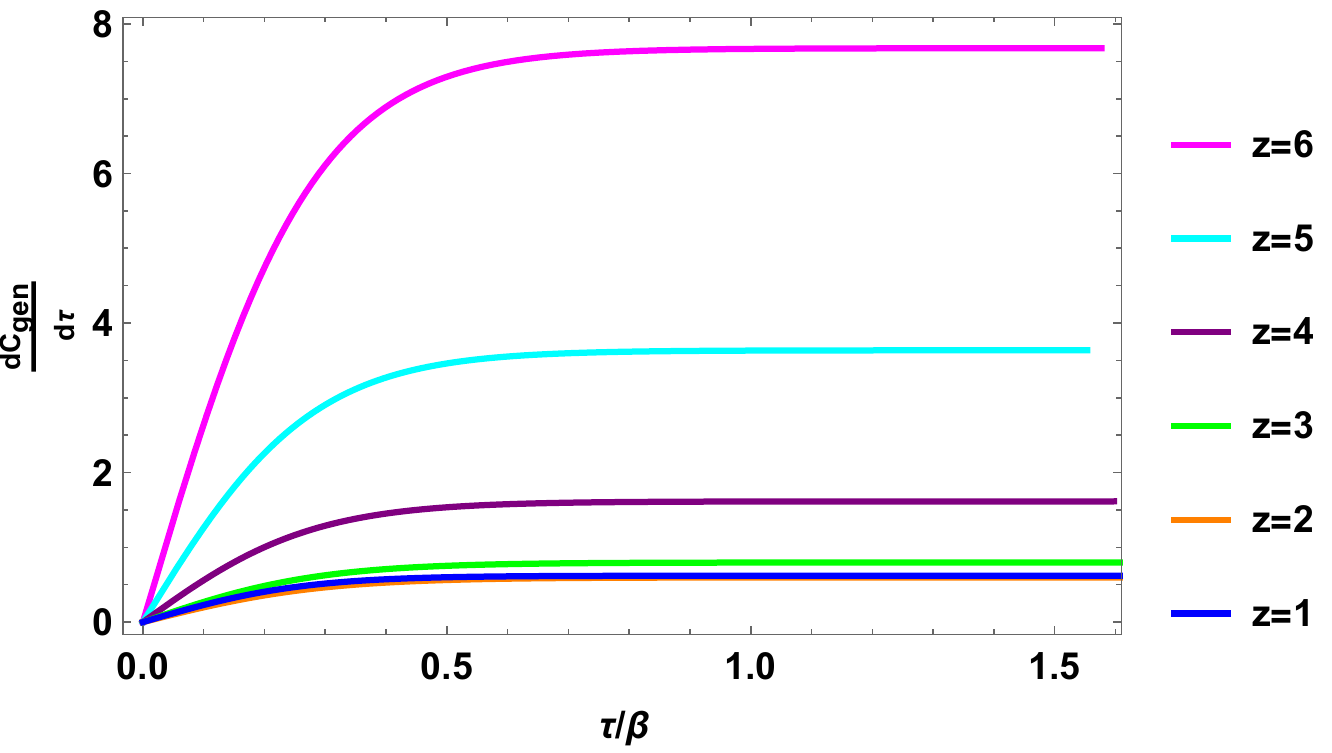}
		%		\\
		%		\includegraphics[scale=0.31]{dCgen-dtau-lambda-neq-0-d-4-theta-0-z.pdf}
		%		\hspace{0.5cm} 
		%		\includegraphics[scale=0.31]{dCgen-dtau-lambda-neq-0-d-5-theta-m1-z.pdf}
	\end{center}
	\caption{ The growth rate $ \frac{d \mathcal{C}_{\rm gen}}{d \tau}$ as a function of $\tau / \beta$ for $a(r)= 1 + \lambda_1 R + \lambda_2 R_{\mu \nu} R^{\mu \nu} + \lambda_3 L^4 R_{\mu \nu \rho \sigma} R^{\mu \nu \rho \sigma}$ and different values of $d$, $\theta$ and $z$:
		{\it Left}) $d=1$, $z=1$, $\lambda_1 = - 10^{-4} $, $\lambda_2= 10^{-5}$ and $\lambda_3= -10^{-5}$.
		{\it Right}) $d=2$, $\theta=1$, $\lambda_1 = - 10^{-2}$, $\lambda_2= - 10^{-2}$ and $\lambda_3= 10^{-2}$.
		%		{\it Down Left}) $d=4$ and $\theta=0$ and $\lambda= 10^{-5}$.
		%		{\it Down Right}) $d=5$ and $\theta=-1$ and $\lambda= 10^{-5}$.
		%	 We also plotted the case $\lambda =0$, i.e. the volume complexity, which is indicated by the dashed black curve. 
	%	We set $ L= r_h =r_F= G_N= V_d= 1$.
	}
	\label{fig: dCgen-dtau-lambda-theta-R-Ric-Riemann}
\end{figure}

In this appendix, we briefly examine other extensions of the generalized volume-complexity where one has
%%%%%%%%%%%%%%%%%%%%%%%%%%%%%
\footnote{We would like to thank the referee very much for her/his helpful suggestions to do these calculations. }
%%%%%%%%%%%%%%%%%%%%%%%%%%%%%
\bea
F_1 (g_{\mu \nu} ; X^{\mu} ( \sigma)) & = & F_2 (g_{\mu \nu} ; X^{\mu} ( \sigma)) =
\cr && \cr 
a(r)  & = & 1 + L^4 \left( \lambda_1 R + \lambda_2 R_{ \mu \nu} R^{ \mu \nu} + \lambda_3 R_{ \mu \nu \rho \sigma} R^{ \mu \nu \rho \sigma} \right).
\label{F1-F2-R2-Ric2-Riemann2}
\eea 
In this case, eqs. \eqref{C-gen-HV} to \eqref{C-2} and eqs. \eqref{C-gen-HV-1} to \eqref{dC-dtau-expansion} are still valid. However, one needs to replace the old $a(r)$ in eq. \eqref{ar} with the new one in eq. \eqref{F1-F2-R2-Ric2-Riemann2}. It is also evident that in this case the effective potential is also changed. Since finding the closed form of $\hat{U}(w)$ for arbitrary values of $d, \theta$ and $z$ is very complicated, we do not write it down here. However, it is easy to find it for each value of these parameters. In figures \ref{fig: dCgen-dtau-lambda-theta-R}, \ref{fig: dCgen-dtau-lambda-theta-Ric} and \ref{fig: dCgen-dtau-lambda-theta-Riemann}
%%%%%%%%%%%%%%%%%%%%%%%%%%
\footnote{In these figure, we set $ L= r_h =r_F= G_N= V_d= 1$.}
%%%%%%%%%%%%%%%%%%%%%%%%%%
we plotted numerically the growth rate of $\mathcal{C}_{\rm gen}$ for the following cases, respectively
\begin{itemize}
	\item $\lambda_1 \neq 0$ and $\lambda_{2,3}=0$
	\item $\lambda_2 \neq 0$ and $\lambda_{1,3}=0$
	\item $\lambda_3 \neq 0$ and $\lambda_{1,2}=0$.
\end{itemize}
%%%%%%%%%%%%%%%%%%%%%%%%%%%%%%%%%
%%%%%%%%%%%%%%%%%%%%%%%%%
%%%%%%%%%%%%%%%%%%%%%%%%%%%%%%%%%
%%%%%%%%%%%%%%%%%%%%%%%%%
%%%%%%%%%%%%%%%%%%%%%%%%%%%%%%%%%
%%%%%%%%%%%%%%%%%%%%%%%%%
Moreover, in Figure \ref{fig: dCgen-dtau-lambda-theta-R-Ric-Riemann}, we plotted the growth rate for the case $\lambda_{1,2,3} \neq 0$. In all figures, it is evident that the growth rate of $\mathcal{C}_{\rm gen}$ at late times is linear in time, if one chooses $\lambda_{1,2,3}$ such that $\hat{U}(w)$ has a maximum inside the horizon. Therefore, all of the corrections to volume-complexity in eq. \eqref{F1-F2-R2-Ric2-Riemann2} behave universally at late times.
%%%%%%%%%%%%%%%%%%%%%%%%%%%%%%%%%
%%%%%%%%%%%%%%%%%%%%%%%%%

%%%%%%%%%%%%%%%%%%%%%%%%%%%%
%\begin{table}
%	\begin{center}
%		\scalebox{0.9}{
%			\begin{tabular}{| l | l | l | l | l | l |}
%				\hline
%				& $d=2$ & $d=3$ & $d=4$ & $d=5$ 
%				\\ \hline
%				$z=1$ & $  \lambda >  \frac{1}{12}  \lambda_{crt,2} $ & $  \lambda >  \frac{1}{72} \lambda_{crt,2} $ & $   \lambda >  \frac{1}{240} \lambda_{crt,2} $ & $  \lambda >  \frac{1}{600} \lambda_{crt,2}$
%				\\ \hline
%				$ z=2$ & There is no $\lambda$ & $\lambda > \frac{655 + 23 \sqrt{805}}{ 5300} $ & $ \lambda >  \frac{2777 + 103 \sqrt{721}}{100224}$ & $ \lambda > \frac{27947 + 323 \sqrt{742s9}}{2655408}$
%				\\ \hline
%				$ z=3$ & There is no $\lambda$ & There is no $\lambda$ & There is no $\lambda$ & $ \lambda > \frac{11391 + 381 \sqrt{889}}{418816}$
%				\\ \hline
%				$ z=4$ & There is no $\lambda$ & There is no $\lambda$ & There is no $\lambda$ & There is no $\lambda$
%				\\ \hline
%				$ z=5$ & There is no $\lambda$ & There is no $\lambda$ & There is no $\lambda$ & There is no $\lambda$
%				\\ \hline
%				$ z=6$ & There is no $\lambda$ & There is no $\lambda$ & There is no $\lambda$ & There is no $\lambda$
%				\\ \hline
%		\end{tabular} }
%		\caption{The appropriate ranges of the coupling constant $\lambda_1$ for which $w_{fB}$ is a maximum inside the horizon.}
%		\label{Table: Max-lambda1}
%	\end{center}
%\end{table}
%%%%%%%%%%%%%%%%%%%%%%%%%%%%

%%%%%%%%%%%%%%%%%%%%%%%%%%%%%%%%%

%%%%%%%%%%%%%%%%%%%%%%%%%%%%%

%%%%%%%%%%%%%%%%%%%%%%%%%%%%%
\end{document}